\documentclass[11pt]{report}
 
\usepackage[margin=1in]{geometry} 
\usepackage{amsmath,amsthm,amssymb}
\usepackage{braket}
\usepackage{dsfont}
\usepackage{hyperref}
\usepackage{bbm}
\usepackage{esint}
\usepackage{undertilde}
\usepackage{tikz}
\usetikzlibrary{shapes,positioning,intersections,quotes,decorations.markings}
\usepackage{tkz-euclide}
\usepackage{titling}
\usepackage{bm}
\usepackage{titlesec}
\usepackage{physics}
\usepackage [autostyle, english=british]{csquotes}
\usepackage[toc,page]{appendix}

\MakeOuterQuote{"}

\titleformat{\subsection}[runin]
  {\normalfont\large\bfseries}{\thesubsection}{1em}{}
  
\titleformat{\subsubsection}[runin]
  {\normalfont\large\bfseries}{\thesubsection}{1em}{} 

\allowdisplaybreaks

\makeatletter
\renewcommand\chapter{\@startsection {chapter}{0}{\z@}%
                                   {-4.5ex \@plus -1ex \@minus -.2ex}%
                                   {3.3ex \@plus.2ex}%
                                   {\normalfont\LARGE\bfseries}}

\newenvironment{remark}[2][Remark]{\begin{trivlist}
\item[\hskip \labelsep {\bfseries #1}\hskip \labelsep {\bfseries #2.}]}{\end{trivlist}}                                 

\newcommand{\A}{\mathcal{A}}
\newcommand{\B}{\mathcal{B}}
\newcommand{\D}{\mathcal{D}}
\newcommand{\F}{\mathcal{F}}
\newcommand{\HH}{\mathcal{H}}
\newcommand{\Lag}{\mathcal{L}}
\newcommand{\M}{\mathcal{M}}
\newcommand{\N}{\mathcal{N}}
\newcommand{\R}{\mathcal{R}}
\newcommand{\del}{\partial}
\newcommand{\bigO}{\mathcal{O}}
\newcommand{\arctanh}{\text{arctanh}}
\newcommand{\im}{\text{im}}
\newcommand{\diag}{\text{diag}}

\begin{document}

\begin{titlepage}
\title{The Exact Low Energy Action of $\N=2$ SYM via Seiberg-Witten Theory and Localisation\\[1cm]}
\author{Robert Pryor\\[1cm]
School of Mathematics and Statisitcs\\[0.25cm]
University of Melbourne\\[0.5cm]
Submitted for the degree of Master of Science\\[0.5cm]}
\date{May 2020}
\end{titlepage}

\begin{minipage}{\textwidth}
\maketitle
\begin{abstract}
In the mid $1990$'s Seiberg and Witten determined the exact low energy effective action of $\N=2$ supersymmetric Yang-Mills theory with gauge group $SU(2)$. Later, in the early $2000$'s Nekrasov calculated this action directly using localisation techniques. This work serves as an introduction to the area, developing both approaches and reconciling their results.
\end{abstract}
\end{minipage}

\pagebreak

\pagenumbering{roman}

\chapter*{Acknowledgements}

I would like to thank my supervisor, the late Professor Omar Foda, for his encouragement, insight and patience. I consider myself very lucky to have had the opportunity to work with someone who cared so much for his field and the people in it. Without his guidance I would have remained unaware of the fascinating world of supersymmetry.

I would also like to thank Doctor Thomas Quella for his support and assistance in the final stage of preparing my thesis. In particular I appreciate both his readiness to review my thesis and his offer to put me in contact with others willing to do the same. Finishing this thesis would have been much more difficult without his help.
 
I would also like to thank Doctor Johanna Knapp and Doctor David Ridout for taking the time to help review my thesis. Their support, insight and willingness to help was invaluable. Their constructive criticism has undoubtedly led to a better finished work than would have otherwise been possible.

Finally, I would like to thank my partner, Gaia, for supporting my choice to pursue mathematics, and without whom none of this would have been possible.
\pagebreak

\tableofcontents{}

\pagebreak

\pagenumbering{arabic}

\chapter*{Introduction}
\addcontentsline{toc}{chapter}{Introduction}
This thesis aims to introduce the fundamentals of supersymmetric quantum field theory and in particular four dimensional $\N=2$ supersymmetric Yang-Mills theory ($\N=2$ SYM). In particular we will find that the low energy effective action of $\N=2$ SYM is completely determined by a certain holomorphic function called the prepotential. Seiberg-Witten theory will be used to calculate the prepotential for gauge group $SU(2)$ \cite{SW:1}. The prepotential will then be calculated again using the localisation techniques of Nekrasov, this time for gauge group $SU(N)$ \cite{Nekrasov:1}.
\section{The Seiberg-Witten approach.}
The Seiberg-Witten approach proceeds by identifying and imposing certain consistency conditions on the form of the prepotential. More precisely, we examine the moduli space of vacua, $\M$ of $\N=2$ SYM and the metric on this space. It turns out that the prepotential can be obtained from this metric \cite{Bilal:1}.

The consistency conditions are obtained by considering certain singularities located in regions of weak and strong coupling. The weak coupling singularity can be analysed directly, while the strong coupling singularities can be understood using Seiberg-Witten duality. This duality maps strongly coupled regions of $\N=2$ SYM to weakly coupled regions and provides a dual set of coordinates on $\M$ which are valid where the original set are not \cite{SW:1}.

Examining the monodromy properties of the coordinates about the singularities of $\M$ proves to be sufficient to obtain the prepotential \cite{SW:1}. We discuss two different ways to do this.

The first method proceeds by identifying the coordinates on $\M$ with solutions to a certain differential equation which is fixed uniquely by the monodromy properties.

The second method is the original approach of Seiberg and Witten \cite{SW:1}. It involves identifying points on $\M$ with certain elliptic curves. The period integrals of these curves are then naturally identified with the coordinates on $\M$. Both these approaches yield the $SU(2)$ prepotential in terms of hypergeometric functions.
\section{The Nekrasov approach.}
The Nekrasov approach uses localisation techniques to construct the instanton partition function, $Z_{\text{inst}}$ explicitly. The prepotential can then be extracted from $Z_{\text{inst}}$ by applying localisation techniques a second time \cite{Nekrasov:1}.

The instanton partition function is defined in terms of a path integral which we will refer to as the partition function path integral. To obtain $Z_{\text{inst}}$ explicitly, we localise this path integral to the instanton moduli space. It turns out that the $k$-instanton moduli space is finite dimensional and can be produced explicitly by the so-called ADHM construction \cite{CG:1}. This procedure yields $Z_{inst}$ as a power series in the vacuum expectation values of certain scalar fields, where each term is given by a certain contour integral.

The instanton partition function can be related to the prepotential by considering $\N=2$ SYM in a certain deformed spacetime known as the $\Omega-$background \cite{Losev}. Upon localisation, this procedure allows us to obtain $Z_{inst}$ in the $\Omega$-background in terms of a two-parameter generalisation of the prepotential. Equating the two results in the undeformed limit of the $\Omega-$background then yields the prepotential explicitly as a power series.
\section{Reconciliation.}
The two approaches discussed above give the prepotential in somewhat different forms. The first in terms of the period integrals of certain elliptic curves, and the second as a power series in certain vacuum expectation values. We aim to reconcile these two results.

The power series form is obtained from the Seiberg-Witten approach by simply expanding the period integrals and performing some power series manipulations \cite{DHoker:1}.

The Seiberg-Witten solution is obtained from the localisation approach by studying the many instanton limit. This limit recovers the Seiberg-Witten differential and thus the elliptic curve approach \cite{NO:1}. 
\section{Reformulation in terms of Young Diagrams.}
The singularities of the contour integrals encountered when calculating $Z_{\text{inst}}$ via localisation are in one to one correspondence with objects called coloured partitions \cite{Nekrasov:1, NO:1}. Reformulating the instanton partition function in terms of these objects gives another interpretation of many of the quantities previously encountered.
\chapter{The Basics of Supersymmetry}
This section provides a brief introduction to supersymmetric quantum field theory (SUSY QFT). We introduce the SUSY Poincar\'e algebra and its representations, the superfield formalism and the construction of supersymmetric Lagrangians. In particular the Lagrangian of $\N=2$ SYM will be obtained. For a more comprehensive introduction to the topic see \cite{Bilal:2, DP:1, West}. For an introduction to ordinary QFT see for example \cite{PS, Schwartz, Srednicki}.
\section{Ordinary Quantum Field Theory}
A quantum field theory can be specified by an action. An action, $S$ is a scalar functional of the fields in the theory. The action is related to the partition function, $Z$ by the path integral approach. The path integral approach consists of taking an often ill-defined integral over an in general uncountably infinite dimensional function space consisting of the fields in the theory:
\begin{equation}
Z=\int\mathcal{D}Xe^{iS[X]}\label{Z}
\end{equation}
This can be viewed as a generalisation of the definition of the ordinary partition function as a sum over states to an infinite dimensional quantum situation.

\begin{remark}{1.1}
Throughout this thesis we set $\hbar=1$ as in \eqref{Z}.
\end{remark}

For now we are mostly concerned with fields defined on flat $4$-dimensional Minkowski spacetime, $\mathbb{R}^{1,3}$. The symmetry group of Minkowski space is known as the Poincar\'e group, and the corresponding Lie algebra is known as the Poincar\'e algebra. Physically, the Poincar\'e group is generated by four translations, three rotations and three boosts. The infinitesimal rotations ($J_{i}$), and the infinitesimal boosts ($K_{i}$), generate a Lie subalgebra of the the Poincar\'e algebra known as the Lorentz algebra. Its Lie bracket is as follows:
\begin{align*}
[J_{i},J_{j}]=i\epsilon_{ijk}J_{k}
\text{, }
\quad
[K_{i},K_{j}]=-i\epsilon_{ijk}J_{k}
\text{, }
\quad
[J_{i},K_{j}]=i\epsilon_{ijk}K_{j}
\text{ , }
\quad
i,j,k\in\{1,2,3\}
\end{align*}
Complexifying the Lorentz algebra by defining $J_{k}^{\pm}=\frac{1}{2}(J_{k}\pm iK_{k})$, leads to the following Lie bracket:
$$
[J_{i}^{\pm},J_{j}^{\pm}]=i\epsilon_{ijk}J_{k}^{\pm}
\text{, }\quad
[J_{i}^{\pm},J_{j}^{\mp}]=0
$$
So the complexified Lorentz algebra is isomorphic to the direct sum of two copies of the simple complex Lie algebra $\mathfrak{sl}(2)$, or equivalently to the non-simple Lie algebra $\mathfrak{so}(4)$. Often the distinction between the complexified Lorentz algebra and the Lorentz algebra itself will be ignored.

To form the full Poincar\'e algebra, we must add the four linearly independent translations $P_{\mu}$. Defining $J_{0i}=K_{i}$, $J_{ij}=\epsilon_{ijk}J_{k}$, and $J_{\mu\nu}=-J_{\nu\mu}$, the resulting commutation relations are \cite{Bilal:2}:
\begin{align*}
[J_{\mu\nu},J_{\rho\sigma}]
&=
ig_{\nu\rho}J_{\mu\sigma}-ig_{\mu\rho}J_{\nu\sigma}-ig_{\nu\sigma}J_{\mu\rho}+ig_{\mu\sigma}M_{\nu\rho}
\\
[J_{\mu\nu},P_{\rho}]
&=
-ig_{\rho\mu}P_{\nu}+ig_{\rho\nu}P_{\mu}
\\
[P_{\mu},P_{\nu}]
&=
0
\end{align*}
where $g_{\mu\nu}=\text{diag}(+1,-1,-1,-1)$ is the Minkowski metric.

One then identifies elementary particles with irreducible representations of the Poincar\'e algebra.

A representation of the Poincar\'e algebra on the space of fields is given by the following differential operators:
\begin{align}
\begin{split}
P_{\mu}
&=
i\del_{\mu}
\\
J_{\mu\nu}
&=
ix_{\mu}\del_{\nu}-ix_{\nu}\del_{\mu}+S_{\mu\nu}
\end{split}\label{Poincare}
\end{align}
where $S_{\mu\nu}$ is a spin operator.
\section{Supersymmetry and the Super Poincar\'e Algebra}
Quantum field theories are very hard to analyse in anything more than the perturbative regime, however it is known that they exhibit many interesting non-perturbative phenomena \cite{SW:1}. In general we expect a system with more symmetries to be easier to analyse. To this end it is interesting to consider the effect of adding additional symmetries to a given QFT.

The Coleman-Mandula theorem states that the spacetime Poincar\'e symmetry of a QFT cannot interact non-trivially with any internal symmetry  \cite{ColemanMandula}. This theorem is however posited upon a Lie algebraic symmetry structure. If we allow instead a Lie superalgebraic symmetry structure, a non-trivial interaction can be obtained. The idea of supersymmetry is to allow such a symmetry structure, it is achieved by introducing extra fermionic generators known as supercharges.

The more supercharges we incorporate, the less realistic but easier to analyse a theory becomes. The number of supercharges is denoted by $\N$. Here we will only be concerned with the cases $\mathcal{N}=1$, and $\mathcal{N}=2$.

Upon introducing fermionic supercharges $Q_{\alpha}^{A}$, and $\bar{Q}_{A,\dot{\alpha}}$ with spinor indices $\alpha=1,2$ and $\dot{\alpha}=\dot{1},\dot{2}$, as well as supercharge indices $A=1,2,\ldots,\mathcal{N}$, it can be shown that we must have the following commutation and anticommutation relations \cite{West}:
\begin{align}
\begin{split}
[P_{\mu},Q_{\alpha}^{A}]&=0
\text{ , }\qquad\qquad\qquad\ \ \ \,\,
[P_{\mu},\bar{Q}_{A,\dot{\alpha}}]=0
\\
[J_{\mu\nu},Q_{\alpha}^{A}]&=i(\sigma_{\mu\nu})_{\alpha}^{\beta}Q_{\beta}^{A}
\text{ , }\qquad\ \ \ \ \,\,
[J_{\mu\nu},\bar{Q}^{\dot{\alpha}}_{A}]=i(\bar{\sigma}_{\mu\nu})^{\dot{\alpha}}_{\dot{\beta}}\bar{Q}^{\dot{\beta}}_{A}
\\
\{Q_{\alpha}^{A},Q_{\beta}^{B}\}&=\epsilon_{\alpha\beta}Z^{AB}\mathcal{Z}
\text{ , }\qquad\
\{\bar{Q}_{A,\dot{\alpha}},\bar{Q}_{N,\dot{\beta}}\}=\epsilon_{\dot{\alpha}\dot{\beta}}Z^{*}_{AB}Z
\\
\{Q_{\alpha}^{A},\bar{Q}_{B,\dot{\beta}}\}&=2(\sigma^{\mu})_{\alpha\dot{\beta}}P_{\mu}\delta_{A}^{B}
\end{split}
\end{align}
In the above, $Z^{AB}$ is an $\mathcal{N}\times\mathcal{N}$ antisymmetric matrix and $\mathcal{Z}$ is a central element called the central charge. The various $\sigma$ matrices and indices are defined and explained in Appendix A. The resulting Lie superalgebra is called the SUSY Poincar\'e algebra, or for short just the SUSY algebra.
\begin{remark}{1.2}
Since $Z^{AB}$ is antisymmetric, if $\N=1$, then $Z^{AB}=0$. Thus for $\N=1$ "unextended", SUSY the central charge is absent.
\end{remark}
\section{Particles and Irreducible Representations}
In ordinary QFT, elementary particles are identified with irreducible representations of the Poincar\'e algebra, so in analogy we seek irreducible representations of the SUSY Poincar\'e algebra. The usual situation is that irreducible representations of the SUSY algebra restrict to reducible representations of the Poincar\'e algebra and thus correspond to several particles rather than to just one. These sets of particles which transform together irreducibly under the SUSY Poincar\'e algebra are called supermultiplets. To build supersymmetric theories, we seek to classify these supermultiplets and the corresponding representations.

\begin{remark}{1.3}
We often abuse terminology and identify supermultiplets with the corresponding representations.
\end{remark}

For our purposes we only need to classify the massless irreducible representations of spin $\leq 1$, since these are the only multiplets involved in pure Yang-Mills theory. The classification is actually rather straightforward but to avoid getting off topic we simply quote the results. More details can be found in \cite{Bilal:2}. 
\begin{remark}{1.4}
Although SYM is built from massless supermultiplets, not all particles will be massless. The Higgs mechanism will still be able to generate masses, but we will not need to work with inherently massive supermultiplets.
\end{remark}
\subsection{$\N=1$ Supermultiplets.}
For $\N=1$ there are only two supermultiplets of spin not exceeding $1$, namely the $\N=1$ chiral and vector multiplets.

The $\N=1$ chiral multiplet corresponds to a Weyl spinor and a complex scalar in an arbitrary representation of the gauge group. For definitions regarding spinors, see Appendix A.

The $\N=1$ vector multiplet corresponds to a massless vector particle (that is to say a massless fundamental representation of the Lorentz group), and a Weyl spinor both necessarily in the adjoint representation of the gauge group.
\subsection{$\N=2$ Supermultiplets.}
For $\N=2$ there are again only two relevant supermultiplets, namely the $\N=2$ vector (or chiral), multiplet, and the $\N=2$ hypermultiplet.

The $\N=2$ vector multiplet consists of a massless vector particle, a complex scalar and two Weyl spinors, all necessarily in the adjoint representation of the gauge group. The $\N=2$ hypermultiplet corresponds to two scalars and two Weyl spinors in an arbitrary representation of the gauge group.
\section{Superspace, Superfields and Supermultiplets}
To build supersymmetric field theories we will require representations of the SUSY algebra on spaces of fields. To this end we introduce the notions of superspaces and superfields.
\subsection{Superspace}
Supersymmetric theories are naturally defined on so-called supermanifolds. A supermanifold is a generalisation of a manifold to include fermionic (i.e. Grassmann valued), coordinates. Here the word superspace will refer to the coordinate space of a supermanifold on which a theory of interest is defined.

Rather than delving deeply into the theory of supermanifolds, for the case at hand we simply take our superspace to be a space having $d$ bosonic spacetime coordinates $x^{\mu}$, $\N$ fermionic "left-handed" Grassmann coordinates $\theta_{A}^{\alpha}$, $\N$ "right-handed" Grassmann coordinates $\bar{\theta}^{A,\dot{\alpha}}$, and finally one more bosonic coordinate $z$. Since each Grassmann coordinate has two components, our superspace has $d+1$ bosonic coordinates and $4\N$ real fermionic coordinates, its superdimension is then denoted $(d+1,4\N)$.

\begin{remark}{1.5}
Since Grassmann numbers anticommute, any product involving more than two instances of the same Grassmann coordinate will vanish.
\end{remark}

A superfield is simply a function defined on superspace. The space of superfields will play the part of our representation space. To this end we write down a set of differential operators which give a (typically reducible), representation of the SUSY algebra on superfields:
\begin{align}
\begin{split}
\mathcal{Z}
&=
i\frac{\del}{\del z}
\\
Q_{\alpha}^{A}
&=
\frac{\del}{\del \theta_{A}^{\alpha}}
+
i\sigma_{\alpha\dot{\beta}}^{\mu}\bar{\theta}^{A,\dot{\beta}}\del_{\mu}
+
\frac{i}{2}\epsilon_{\alpha\beta}Z^{AB}\theta^{\beta}_{B}\frac{\del}{\del z}
\\
\bar{Q}_{A,\dot{\alpha}}
&=
\frac{\del}{\del \bar{\theta}^{A,\dot{\alpha}}}
+
i\theta_{A}^{\beta}\sigma_{\beta\dot{\alpha}}^{\mu}\del_{\mu}
+
\frac{i}{2}\epsilon_{\dot{\alpha}\dot{\beta}}Z^{*}_{AB}\bar{\theta}^{B,\dot{\beta}}\frac{\del}{\del z}
\end{split}
\end{align}
where the rest of the representation is given by \eqref{Poincare}.
\subsection{SUSY covariant derivatives.}
Our first goal is to write $\N=1$ and $\N=2$ supersymmetric actions; actions which transform trivially under the SUSY algebra. To do so, the irreducible representations of the SUSY algebra must be realised on superspace. To this end it will be helpful to introduce the SUSY covariant derivatives:
\begin{align}
\begin{split}
\D_{\mu}&=i\partial_{\mu}\text{, }\qquad\D_{z}=i\frac{\del}{\del z}
\\
\D_{\alpha}^{A}
&=
\frac{\del}{\del \theta_{A}^{\alpha}}
-
i\sigma_{\alpha\dot{\beta}}^{\mu}\bar{\theta}^{A,\dot{\beta}}\del_{\mu}
-
\frac{i}{2}\epsilon_{\alpha\beta}Z^{AB}\theta^{\beta}_{B}\frac{\del}{\del z}
\\
\bar{\D}_{A,\dot{\alpha}}
&=
\frac{\del}{\del \bar{\theta}^{A,\dot{\alpha}}}
-
i\theta^{\beta}_{A}\sigma_{\beta\dot{\alpha}}^{\mu}\del_{\mu}
-
\frac{i}{2}\epsilon_{\dot{\alpha}\dot{\beta}}A_{AB}^{*}\bar{\theta}^{B,\dot{\beta}}\frac{\del}{\del z}
\end{split}
\end{align}
The SUSY covariant derivatives are derived by transforming an ordinary derivative with respect to a supercoordinate, then shifting by the non-covariant part of the transformation \cite{Shadchin}.

It is straightforward to show that the SUSY covariant derivatives anticommute with the supercharges.

An important property of the SUSY covariant derivative is that the quantity $y^{\mu}=x^{\mu}-i\theta_{A}\sigma^{\mu}\bar{\theta}^{A}$ is covariantly constant in the $\bar{\theta}^{A,\dot{\alpha}}$ and $z$ directions:
$$
\bar{\D}_{A,\dot{\alpha}}y^{\mu}
=
0
-
i
\left(
\frac{\del}{\del\bar{\theta}^{A,\dot{\alpha}}}(\theta_{B}^{\gamma}\sigma^{\mu}_{\gamma\dot{\gamma}}\bar{\theta}^{B,\dot{\gamma}})
+
\theta_{A}^{\beta}\sigma_{\beta\dot{\alpha}}^{\nu}\del_{\nu}x^{\mu}
\right)
=
-\theta_{A}^{\gamma}\sigma^{\mu}_{\gamma\dot{\alpha}}
+
\theta_{A}^{\beta}\sigma^{\mu}_{\beta\dot{\alpha}}
=
0
$$
This constraint will allow us to explicitly determine some supermultiplets. In preparation, note that for the $\N=1$ case Remarks $1.2$ and $1.5$ imply that a general scalar superfield can be expanded as a finite power series in the Grassmann coordinates:
\begin{equation}
F(x,\theta,\bar{\theta})
=
f(x)+\theta\psi(x)+\bar{\theta}\bar{\chi}(x)+\theta^{2}m(x)+\bar{\theta}^{2}n(x)+\theta\sigma^{\mu}\bar{\theta}A_{\mu}(x)+\theta^{2}\bar{\theta}\bar{\lambda}(x)+\bar{\theta}^{2}\theta\rho(x)+\theta^{2}\bar{\theta}^{2}D(x)\label{gensfield}
\end{equation}
The spinor indices on $\theta$ and $\bar{\theta}$ have been suppressed in the above expression.
\subsection{The $\N=1$ chiral multiplet.}
To construct the $\N=1$ chiral multiplet, consider a scalar function on superspace $\Phi(x,\theta,\bar{\theta})$, as in \eqref{gensfield}. This provides an irreducible scalar representation of the Lorentz algebra. However, since the covariant derivative anticommutes with the supercharges, Schur's Lemma says that the representation of the SUSY algebra generated by $\Phi$ is reducible. To get an irreducible representation, we fix the constraint $\bar{\mathcal{D}}_{\dot{\alpha}}\Phi(x,\theta,\bar{\theta})=0$. Using that $y$ is covariantly constant this constraint is easy to solve:
$$
\Phi(y,\theta)=H(y)+\sqrt{2}\theta\psi(y)+\theta^{2}f(y)
$$
In the above expression $H$ is a complex scalar field, $\psi$ is a Weyl spinor, and $f$ is an auxiliary scalar field which will be integrated out. The field content of $\Phi$ is thus a complex scalar and a Weyl spinor. This is exactly the field content of the chiral scalar multiplet, so we conclude that $\Phi$ yields an irreducible representation of this multiplet. Note that $\Phi$ can in general be in any representation of the gauge group.
\subsection{The $\N=1$ vector multiplet.}
To construct the $\N=1$ vector multiplet, we once again start with a general scalar function \eqref{gensfield} on $\N=1$ superspace and this time impose the reality condition $V(x,\theta,\bar{\theta})=V(x,\theta,\bar{\theta})^{\dagger}$. 
This leads to the following conditions on the component fields:
$$
f=f^{*}\text{ , }
\quad
\psi=\chi\text{ , }
\quad
m=n^{*}\text{ , }
\quad
A_{\mu}^{\dagger}=A_{\mu}\text{ , }
\quad
\lambda=\rho\text{ , }
\quad
D=D^{*}
$$
We now plug these results back into $V(x,\theta,\bar{\theta})$. Upon rescaling and shifting some of the resulting terms for future convenience, we end up with the following expansion:
\begin{align*}
V(x,\theta,\bar{\theta})
=&
f
+
i\theta\chi
-
i\bar{\theta}\bar{\chi}
+
\theta\sigma^{\mu}\bar{\theta}A_{\mu}
+
\frac{i}{2}\theta^{2}(M+iN)
-\frac{i}{2}\bar{\theta}^{2}(M-iN)
+
i\theta^{2}\bar{\theta}\left(\bar{\lambda}+\frac{i}{2}\bar{\sigma}^{\mu}\del_{\mu}\chi\right)
\\
&-
i\bar{\theta}^{2}\theta\left(\lambda-\frac{i}{2}\sigma^{\mu}\del_{\mu}\bar{\chi}\right)
+
\frac{1}{2}\theta^{2}\bar{\theta}^{2}\left(D-\frac{1}{2}\del^{2}f\right)
\end{align*}
where $-\frac{1}{2}N:=\Re(m)$, and $\frac{1}{2}M:=\Im(m)$.

We claim that the corresponding representation is still reducible. To see why, note that $V$ contains 8 bosonic and 8 fermionic components. The classification of irreducible representations of the SUSY algebra (Section $1.3$), says that this is impossible for an irreducible representation. We thus seek to gauge out some components. This is done by making the following transformation:
$$
e^{2V}\mapsto e^{-i\Lambda^{\dagger}}e^{2V}e^{i\Lambda}
$$
where $\Lambda(y,\theta)$ is a chiral superfield. Under such a transformation $A_{\mu}(x)\mapsto A_{\mu}(x)-\nabla_{\mu}\Re\alpha(x)$, where $\alpha(x)$ is the scalar component of $\Lambda$, and $\nabla_{\mu}:=\partial_{\mu}-i[A_{\mu}(x),\cdot]$ is the gauge covariant derivative in the adjoint representation. So indeed this is the supersymmetric generalisation of a gauge transformation. Demanding that this gauge transformation is a symmetry of the action sets $\chi,f,M$, and $N$ to zero. This is known as the Wess-Zumino gauge. It results in the following superfield:
$$
V_{WZ}(x,\theta,\bar{\theta})=\theta\sigma^{\mu}\bar{\theta}A_{\mu}(x)-i\bar{\theta}^{2}\theta\lambda(x)+i\theta^{2}\bar{\theta}\bar{\lambda}(x)+\frac{1}{2}\theta^{2}\bar{\theta}^{2}D(x)
$$
Since $D$ is an auxiliary field which will be integrated out, $V_{WZ}$ has the correct 6 degrees of freedom. We thus identify it with the abstract $\N=1$ vector multiplet. Note also that in this expression each component is necessarily in the adjoint representation of the gauge group $G$, so $V=V_{a}T^{a}$, where the $T^{a}$ are a basis for $\text{Lie}(G)$.
\begin{remark}{1.6}
The field $V_{WZ}$ has the useful property that each term is of degree at least one in $\theta$ and $\bar{\theta}$, so that  $V_{WZ}^{n}=0$, for $n\geq 3$. In particular, $e^{V_{WZ}}=1+V_{WZ}+\frac{1}{2}V_{WZ}^{2}$. The subscript $WZ$ will be suppressed from now on.
\end{remark}
According to the classification theorem of Section $1.3$, we have now constructed superspace representations of all $\N=1$ supermultiplets of spin not exceeding $1$.
\section{Renormalizable Supersymmetric Actions}
In this section we will obtain the most general $\N=1$ supersymmetric Lagrangian for a single chiral multiplet and for a single vector multiplet. Minimal coupling of these theories will then lead to $\N=2$ SYM.
\subsection{A single $\N=1$ chiral multiplet.}
Let $F$ and $W$ be gauge invariant superfields and in addition suppose $W$ is chiral. It can be shown that the following Lagrangian is SUSY and gauge invariant  \cite{Bilal:2}:
\begin{equation}
\Lag=\int d^{2}\theta d^{2}\bar{\theta}F(x,\theta,\bar{\theta})+\int d^{2}\theta W(\Phi)+\int d^{2}\bar{\theta} W(\Phi)^{\dagger}\label{Chiral}
\end{equation}
The superfield $W$ is known as a superpotential. For a renormalizable theory $W$ must be a polynomial of degree less than 3 in $H$, and $F$ must be of the form $\Phi^{\dagger}K\Phi$ for a Hermitian matrix $K$. We can thus take:
\begin{align*}
\Lag_{\N=1\text{, matter}}
&=
\text{Tr}\left(
\int d^{2}\theta d^{2}\bar{\theta}\Phi^{\dagger}\Phi
+
\left(
\int d^{2}\theta W(\Phi)+h.c.
\right)
\right)
\\
&=
\text{Tr}\left(
|\partial_{\mu}H|^{2}
-
i\psi\sigma^{\mu}\partial_{\mu}\bar{\psi}
+
\left(
\int d^{2}\theta W(\Phi)+h.c.
\right)
\right)
\end{align*}
where the trace is over the gauge group indices which have been suppressed.

One recognizes the usual kinetic terms for a complex scalar and a spinor as well as possibly some interaction governed by $W(\Phi)$, so this is indeed a supersymmetric theory of matter.
\subsection{A single $\N=1$ vector multiplet.}
To construct SUSY invariant Lagrangians using the $\N=1$ vector superfield we define a spinorial superfield, the SUSY field strength:
$$
W_{\alpha}(x,\theta,\bar{\theta})=-\frac{1}{4}\bar{\D}_{\dot{\alpha}}\bar{\D}^{\dot{\alpha}}e^{-2V_{WZ}}\D_{\alpha}e^{2V_{WZ}}
$$
\begin{remark}{1.7}
Since $\bar{\D}^{3}=0$, we have that $\bar{\D}W_{\alpha}=0$ and so $W_{\alpha}$ is a chiral superfield.
\end{remark}
Recalling the definition of field strength $F_{\mu\nu}=\partial_{\mu}A_{\nu}-\partial_{\nu}A_{\mu}-i[A_{\mu},A_{\nu}]$, and expanding $W_{\alpha}$ in terms of the component fields yields the following expression:
$$
W_{\alpha}(y,\theta)
=
-i\lambda_{\alpha}(y)
+
\theta_{\alpha}D(y)
-
i(\sigma^{\mu\nu})_{\alpha}^{\beta}\theta_{\beta}F_{\mu\nu}(y)
-
\theta^{\beta}\theta_{\beta}\sigma^{\mu}_{\alpha\dot{\beta}}\nabla_{\mu}\bar{\lambda}^{\dot{\beta}}(y)
$$
Rescaling the fields by a real coupling constant $g$ and defining the complexified coupling constant $\tau=\frac{\Theta}{2\pi}+\frac{4\pi i}{g^{2}}$, the most general renormalizable SUSY and gauge invariant 
Lagrangian for a single vector superfield is (up to normalization) \cite{Bilal:2}:
$$
\Lag_{\N=1\text{, gauge}}
=
\frac{1}{32\pi}\Im\tau\int d^{2}\theta \text{Tr} W^{\alpha}W_{\alpha}
=
\text{Tr}\left(-\frac{1}{4}F_{\mu\nu}F^{\mu\nu}-i\lambda\sigma^{\mu}\nabla_{\mu}\bar{\lambda}+\frac{1}{2}D^{2}\right)+\frac{\Theta}{32\pi^{2}}g^{2}\text{Tr} F_{\mu\nu}(\star{F})^{\mu\nu}
$$
where $(\star{F})^{\mu\nu}:=\frac{1}{2}\epsilon^{\mu\nu\rho\sigma}F_{\rho\sigma}$ is the dual field strength tensor.

Note that in the above expression, the trace is over the gauge group indices, which we have suppressed.

The quantity $\Theta$ is known as the the instanton angle. It is a real parameter which multiplies the topological part of the action. This part of the action corresponds to instanton configurations, field configurations which obey the classical equations of motion and give a finite non-zero contribution to the action \cite{Coleman:1}. Instantons will be very important later when we discuss localisation.
\subsection{Minimal coupling.}
With Lagrangians for both $\N=1$ multiplets at hand, we consider minimal coupling of these theories. Minimal coupling amounts to simply putting the chiral multiplet in some representation of the gauge group (not necessarily the adjoint), and taking $\Phi^{\dagger}\Phi\rightarrow \Phi^{\dagger}e^{2g  V}\Phi$, which swaps ordinary derivatives for gauge covariant derivatives and introduces the minimal interaction terms necessary for SUSY invariance \cite{Bilal:2}:
\begin{align*}
\Lag_{\N=1\text{, coupled}}
&=
\text{Tr}\left(
\int d^{2}\theta d^{2}\bar{\theta}\Phi^{\dagger}e^{2gV_{WZ}}\Phi
+
\int d^{2}\theta W(\Phi)
+
\int d^{2}\theta W(\Phi)^{\dagger}
\right)
\\
&=
\text{Tr}
\biggr(
|\nabla_{\mu}H|^{2}
-
i\bar{\psi}\bar{\sigma}^{\mu}\nabla_{\mu}\psi
+
f^{\dagger}f
-
gH^{\dagger}[D,H]
-
i\sqrt{2}gH^{\dagger}\{\lambda,\psi\}
\\&
+
i\sqrt{2}g\bar{\psi}[\bar{\lambda},H]
+
\int d^{2}\theta W(\Phi)
+
\int d^{2}\theta W(\Phi)^{\dagger}
\biggr)
\\
&=
\text{Tr}\left(
|\nabla_{\mu}H|^{2}-i\psi\sigma^{\mu}\nabla_{\mu}\bar{\psi}
+
i\sqrt{2}gH^{\dagger}\lambda\psi
-
i\sqrt{2}g\bar{\psi}\bar{\lambda}\psi
+
\int d^{2}\theta W(\Phi)
+
\int d^{2}\theta W(\Phi)^{\dagger}
\right)
\end{align*}
Here the last equality is up to a total derivative and has been obtained using the equations of motion for $f$ and $D$. Note that Lagrangians which differ by a total derivative give the same action provided the fields decay sufficiently quickly at spatial infinity.
\subsection{$\N=2$ SYM.}
According to the classification theorem of Section $1.3$, the $\N=2$ chiral multiplet has the same field content as the combination of an $\N=1$ vector and an $\N=1$ chiral scalar multiplet. Thus we might hope that some linear combination of $\Lag_{\N=1,\text{coupled}}$ and $\Lag_{\N=1,\text{gauge}}$ will have $\N=2$ supersymmetry. This turns out to be the case, yielding the Lagrangian of  $\N=2$ Yang Mills theory.

Up to normalisation, the unique linear combination possessing $\N=2$ supersymmetry is: 
\begin{equation}
\Lag_{\N=2\text{, YM}}:=\frac{1}{g^{2}}\Lag_{\N=1\text{, coupled}}+\Lag_{\N=1\text{, gauge}}
\label{LYM2}
\end{equation}
with $W(\Phi)=0$, and $\Phi$ in the adjoint representation \cite{Bilal:2}. To make it more clear that $\Lag_{\N=2,YM}$ is in fact an $\N=2$ SUSY invariant Lagrangian, we introduce the $\N=2$ chiral superfield in the superspace formalism.

Similarly to the case of the chiral $\N=1$ superfield, the $\N=2$ chiral superfield is defined by the condition that $\bar{\D}_{A,\dot{\alpha}}\Psi(x,\theta,\bar{\theta},z)=0$. It can be shown that this condition ensures $\Psi$ is in fact independent of $z$, then expanding as a truncated power series and regrouping terms we find that $\Psi$ can be written in terms of a pair of $\N=1$ chiral multiplets $\Phi$ and $\mathcal{G}$, as well as the supersymmetric field strength:
$$
\Psi(y,\theta)=\Phi(y,\theta_{1})+i\sqrt{2}\theta_{2}W(y,\theta_{1})+\theta_{2}^{2}\mathcal{G}(y,\theta_{1})
$$
where $\Phi$ and $\mathcal{G}$ are related by 
$
\mathcal{G}(y,\theta)
=
-\frac{1}{2}\int d^{2}\bar{\theta}
\Phi^{\dagger}(y-2i\theta\sigma\bar{\theta},\bar{\theta})e^{2V(y,\theta,\bar{\theta})}
$
\cite{Shadchin}.

With this notation we can then rewrite $\Lag_{YM}$ up to normalisation as:
\begin{equation}
\Lag_{\N=2\text{, YM}}=\Im\tau\int d^{2}\theta_{1}d^{2}\theta_{2}\text{Tr}\Psi^{2}\label{LYM}
\end{equation}
Since $\Psi^{2}$ is a chiral superfield, the above expression is clearly $\N=2$ SUSY and gauge invariant. The expressions \eqref{LYM2} and \eqref{LYM} can be reconciled by expanding out \eqref{LYM2} explicitly and carrying out one of the $\theta$ integrations in \eqref{LYM} \cite{Bilal:1}.
\begin{remark}{1.8}
Unlike the $\N=1$ case, for $\N=2$ no non-trivial superpotential is permitted. This is due to the uniqueness of \eqref{LYM2}.
\end{remark}
\section{$\N=2$ SYM From Dimensional Reduction}
There is another way to obtain the $\N=2$ SYM action which will be utilised later when we discuss Lorentz deformation and the $\Omega-$background. This approach consists of reducing an $\N=1$ theory in six dimensions to an $\N=2$ theory in four dimensions.

Consider $\N=1$ SYM in six dimensions. We compactify spacetime as $\mathbb{R}^{3,1}\times\mathbb{T}^{2}$ by taking the compactified coordinates to be $x^4$ and $x^5$ with radii of compactification $R_4$ and $R_5$ respectively.

Up to normalisation, the $\N=1$ $d=6$ SYM action is \cite{Shadchin}:
\begin{equation}
S_{\N=1\text{, }d=6}
=
\frac{1}{g^{2}}
\int d^{4}x\text{Tr}
\left(
-\frac{1}{4}F_{IJ}F^{IJ}
+
\frac{i}{2}\bar{\Psi}_{A}\Gamma^{I}\nabla_{I}\Psi^{A}
\right)\label{SYMd6}
\end{equation}
where $\Psi^{A}$ is a six dimensional Weyl spinor and the $\Gamma^{I}:=\gamma_{6}^{I}$ are certain $8\times 8$ matrices defined in appendix $A$.

To compactify the theory we assume that the radii $R_{4,5}$ are so small that all fields are independent of the corresponding coordinates. This allows the field strength tensor to be simplified as follows:
$$
F_{\mu 4}
=
\partial_{\mu}A_{4}-\partial_{4}A_{\mu}-i[A_{\mu},A_{4}]
=
\partial_{\mu}A_{4}-i[A_{\mu},A_{4}]
=
\nabla_{\mu}A_{4}
$$
And likewise $F_{\mu 5}=\nabla_{\mu}A_{5}$.

We suggestively define the following complex scalar field: $H=\frac{1}{\sqrt{2}}(A_{4}+iA_{5})$. This definition allows  another component of $F_{IJ}$ to be simplified:
$$
F_{45}
=
\partial_{4}A_{5}-\partial_{5}A_{4}-i[A_{4},A_{5}]
=
0-i\biggr[\frac{1}{\sqrt{2}}(H+H^{\dagger}),\frac{1}{i\sqrt{2}}(H-H^{\dagger})\biggr]
=
-\frac{1}{2}[H+H^{\dagger},H-H^{\dagger}]
=
[H,H^{\dagger}]
$$
The gauge kinetic term of the Lagrangian can now be written as follows:
\begin{align*}
-\frac{1}{4}F_{IJ}F^{IJ}
&=
-\frac{1}{4}
\left(
F_{\mu\nu}F^{\mu\nu}
+
2F_{\mu 4}F^{\mu 4}
+
2F_{\mu 5}F^{\mu 5}
+
2F_{45}F^{45}
+
0
\right)
\\
&=
-\frac{1}{4}
F_{\mu\nu}F^{\mu\nu}
-
\frac{1}{2}
(
-\nabla_{\mu}(A_{4})\nabla^{\mu}(A_{4})
-
\nabla_{\mu}(A_{5})\nabla^{\mu}(A_{5})
)
-
\frac{1}{2}[H,H^{\dagger}]^{2}
\\
&=
-\frac{1}{4}
F_{\mu\nu}F^{\mu\nu}
+
\nabla_{\mu}H\nabla^{\mu}H^{\dagger}
-
\frac{1}{2}[H,H^{\dagger}]^{2}
\end{align*}
Upon identifying $H$ with the scalar field of $\N=2$, $d=4$ SYM we see that this term is exactly the bosonic part of the $\N=2$, $d=4$ SYM action!

The spinorial part of $S_{\N=2\text{, }d=6}$ must also be considered. Note that since the fields are assumed independent of $x^4$ and $x^5$ we can take the corresponding components of the spinors to vanish and write
$
\Psi^{A}
=
(\psi^{A}_{\alpha},\chi^{A}_{\alpha},0,0)^{T}
$.
Then due to some general properties of spinors in six dimensions (see for example Appendix A of \cite{Shadchin}), it turns out that $\chi^{A,\dot{\alpha}}=\epsilon^{AB}\epsilon^{\dot{\alpha}\dot{\beta}}\psi_{B,\dot{\beta}}$, allowing the action to be expressed solely in terms of the Weyl spinors $\psi^{A}$ and their conjugates:
\begin{align*}{}
\frac{i}{2}
\bar{\Psi}_{A}\Gamma^{I}\nabla_{I}\Psi^{A}
&=
\frac{i}{2}
\bar{\Psi}^{A}\Gamma^{\mu}\nabla_{\mu}\Psi^{A}
+
\frac{i}{2}
\bar{\Psi}^{A}\Gamma^{4}\nabla_{4}\Psi^{A}
+
\frac{i}{2}
\bar{\Psi}^{A}\Gamma^{5}\nabla_{5}\Psi^{A}
\\
&=
\frac{i}{2}
(0,0,\psi_{A},\bar{\psi}_{A})
\begin{pmatrix}
0_{4}&
\begin{pmatrix}
0&\sigma^{\mu}\\
\bar{\sigma}^{\mu}&0
\end{pmatrix}
\\
\begin{pmatrix}
0&\sigma^{\mu}\\
\bar{\sigma}^{\mu}&0
\end{pmatrix}
&0_{4}\\
\end{pmatrix}
\begin{pmatrix}
\nabla_{\mu}\psi^{A}\\
\nabla_{\mu}\bar{\psi}^{A}\\
0_{2}\\
0_{2}
\end{pmatrix}
\\
&\qquad
+
\frac{i}{2}
\bar{\Psi}^{A}\Gamma^{4}\nabla_{4}\Psi^{A}
+
\frac{i}{2}
\bar{\Psi}^{A}\Gamma^{5}\nabla_{5}\Psi^{A}
\\
&=
\frac{i}{2}
(
\psi_{A}\sigma^{\mu}\nabla_{\mu}\bar{\psi}^{A}
+
\bar{\psi}_{A}\sigma^{\mu}\nabla_{\mu}\psi^{A}
)
+
\frac{1}{2}
(
-
\psi_{A}[A_{4},\psi^{A}]
+
\bar{\psi}^{A}[A_{4},\bar{\psi}^{A}]
)
\\
&
\qquad\quad
-
\frac{1}{2}
(
\psi_{A}[A_{5},\psi^{A}]
+
\bar{\psi}_{A}[A_{5},\bar{\psi}^{A}]
)
\\
&=
\frac{i}{2}
(
\psi_{A}\sigma^{\mu}\nabla_{\mu}\bar{\psi}^{A}
+
\bar{\psi}_{A}\sigma^{\mu}\nabla_{\mu}\psi^{A}
)
-
\frac{i}{\sqrt{2}}\psi_{A}[H^{\dagger},\psi^{A}]
+
\frac{i}{\sqrt{2}}\bar{\psi}_{A}[H,\bar{\psi}^{A}]
\\
&=
i\psi^{A}\sigma^{\mu}\nabla_{\mu}\bar{\psi}_{A}
-
\frac{i}{\sqrt{2}}\psi_{A}[H^{\dagger},\psi^{A}]
+
\frac{i}{\sqrt{2}}\bar{\psi}^{A}[H,\bar{\psi}_{A}]
\end{align*}
where we have used that in six dimensions the Dirac adjoint swaps components.

Compactification has thus reproduced the spinorial part of the $\N=2$ SYM action, so indeed compactifying two directions of $\N=1$, $d=6$ SYM results in $\N=2$, $d=4$ SYM.
\section{Non-renormalizable Supersymmetric Actions}
So far, the form of our SUSY actions have been constrained by renormalizability. We now discuss what happens when this constraint is dropped. Renormalizability is not an issue in the low energy regime, so non-renormalizable actions may be used as so-called effective theories. In this section the construction of such actions for the $\N=2$ case will be briefly discussed.

It was previously mentioned that without regard for renormalizability the most general $\N=1$ supersymmetric Lagrangian is given by \eqref{Chiral}. On the other hand, without regard for renormalizability the most general gauge field Lagrangian is:
$$
\Lag_{\text{NR, gauge}}=\frac{1}{16g^{2}}\int d^{2}\theta f_{ab}(\Phi)W^{a\alpha}W^{b}_{\alpha}+h.c.
$$
where $f_{ab}$ depends on $\Phi$ only and is thus holomorphic, and $a,b$ are gauge group Lie algebra indices \cite{Bilal:2}.

As in the renormalizable case, to obtain the most general non-renormalizable $\N=2$ Lagrangian from \eqref{Chiral} we must include gauge fields. Schematically this can be achieved by adding the most general kinetic terms for gauge fields as well as converting all derivatives to gauge covariant ones. In practice this can be accomplished in two steps. First we swap $\Phi^{\dagger}$ for $\phi^{\dagger}e^{2gV}$ in the argument of $F$. Secondly we add an appropriate linear combination of $\Lag_{\text{NR, gauge}}$.

Since the matter and gauge Lagrangians must be related in an $\N=2$ invariant theory, the functions $F$ and $f_{ab}$ must be related for the resulting theory to have $\N=2$ supersymmetry. One can show that the correct relation is given by taking $w=0$, and setting \cite{Bilal:2}:
\begin{align*}
\frac{16\pi}{4g^{2}}f_{ab}
&=:
-i\frac{\partial^2}{\partial\phi^2}\F
\\
\frac{16\pi}{4g^2}F
&=:
-\frac{i}{2}\phi^{\dagger}\frac{\partial}{\partial\phi}\F+h.c.
\end{align*}
The holomorphic quantity $\F$ is called the $\N=2$ prepotential.

This action can be rewritten conveniently in $\N=2$ superspace language as \cite{Bilal:2}:
\begin{equation}
S_{\text{eff}}
=
\frac{1}{8\pi^{2}i}\Im\int d^4xd^2\theta\F(\Psi)\label{NRaction}
\end{equation}
Or in $\N=1$ language:
\begin{equation}
S_{\text{eff}}
=
\frac{1}{16\pi}\Im \int d^{4}x\left[\frac{1}{2}\int d^{2}\theta \F_{ab}(\Phi)W^{a\alpha}W^{b}_{\alpha}+\int d^{2}\theta d^{2} \bar{\theta}(\Phi^{\dagger}e^{V})^{a}\F_{a}(\Phi)\right]\label{Seffgen}
\end{equation}
Upon comparison with \eqref{LYM}, we see that in the renormalizable case $\F\propto\Psi^{2}$.
\begin{remark}{1.9}
From \eqref{NRaction} it is clear that $\F$ completely determines the low energy effective action of $\N=2$ SYM. Amazingly $\F$ can be calculated exactly. Doing so in two different ways is the major goal of this thesis.
\end{remark}
\chapter{Seiberg-Witten Theory}
Seiberg-Witten theory provides a way to calculate the $\N=2$ prepotential exactly. We present here in detail the original Seiberg-Witten approach for pure SYM with gauge group $SU(2)$ \cite{SW:1}. Generalisations to include matter and different gauge groups are well known \cite{Argyres, APS:1, Danielsson, Minahan, SW:2}. Introductions to this area include \cite{AG:1, Bilal:2}.
\section{The Moduli space of vacua} 
The first important object to introduce is $\M$, the moduli space of vacua. Points in $\M$ correspond to gauge inequivalent vacua of $\N=2$ SYM, that is to gauge inequivalent Poicar\'e invariant field configurations which minimise the action.

It is well known from ordinary QFT that Lorentz invariance implies all non-scalar fields and all spacetime derivatives must have a vanishing vaccum expectation value (VEV), however a scalar field can have a non-zero VEV. We now discuss the vacuum configurations of $\N=2$ SYM. From \eqref{LYM2}, the full $\N=2$ SYM action can be expanded out to give \cite{Bilal:2}:
\begin{align}
\begin{split}
S_{\N=2\text{ YM}}
&=
\int d^{4}x
\text{Tr}\biggr(-\frac{1}{4}F_{\mu\nu}F^{\mu\nu}-i\lambda\sigma^{\mu}\nabla_{\mu}\bar{\lambda}-i\psi\sigma^{\mu}\nabla_{\mu}\bar{\psi}+|\nabla H|^{2}+\frac{\Theta}{32\pi^{2}}g^{2}F_{\mu\nu}(\star{F})^{\mu\nu}
\\
&\quad    
+
\frac{1}{2}D^{2}+f^{\dagger}f+i\sqrt{2}gH^{\dagger}\{\lambda,\psi\}-i\sqrt{2}g\{\bar{\psi},\bar{\lambda}\}H+gD[H,H^{\dagger}]\biggr)
\end{split}\label{LYMexpanded}
\end{align}
The corresponding scalar potential is thus $V=-\text{Tr}(\frac{1}{2}D^{2}+f^{\dagger}f+gD[H,H^{\dagger}])$. Since $f$ and $D$ are auxiliary fields we can easily integrate them out by solving their equations of motion. The Euler-Lagrange equations give the following equations of motion:
$$
D+g[H,H^{\dagger}]=f=f^{\dagger}=0
$$
So the scalar potential is:
$$
V
=
\frac{1}{2}g^{2}\text{Tr}([H,H^{\dagger}])^{2}
$$
\subsection{Parametrisation of $\M$.}
By definition, a vacuum minimises the action, and thus the scalar potential $V$. Clearly $V\geq 0$, so any minimum $H_{0}$ has $V(H_{0})\geq 0$. In fact, unbroken SUSY requires $V_{0}=0$, so the possible vacua are parametrised by the solutions of the equation $[H,H^{\dagger}]=0$  \cite{Bilal:2}. The moduli space of vacua $\M$ is then this space considered up to gauge transformations. It turns out that the prepotential is closely related to the metric on this space.

For pure SYM, all fields are necessarily in the adjoint representation of the gauge group. This means that the fields are $\text{Lie}(G)$ valued functions, and thus for the case of $SU(2)$ can be expanded in terms of the Pauli matrices: $H(x)=\sum_{j=1}^{3}(a_{j}(x)+ib_{j}(x))\tau_{j}$. We assume without loss of generality that not all of the $a_{j}(x)$ vanish.

In the adjoint representation the gauge group acts as $G\times \text{Lie(G)}\rightarrow \text{Lie(G)}\text{, }\phi\mapsto g\phi g^{-1}$. Such a gauge transformation can be used to set $a_1(x)=a_2(x)=0$, then $[H,H^{\dagger}]=0$ enforces that $b_1(x)=b_2(x)=0$. So without loss of generality we can write $H(x)=\frac{1}{2}a(x)\tau_{3}$, with $a(x):=a_3(x)+ib_3(x)$. Let $a$ denote the VEV of $a(x)$ with respect to a particular vacuum. Then $a$ is a parameter labelling the different vacua of the theory.
\begin{remark}{2.1}
The condition $[H,H^{\dagger}]=0$ says that $H$ is an element of a Cartan subalgebra of $\text{Lie}(G)$. This observation proves useful for the case of gauge group $SU(N)$ \cite{SW:2}.
\end{remark}
Noting that
$
\begin{pmatrix}
0&-1\\
1&0
\end{pmatrix}
\in SU(2)
$
sends $H\rightarrow -H$ and thus $a\rightarrow -a$, we see that $a$ and $-a$ are also gauge equivalent. Thus gauge inequivalent vacua can be labelled by the gauge invariant parameter $\text{Tr} H^{2}$ which is given by $\frac{1}{2}a^{2}$ in the vacuum. In general, we define $u:=\expval{\text{Tr} H^{2}}$, and $\expval{H}=:\frac{1}{2}a\tau_{3}$, then classically $u=\frac{1}{2}a^{2}$. The parameter $u$ then labels gauge inequivalent vacua in the full quantum theory and is thus a coordinate on $\mathcal{M}$.
\subsection{Gauge symmetry breaking and the effective theory.}
For $\expval{H}\neq 0$, the $SU(2)$ gauge symmetry of $\N=2$ SYM is broken, causing the $a=1,2$ components of the fields to develop masses. This is the well-known Higgs mechanism of ordinary QFT, and can be seen by writing
$
H(x)=H'(x)+H_{0}
=
(
0,
0,
\frac{1}{2}(a(x)+a)
)^{T}
$, where $H_{0}$ is the VEV of $H(x)$, then expanding the $|\nabla H|^{2}$ term of \eqref{LYMexpanded}.

In fact, the gauge symmetry breaks to $U(1)$, so at low energies the theory is described by an $\N=2$ theory with gauge group $U(1)$. To see this, note that the vacuum vector is only invariant under the $U(1)$ subgroup of $SU(2)$:
\begin{align*}
H_{0}
\mapsto
U H_{0}U^{\dagger}
=
\frac{1}{2}a
\begin{pmatrix}
\alpha&-\bar{\beta}\\
\beta&\bar{\alpha}
\end{pmatrix}
\begin{pmatrix}
1&0\\
0&-1
\end{pmatrix}
\begin{pmatrix}
\bar{\alpha}&\bar{\beta}\\
-\beta&\alpha
\end{pmatrix}
=
\frac{1}{2}a
\begin{pmatrix}
|\alpha|^{2}-|\beta|^{2}&2\alpha\bar{\beta}\\
2\bar{\alpha}\beta&-(|\alpha|^{2}-|\beta|^{2})
\end{pmatrix}
\end{align*}
which is equal to $H_{0}$ if and only if $\beta=0$ and $|\alpha|^{2}=1$, that is if $U\in U(1)\subset SU(2)$.

We have shown that at low energies the theory has a $U(1)$ gauge symmetry and each field has only a single gauge group index. So from the general non-renormalizable $\N=2$ action \eqref{Seffgen} we have that the effective Lagrangian has the form:
\begin{equation}
S_{\text{eff}}=\frac{1}{16\pi}\Im \int d^{4}x\left[\frac{1}{2}\int d^{2}\theta \F''(\Phi)W^{\alpha}W_{\alpha}+\int d^{2}\theta d^{2} \bar{\theta}\Phi^{\dagger}\F'(\Phi)\right]\label{Seff}
\end{equation}
for some holomorphic function $\F$.
\begin{remark}{2.2}
In the above expression, the $e^{V}$ term has been replaced by $1$. To see why, recall that $e^{V}=1+V+\frac{1}{2}V^{2}$, but since $V$ is adjoint valued and $U(1)$ is abelian, only the $1$ term remains in a $U(1)$ theory.
\end{remark}
A less abstract interpretation of the function $\F$ is given as follows. If we expand the effective action term by term we see that $\Im \F''$ plays the role of a metric in field space:
$$
S_{\text{eff}}
\propto
\int d^{4}x\left[\Im\F''(\Phi)(|\partial_{\mu}H|^{2}-i\psi\sigma^{\mu}\partial_{\mu}\bar{\psi}-\frac{1}{4}F_{\mu\nu}(F^{\mu\nu}-i(\star{F})^{\mu\nu})-i\lambda\sigma^{\mu}\partial\bar{\lambda}+\ldots)\right]
$$
By passing to the moduli space (effectively replacing fields by their VEVs), it is clear that the metric on $\M$ is given by $ds^{2}=\Im(\F''(a))dad\bar{a}=\Im\tau(a)dad\bar{a}$, where $\tau(a):=\F''(a)$ is the complexified effective coupling. So if we can determine the metric on $\M$ we have in principle determined $\F$.
\section{Seiberg-Witten Duality}
An obvious consistency condition is that the metric on $\M$ must be positive definite: $\Im\tau(a)>0$ for all $a$. However this cannot be the case on all of $\M$, since $\F$ is holomorphic so $\Im(\tau)=\Im(\F'')$ is harmonic. Thus $\Im(\tau)$ cannot have a minimum on $\mathbb{C}$, and so we cannot have $\Im(\tau)>0$ everywhere. We conclude that the description of $\M$ in terms of $\tau$ cannot be valid everywhere, that is when $\Im(\tau)$ approaches zero we must switch to a different set of coordinates, $a_{D}$ and $\tau_{D}$. These dual quantities are provided by Seiberg-Witten duality \cite{SW:1}.
\subsection{The duality transformation.}
Following Seiberg and Witten, we define a dual superfield $\Phi_D$ and a dual prepotential $\F_D$ as the Legendre transform of $\Phi$ and $\F$:
\begin{equation}
\Phi_D:=\F'(\Phi)\quad\F_{D}'(\Phi_D)=:-\Phi\label{duality}
\end{equation}
The form of the second term of the effective action action \eqref{Seff} is easily seen to be invariant under this transformation since:
$$
\Im \Phi^{\dagger}\F'(\Phi)=-\Im((\Phi_{D}^{\dagger}\F'_{D}(\Phi_D))^{\dagger})=\Im\Phi_{D}^{\dagger}\F'_{D}(\Phi_{D})
$$
It is less easy to show that the form of the first term is also invariant. To do so we perform a change of variables in the path integral defining the effective partition function and show that this leads to an action of the same form but in the dual variables.

As a preliminary step, note that due to $U(1)$ symmetry the field strength is simply $F_{\mu\nu}=\del_{\mu}A_{\nu}-\partial_{\nu}A_{\mu}$, so it obeys the Bianchi Identity: $\del_{\lambda}F_{\mu\nu}+\del_{\mu}F_{\nu\lambda}+\del_{\nu}F_{\lambda\mu}=0$. Upon contracting with $\epsilon^{\mu\nu\rho\sigma}$ it is clear that $\frac{1}{2}\epsilon^{\mu\nu\rho\sigma}\del_{\nu}F_{\rho\sigma}=0$. This identity is equivalent to the reality condition $\Im D_{\alpha}W^{\alpha}=0$ \cite{Bilal:1}.

To show invariance of the first term of \eqref{Seff} under duality we change variables in the path integral and enforce the reality condition with a real Lagrange multiplier superfield, $V_{D}$. The relevant part of the path integral is
$
Z_{eff}
=
\int \mathcal{D} V \exp\left(\frac{i}{32\pi}\Im\int d^{2}\theta \F''(\Phi)W^{\alpha}W_{\alpha}\right)
$
. 
For the Lagrange multiplier we add $\frac{i}{64\pi}\Im\int d^{2}\theta d^{2} \bar{\theta}V_{D}D_{\alpha}W^{\alpha}$ to $S_{\text{eff}}$, yielding the following equality:
\begin{align*}
Z_{eff}
&=
\int \mathcal{D} V \exp\left({\frac{i}{32\pi}\Im\int d^{2}\theta \F''(\Phi)W^{\alpha}W_{\alpha}}\right)
\\
&=
\int \mathcal{D} W \mathcal{D} V_{D}\exp\left\{\frac{i}{32\pi}\Im\int d^{4}x\int d^{2}\theta\left(\F''(\Phi)W^{\alpha}W_{\alpha}+\frac{1}{2}d^{2} \bar{\theta}V_{D}D_{\alpha}W^{\alpha}\right)\right\}
\end{align*}
Integrating by parts following \cite{Bilal:1} and defining $(W_{D})_{\alpha}=-\frac{1}{4}\bar{\D}^{2}e^{-2V_{D}}\D_{\alpha}e^{2V_{D}}$ yields:
$$
\int d^{2}\theta d^{2} \bar{\theta}^{2}V_{D}D_{\alpha}W^{\alpha}
=
-\int d^{2}\theta d^{2} \bar{\theta}^{2}D_{\alpha}V_{D}W^{\alpha}
=
\int d^{2}\theta \bar{D}^{2}(D_{\alpha}V_{D}W^{\alpha})
=
-4\int d^{2}\theta (W_{D})_{\alpha}W^{\alpha}
$$
where we have used that $\bar{D}W_{\alpha}=0$, since up to a total spacetime derivative:
\begin{align*}
\bar{D}^{2}f(x,\theta,\bar{\theta})
&=
\epsilon^{\dot{\alpha}\dot{\beta}}\frac{\del}{\del\bar{\theta}^{\dot{\alpha}}}\frac{\del}{\del\bar{\theta}^{\dot{\beta}}}f(x,\theta,\bar{\theta})
+
\del_{v}
\left(
2i\theta^{\alpha}(\sigma^{\nu})_{\alpha}^{\dot{\alpha}}\frac{\del f(x,\theta,\bar{\theta})}{\del\bar{\theta}^{\dot{\alpha}}}
\right)
\\
&\qquad+
\del_{\mu}\del_{\nu}
\left(
\frac{1}{2}\theta^{2}(\bar{\sigma}^{\mu})^{\alpha\dot{\beta}}(\sigma^{v})_{\alpha\dot{\beta}}f(x,\theta,\bar{\theta})
\right)
\\
&=
\epsilon^{\dot{\alpha}\dot{\beta}}\frac{\del}{\del\bar{\theta}^{\dot{\alpha}}}\frac{\del}{\del\bar{\theta}^{\dot{\beta}}}f(x,\theta,\bar{\theta})
\\
&=
-\int d^{2}\bar{\theta}f(x,\theta,\bar{\theta)}
\end{align*}
Completing the square allows the $W$ integral to be carried out explicitly:
\begin{align*}
\int d^{2}\theta\F''(\Phi)W^{\alpha}W_{\alpha}
+
&
\frac{1}{2}\int d^{2}\theta d^{2} \bar{\theta}V_{D}D_{\alpha}W^{\alpha}
=
\int d^{2}\theta\left(\F''(\Phi)W^{\alpha}W_{\alpha}-2(W_{D})_{\alpha}W^{\alpha}\right)
\\
&=
\int d^{2}\theta\left\{\F''(\Phi)\left(W^{\alpha}-\frac{(W_{D})^{\alpha}}{\F''(\Phi)}\right)\left(W_{\alpha}-\frac{(W_{D})_{\alpha}}{\F''(\Phi)}\right)-\frac{(W_{D})^{\alpha}(W_{D})_{\alpha}}{\F''(\Phi)}\right\}
\end{align*}
This results in the following expression for $Z_{eff}$:
\begin{align*}
Z_{eff}
&=
\int\D W\D V_{D}
\exp
\left\{
\frac{i}{16\pi}\Im \int d^{4}x\int d^{2}\theta\F''(\Phi)\left(W^{\alpha}-\frac{(W_{D})^{\alpha}}{\F''(\Phi)}\right)\left(W_{\alpha}-\frac{(W_{D})_{\alpha}}{\F''(\Phi)}\right)
\right\}
\\
&
\qquad
\exp
\left\{
\frac{i}{16\pi}\Im \int d^{4} x \int d^{2}\theta\left(-\frac{1}{\F''(\Phi)}(W_{D})^{\alpha}(W_{D})_{\alpha}\right)
\right\}
\\
&=
\int\D V_{D}
\exp
\left\{
\frac{i}{16\pi}\Im \int d^{4} x \int d^{2}\theta\left(-\frac{1}{\F''(\Phi)}(W_{D})^{\alpha}(W_{D})_{\alpha}\right)
\right\}
\end{align*}
where we have used that the $W$ integral is Gaussian and thus evaluates to a constant which by appropriate normalisation we can take to be unity.

So indeed the form of this part of the action is invariant up to replacing the effective coupling $\F''(\Phi)$ with $-\frac{1}{\F''(\Phi)}=\F_{D}(\Phi_{D})$.

In summary we have shown that under the duality transformation \eqref{duality}, $S_{\text{eff}}$ becomes:
$$
\frac{1}{16\pi}\Im \int d^{4}x\left[\frac{1}{2}\int d^{2}\theta \F_{D}''(\Phi_{D})W_D{}^{\alpha}W_{D\alpha}+\int d^{2}\theta d^{2} \bar{\theta}\Phi_{D}^{\dagger}\F_{D}'(\Phi_{D})\right]
$$
That is to say the form of $S_{\text{eff}}$ is duality invariant.

Defining the dual coupling, $\tau_D=-\frac{1}{\tau}$ we see that the duality transformation maps strongly coupled regions of $\N=2$ SYM to weakly coupled regions of $\N=2$ SYM and vice versa. It is thus an example of a so called $S$-duality. Furthermore, as $\Im \tau \rightarrow 0$, $\tau _D \rightarrow \infty$, so indeed the dual description should yield extended coordinates on $\M$.
\subsection{The full duality group.}
The full group of duality transformations is in fact larger than that derived in the previous section. To see this, we use the dual variables to rewrite $S_{\text{eff}}$ in a more symmetric form:
\begin{equation}
S_{\text{eff}}
=
\frac{1}{32\pi}\Im\int d^{4}xd^{2}\theta
\frac{d\Phi_{D}}{d\Phi}W^{\alpha}W_{\alpha}
+
\frac{1}{32\pi i}
\int d^{4} x d^{2} \theta d^{2} \bar{\theta}
(
\Phi^{\dagger}\Phi_{D}
-
\Phi_{D}^{\dagger}\Phi
)
\end{equation}
It can now be shown that $S_{\text{eff}}$ is invariant under $\Phi\rightarrow \Phi$, $\Phi_{D}\rightarrow \Phi_{D}+b\Phi$, where $b\in \mathbb{Z}$. This is easy for the second term. For the first term we have:
$$
\frac{1}{16\pi}\Im\int d^{4}xd^{2}\theta
\frac{d\Phi_{D}}{d\Phi}W^{\alpha}W_{\alpha}
\rightarrow
\frac{1}{32\pi}\Im\int d^{4}xd^{2}\theta
\frac{d\Phi_{D}}{d\Phi}W^{\alpha}W_{\alpha}
+
\frac{b}{32\pi}\Im\int d^{4}xd^{2}\theta
W^{\alpha}W_{\alpha}
$$
But
$
\Im\int d^{2}\theta
W^{\alpha}W_{\alpha}
=
-
F_{\mu\nu}(\star{F})^{\mu\nu}
$
so:
$$
\frac{b}{32\pi}\Im\int d^{4}xd^{2}\theta
W^{\alpha}W_{\alpha}
=
-2\pi b\frac{1}{32\pi}\int d^{4}x
F_{\mu\nu}(\star{F})^{\mu\nu}
=
-2\pi b k\in2\pi\mathbb{Z}
$$
In the above calculation we have used the fact that $k:=\frac{1}{32\pi}\int d^{4}x
F_{\mu\nu}(\star{F})^{\mu\nu}$ is the instanton number, and is thus an integer \cite{Coleman:1}. Since $Z_{eff}=e^{iS_{\text{eff}}}$, under this transformation $Z_{eff}\rightarrow Z_{eff}$, so indeed this is a symmetry of the theory.

The above transformation can be written as 
$
\begin{pmatrix}
\Phi_D\\
\Phi
\end{pmatrix}
\rightarrow
\begin{pmatrix}
1&b\\
0&1
\end{pmatrix}
\begin{pmatrix}
\Phi_D\\
\Phi
\end{pmatrix}
$
, while the original duality transformation $(\Phi\rightarrow \Phi_{D}\text{ , }\F\rightarrow \F_{D})$, can be written as 
$
\begin{pmatrix}
\Phi_D\\
\Phi
\end{pmatrix}
\rightarrow
\begin{pmatrix}
0&-1\\
1&0
\end{pmatrix}
\begin{pmatrix}
\Phi_D\\
\Phi
\end{pmatrix}
$
.
Recalling that 
$
SL(2;\mathbb{Z})
=
\left<
\begin{pmatrix}
1&b\\
0&1
\end{pmatrix}
,
\begin{pmatrix}
0&-1\\
1&0
\end{pmatrix}
\right>
$
, it is clear that the group of duality transformations is at least $SL(2;\mathbb{Z})$. In fact it can be shown that this is the full duality group \cite{SW:1}.

Seiberg-Witten duality also descends to the level of metrics. Taking expectation values we define $a_{D}=\frac{\del \F(a)}{\del a}$ and note $\expval{H_{D}}=\frac{1}{2}a_{D}\tau_{3}$. Then $da_{D}=\frac{\partial a_{D}}{\partial a}da=\F''(a)da$, so $ds^{2}=\Im(da_{D}d\bar{a})=\frac{i}{2}(dad\bar{a}_{D}-da_{D}d\bar{a})$. From here it is clear that $ds^{2}$ is also $SL(2;\mathbb{Z})$ invariant.
\section{The BPS Mass Formula}
In this section we introduce the BPS mass formula, named for Bogomolny Prasad and Sommerfield. It relates the masses of particles to their magnetic and electric charges. This will prove useful since Seiberg-Witten duality maps electrically charged states to solitonic magnetic monopoles \cite{SW:1}. The BPS mass formula will thus provide a physical interpretation to many of the following arguments. Unfortunately some of the statements required to obtain the BPS mass formula cannot be proven here without going significantly off track. For some additional details see \cite{SW:1, WO:1}.
 
First recall that the helicity operator is the projection of the spin operator onto the momentum operator, or physically that the helicity of a particle is the component of its spin along its direction of travel. It can be shown that representations of the SUSY algebra can be split into two types; so called short and long multiplets. The long multiplets contain $16$ helicity states while the short ones contain $4$ \cite{Bilal:2}. Massless states necessarily belong to short multiplets, and since the massive states in our theory obtain their mass via the Higgs mechanism, so do they.

Next, it can be shown that states belong to short multiplets if and only if their mass is given by $m^{2}=2|\mathcal{Z}|^{2}$, where $\mathcal{Z}$ is the central charge of the SUSY algebra, while states in long multiplets have $m^{2}>2|\mathcal{Z}|^{2}$ \cite{Bilal:2, WO:1}. This inequality is known as the BPS bound, and states for which $m^{2}=|\mathcal{Z}|^{2}$ are called BPS saturated. The BPS mass formula is then obtained by relating the central charge of a state to its electric and magnetic charge.

Finally, a purely electrically charged state has $\mathcal{Z}=an_e$ where $n_{e}\in\mathbb{Z}$, so by duality, a purely magnetic state has $Z=a_Dn_m$ with $n_m\in\mathbb{Z}$ \cite{SW:1}. Since the central charge is additive, a general state has $\mathcal{Z}=an_{e}+a_{D}n_{m}$. In summary, the BPS mass formula is as follows:
\begin{equation}
m^{2}=2|\mathcal{Z}|^{2}
\text{, where }
\mathcal{Z}=an_{e}+a_{D}n_{m}=(n_{m},n_{e})(a_{D},a)^{T}
\text{.}
\label{BPS}
\end{equation}
\begin{remark}{2.3}
Acting on $a$ and $a_D$ with $SL(2;\mathbb{Z})$ is equivalent to transforming the charge vector $(n_m, n_e)$ by right multiplication. For a sanity check note that in this notation, Seiberg-Witten duality indeed maps $(1,0)$ to $(0,1)$.
\end{remark}
\section{The Weak and Strong Coupling Limits}
The weak and strong coupling limits of $\N=2$ SYM can now be examined. More precisely, we study the monodromy properties of $a(u)$ and $a_D(u)$ as they encircle certain points of $\M$. The points of $\M$ for which $a(u)$ and $a_D(u)$ do not return to their original values upon encircling will be referred to as singularities. This analysis will yield several more consistency conditions for the prepotential. Along with Seiberg-Witten duality, these conditions will then  fix $\F$ entirely.
\subsection{The $u\rightarrow\infty$ limit.}
To study the $u\rightarrow\infty$ limit and the associated monodromy we quote several results without proof. None of these results are particularly out of reach but their derivations would take us quite a long way off topic.

The first result is the tree-level and 1-loop corrections to $\F_{class}=\frac{1}{2}\tau_{0}\Psi^{2}$, which were determined by Seiberg \cite{Seiberg:1}. The result is $\F_{\text{pert}}(\Psi)=\frac{i}{2\pi}\Psi^{2}\log\left(\frac{\Psi^{2}}{\Lambda^{2}}\right)$, where $\Lambda$ is the dynamically generated energy scale of the theory. In short, this is obtained by using the holomorphicity of $\F$ and noting that $\F_{\text{pert}}$ must be invariant under $U(1)$ gauge transformations.

The next result is the SUSY non-renormalization theorem. In SUSY QFTs it can be shown that due to fermionic/bosonic cancellations certain quantities are either not renormalised or they are not renormalised beyond 1-loop level. This can be done either by exploiting holomorphicity and symmetries, or by the analysis of Feynman diagrams in superspace. For more details and some applications of this fact see \cite{Seiberg:1, Seiberg:2}. In particular, for $\N=2$ SUSY the prepotential is not renormalised beyond 1-loop level, and thus $\F_{\text{pert}}$ is the full perturbative result.

We can now split $\F$ into a perturbative and a non-perturbative part: $\F=\F_{\text{pert}}+\F_{\text{inst}}$. The perturbative part has been obtained without any trouble, it is the non-perturbative instanton contribution $\F_{\text{inst}}$ which is the tough bit!

In the high energy limit, the full $SU(2)$ theory is known to be asymptotically free. Here the momentum scale $p$ is large and thus so is the mass scale $m$. The BPS mass formula then implies that this is a region of large $a$.

As detailed in \cite{Bilal:1}, the effective action is an integral of the full action over heavy modes so the dominant contribution is from regions of large $p$ and thus large $a$. Asymptotic freedom ensures that in this region the full theory is perturbative, so the integral defining the effective action is dominated by perturbative contributions. This means that $\F_{\text{pert}}\rightarrow\F$ as $u\rightarrow\infty$. Furthermore in this limit we must have that $u$ approaches its classical value: $u\rightarrow \frac{1}{2}a^{2}$. 

Since the full prepotential approaches its perturbative value in the $u\rightarrow \infty$ limit we have that here, $\F(a)\rightarrow \frac{i}{2\pi}a^{2}\log\left(\frac{a^{2}}{\Lambda^{2}}\right)$, and as $\tau(a)=\F''(a)$, $\tau(a)\rightarrow \frac{i}{\pi}\left(\log\left(\frac{a^{2}}{\Lambda^{2}}\right)+3\right)$.

As a sanity check note that in this limit the metric is given by $\Im(\tau(a))\approx\frac{2}{\pi}\log|\frac{a}{\Lambda}|$, which is indeed positive definite and single valued. Furthermore, since $a_D=\frac{\partial \F(a)}{\partial a}$ we can compute $a_D(a)=\frac{i}{\pi}a\left(\log\left(\frac{a^{2}}{\Lambda^{2}}\right)+1\right)$.

In summary, as $u\to\infty$ we have:
\begin{align}
\begin{split}
&
a(u) \to \sqrt{2u}
\\
&
a_{D}(u) \to \frac{i}{\pi}\sqrt{2u}\left(\log\left(\frac{2u}{\Lambda^{2}}\right)+1\right)
\end{split}
\label{asym1}
\end{align}

The monodromy properties of $a$ and $a_D$ can be examined in this limit by encircling $\infty$ with a circular anti-clockwise contour on the Riemann sphere. To do so, we take $u\rightarrow e^{2\pi i t}u$, $t\rightarrow 1$ and find that $a_{D}\rightarrow -a_{D}+2a$. Similarly, we find that $a\rightarrow -a$. The monodromy transformation at infinity is thus given by 
$
M_{\infty}
=
\begin{pmatrix}
-1&2\\
0&-1
\end{pmatrix}
$
. Since this matrix is non trivial, the point $u=\infty$ is indeed a singularity of $\M$ and in particular a branch point. Branch cuts must end somewhere, so $\M$ must have at least one more singularity. In fact we will show that it has least three singularities and that these singularities come in pairs.
\subsection{$\R$-Symmetry in general.}
To show that the singularities of $\M$ come in pairs we must discuss a certain global symmetry of the action known as $\R-$symmetry. In general $\R-$symmetry refers to a global symmetry of a supersymmetric theory which acts on the supercharges. Symmetries of the action are symmetries of the corresponding classical theory and as such may become anomalous in the quantum regime with only a partial symmetry remaining unbroken. We now discuss the $\R$-symmetry of $\N=2$ SYM and its breaking pattern as well as the implications for the structure of $\M$.

The $\R-$symmetry group of $\N=2$ SYM is $U(1)_{\R}\times SU(2)_{\R}$, where the $SU(2)_{\R}$ subgroup rotates the two supercharges, while the $U(1)_{\R}$ subgroup acts on the Grassmann coordinates and fields in the following way:
$H$ and $\Phi$ have charge 2, $W,\theta$, and $\bar{\theta}$ have charge 1, and $d^{2}\theta$ and $d^{2}\bar{\theta}$ have charge $-2$ \cite{Bilal:2}. The $U(1)_{\R}$ subgroup turns out to be anomalous, it is broken both perturbatively and non-perturbatively.

For a general simple gauge group $G$ and a general $\N=2$ theory, $U(1)_{\R}$ is broken to a discrete $\mathbb{Z}_{\beta}$, where $\beta$ is the leading contribution of the $\beta$ function  \cite{Shadchin}. In a more abstract context $\beta$ is the following quantity (see for example Appendix B of \cite{Shadchin}):
$$
\beta=\zeta\left(l_{adj}-\sum_{\rho}l_{\rho}\right)
$$
where $l_{\rho}$ is defined in terms of the trace form $\text{Tr}_{\rho}$ and the Killing form $\text{Tr}_{adj}$ as follows: $l_{adj}\text{Tr}_{adj}=l_{\rho}\text{Tr}_{\rho}$, where $l_{fund}$ is normalised to unity. The factor $\zeta$ depends on the group in question (it is $1$ for $SU(N)$), and the sum is over the representations of the matter hypermultiplets included in the theory. Fo the case at hand there are no hypermultiplets and the gauge group is $SU(2)$, so as $l_{adj}=2N+2$ for $SU(N+1)$, we have that $\beta=4$, and thus $U(1)_{\R}$ is broken to $\mathbb{Z}_{4}$.
\subsection{$\R-$Symmetry for $\N=2$ SYM.}
The previous discussion is a bit abstract and we haven't proven anything. Luckily for the case at hand, the breaking pattern can be obtained without appealing to general results.

Under $U(1)_{\R}$ we have that $\F_{\text{pert}}$ transforms as $\F_{\text{pert}}\rightarrow e^{4i\alpha}\left(\frac{i}{2\pi}a^{2}\log\frac{a^{2}}{\Lambda^{2}}-a^{2}\frac{2\alpha}{\pi}\right)$, so $\delta S_{\text{eff}}=2\pi k\frac{4\alpha}{\pi}$. Thus by the same path integral argument as used in Section $2.2.2$, the action is invariant if and only if $\alpha=\frac{2\pi n}{8}$ for some $n\in\mathbb{Z}$.

We now discuss the transformation properties of $\F_{\text{inst}}$. To do so, note that the instanton part of the prepotential can be represented as an infinite sum of the form:
\begin{equation}
\F_{\text{inst}}(a,\Lambda)=\sum_{k=0}^{\infty}\Lambda^{\beta k}\F_{k}(a)
\end{equation}
for as of yet undetermined coefficients $\F_{k}\propto a^{2-4k}$. This follows from the renormalization group equations and invariance under the residual $\R$-symmetry \cite{Seiberg:1}.

The above series transforms under the $U(1)_{\R}$ symmetry as
$\F_{\text{inst}}\rightarrow a^{2}\sum_{n=1}^{\infty}c_{k}e^{8i\alpha(1-k)}(\frac{\Lambda}{a})^{4k}$, and so is once invariant if and only if $\alpha=\frac{2\pi n}{8}$ for some $n\in\mathbb{Z}$. This shows that only the $\mathbb{Z}_{8}$ subgroup of $U(1)_{\R}$ is a symmetry of the quantum theory.

Under the residual $\mathbb{Z}_{8}$ symmetry, $H\rightarrow e^{i\pi n}H$, for $n\in\mathbb{Z}$ so that for odd $n$, $H^{2}\rightarrow -H^{2}$. This means that for $u=\expval{\text{Tr} H^{2}}\neq 0$, $\mathbb{Z}_{8}$ is broken further to $\mathbb{Z}_{4}$, as claimed.
\subsection{Singularity Counting.}
We have shown that for a generic vacuum the residual $\R$-symmetry is $\mathbb{Z}_{4}$, while on $\M$ itself we have a full $\mathbb{Z}_{8}$ symmetry under which $u\rightarrow -u$. This implies that each singularity of $\M$ has a partner under $\R$-symmetry, with the only exceptions being the fixed points of this map, namely $0$ and $\infty$.

The only way to have just one additional singularity is if the second is at $u=0$, and in this case $0$ and $\infty$ must have identical monodromies. So since $a^{2}$ was not affected by $M_{\infty}$ it is not affected by any monodromy, and thus is a valid global coordinate. By the harmonic function argument of Section $2.2$ we see that this results in a contradiction, so indeed $\M$ must have at least three singularities.

From now on we assume that $\M$ has has exactly three singularities. In this case, there must be a pair of non-zero singularities $\pm u_{0}$ which are interchanged by the global $\mathbb{Z}_{8}$ symmetry. Note that in particular $u=0$ cannot be a singularity of $\M$.
\subsection{The $u\rightarrow\pm u_{0}$ limits.}
For points of $\M$ (i.e. vacua), with enhanced symmetry, the $SU(2)$ gauge symmetry of the full theory does not break all the way to $U(1)$, and thus the effective description as a $U(1)$ gauge theory as per \eqref{Seff} breaks down. Such points are thus singularities of $\M$. To detect points of enhanced symmetry, note that at these points the Higgs mechanism partially "turns off", resulting in extra massless particles. We can thus regard singularities of $\M$ as being caused by generically massive particles becoming massless.

Seiberg and Witten argue on general grounds that unlike the classical case, the strong coupling singularities cannot be due to massless gauge bosons \cite{SW:1}. Instead we assume that since they are the only other generically massive states in the theory, these singularities occur at points of $\M$ for which massive dyons become massless. Recall that an $(n_{m},n_{e})$-dyon is a soliton of magnetic charge $n_{m}$ and electric charge $n_{e}$.

To begin with assume that a $(1,0)$-dyon, that is a magnetic monopole, becomes massless. In this case \eqref{BPS} implies that $m^{2}=2|a_D|^{2}$, so this corresponds to $a_{D}=0$. We call the point where this occurs $u_{0}$.

Near $u_{0}$ the theory (in the dual description), consists of a massive but light hypermultiplet corresponding to the magnetic monopole coupled locally to the fundamental chiral multiplet \cite{SW:1}, this is exactly $\N=2$ SUSY QED for which the $\beta-$function is known: $\mu\frac{d}{d\mu}g_D=\frac{g_{D}^{3}}{8\pi^{2}}$ \cite{Bilal:2}. The energy scale $\mu$ is proportional to the mass of our monopole and thus to $a_D$, so using that $\Theta=0$ for SUSY QED we have that as $u\rightarrow u_{0}$:
$$
a_D\frac{d}{da_{D}}\tau_D=-\frac{i}{\pi}
$$
Using that $\tau_D=-\frac{da}{da_{D}}$ this ODE can be solved to find that to leading order, $a\approx a_{0}+\frac{i}{\pi}a_D\log a_D$. Since $a$ is singular near $u_{0}$, Seiberg-Witten duality says that $a_{D}$ should be a good coordinate there, and thus depend linearly on $u$:
\begin{align*}
&
a_{D}\approx c_{0}(u-u_{0})
\\
&
a\approx a_{0}+\frac{i}{\pi}(u-u_0)\log(u-u_0)
\end{align*}
The constants $a_{0}$ and $c_{0}$ will be determined later. Taking $(u-u_0)\rightarrow e^{2\pi i}(u-u_0)$, the corresponding monodromy matrix can be read off:
$
M_{u_{0}}
=
\begin{pmatrix}
1&0\\
-2&1
\end{pmatrix}
$
.

To find $M_{-u_0}$, we simply note that since $\M$ is assumed to have exactly three singularities, a contour about $\infty$ can be deformed to two contours encircling $u_0$ and $-u_0$. This situation is shown in Figure $2.1$ and gives the factorisation condition $M_{\infty}=M_{u_{0}}M_{-u_{0}}$ (up to a choice of base point, $P$), which is easily solved to find
$
M_{-u_{0}}
=
\begin{pmatrix}
-1&2\\
-2&3
\end{pmatrix}
$.
\begin{figure}
\centering
\begin{tikzpicture}
\draw (0,0) rectangle (6,6);
\draw[
        decoration={markings, mark=at position 0.75 with {\arrow[line width = 2pt]{>}}},
        postaction={decorate}
        ]
        (3,3) ellipse (2.2cm and 1.8cm);
\filldraw (3,4.8) circle[radius=1.5pt];
\node[above=1pt of {(3,4.8)}, outer sep=1pt] {$P$};
\draw[
        decoration={markings, mark=at position 0.25 with {\arrow[line width = 2pt]{>}}},
        postaction={decorate}
        ]
        (3,4.8) .. controls (3,1.5) and (7,1.5) .. cycle;
\filldraw (2,3) circle[radius=1.5pt];
\node[above=1pt of {(2,3)}, outer sep=1pt] {$-1$};
\draw[
        decoration={markings, mark=at position 0.25 with {\arrow[line width = 2pt]{>}}},
        postaction={decorate}
        ]
        (3,4.8) .. controls (-1,1.5) and (3,1.5) .. cycle;
\filldraw (4,3) circle[radius=1.5pt];
\node[above=1pt of {(4,3)}, outer sep=1pt] {$+1$};
\node[above=1pt of {(2,1.75)}, outer sep=1pt] {$M_{-1}$};
\node[above=1pt of {(4,1.75)}, outer sep=1pt] {$M_{+1}$};
\node[above=1pt of {(3,0.25)}, outer sep=1pt] {$M_{\infty}$};
\end{tikzpicture}
\caption{Monodromy factorisation on $\M$}
\end{figure}
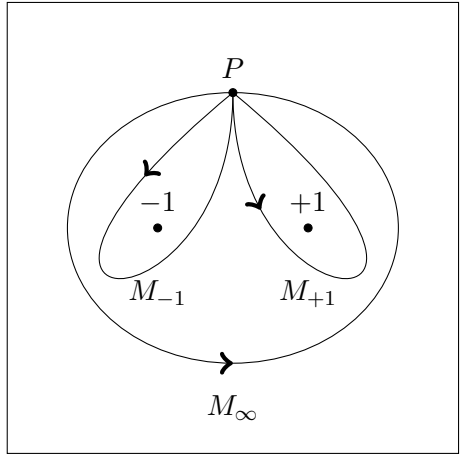

In summary, the monodromies associated to all three singularities are:
\begin{equation}
M_{\infty}
=
\begin{pmatrix}
-1&2\\
0&-1
\end{pmatrix}
\text{, }
\quad
M_{u_{0}}
=
\begin{pmatrix}
1&0\\
-2&1
\end{pmatrix}
\text{, }
\quad
M_{-u_{0}}
=
\begin{pmatrix}
-1&2\\
-2&3
\end{pmatrix}
\label{mono}
\end{equation}
We would like a physical interpretation for the singularity at $-u_{0}$. To this end, note that since mass is a physical observable, the BPS mass formula should be invariant under monodromy. Hence as $\mathcal{Z}=(n_{m},n_{e})(a_{D},a)^{T}$, the monodromy transformation $(a_{D},a)^{T}\to M(a_{D},a)^{T}$ can instead be interpreted as transforming the charge vector as $(n_{m},n_{e})\to (n_{m},n_{e})M$.

A state of zero mass should be invariant under the corresponding monodromy, so for such a state $(n_{m},n_{e})M=(n_{m},n_{e})$, i.e. the charge vector of such a state should be the left eigenvector of the corresponding monodromy matrix. We can thus identify the massless state responsible for $-u_{0}$'s singular behaviour with the left eigenvector $(1,-1)$ of $M_{-u_{0}}$. The physical interpretation is that this singularity is due to a Dyon with charge $(n_{e},n_{m})=(1,-1)$ becoming massless.

We finish this section by briefly stating what happens if $\M$ is assumed to have more singularities. If $\M$ has $p$ singularities it can be shown by a similar argument to the above that they must factorise as $M_{\infty}=M_{u_1}\ldots M_{u_p}$, with 
$
M_{u_i}
=
\begin{pmatrix}
1+2n_m n_e&2n_e^2\\
-2n_m^2&1-2n_m n_e
\end{pmatrix}
$
, and $(n_m,n_e)\in\mathbb{Z}^{2}$. It is considered likely that this system has no solutions for $p>3$ \cite{Bilal:1}.
\section{The Solution}
With the three monodromy matrices \eqref{mono} at hand, the prepotential can now be determined. We will do so in two ways. For the first method we will identify $a$ and $a_D$ with the solutions of a certain differential equation \cite{Bilal:1}. The second method involves identifying $\M$ with a certain Riemann surface for which $a$ and $a_D$ are the periods \cite{SW:1}.
\begin{remark}{2.4}
From now on we take $\pm u_0=\pm 1$. This corresponds to a specific choice of $\Lambda$ and otherwise leaves the discussion unaffected.
\end{remark}
\subsection{The differential equation approach.}

It is a well known fact that functions with non-trivial constant monodromies arise from ODEs with periodic (in the real case), or meromorphic (in the complex case), coefficients and at most regular singular points. For example take the ODE $[\del_{uu}+V(u)]\psi(u)=0$ and fix two linearly independent solutions $\psi_{1,2}(u)$. If $V$ is meromorphic, then encircling any singularity $u_{i}$ leaves the ODE invariant so the rotated solutions must be some linear combination of the non-rotated solutions:
$$
\begin{pmatrix}
\psi_1\\
\psi_2
\end{pmatrix}
(u+e^{2\pi i}(u-u_{i}))=M_{i}
\begin{pmatrix}
\psi_1\\
\psi_2
\end{pmatrix}
(u)
$$
Where $M_i$ is some monodromy matrix. It is well known that these monodromies are non-trivial and constant if $V$ has only regular singular points.
\subsubsection{The form of $V(u)$.}
We now assume that $a$ and $a_D$ are the solutions to a differential equation of the form $[\del_{uu}+V(u)]\psi(u)=0$. For justification, recall that any ODE with 3 regular singular points can be transformed into some hypergeometric equation. Then since with an appropriate change of variables and choice of $V$ the ODE $[\del_{uu}+V(u)]\psi(u)=0$ becomes an arbitrary hypergeometric equation (as we will see later), the assumption holds.

The known monodromies and the assumption that the ODE has three regular singular points $u_{i}\in\{\pm 1,\infty\}$ leads to severe constraints on the form of $V$. Firstly changing variables to $w=1/u$ we find:
$$
\psi''(w)+\frac{2}{w}\psi'(w)-\frac{1}{w^4}V\left(\frac{1}{w}\right)\psi(w)=0
$$
So for $u=\infty$ to be a regular singular point, $V(\frac{1}{w})$ must be $\bigO(w^2)=\bigO(\frac{1}{u^2})$. This shows that first order poles (unless they are in a product such as the term $\frac{1}{(u-1)(u+1)}$), and regular points lead to essential singularities at infinity and so must be excluded. Poles of degree greater than 2 are also excluded since they correspond to essential singularities at $u=\pm 1$. Thus $V$ must be of the form:
$$
V(u)
=
-\frac{1}{4}
\left[
\frac{1-\lambda_{1}^{2}}{(u+1)^{2}}
+
\frac{1-\lambda_{2}^{2}}{(u-1)^{2}}
-
\frac{1-\lambda_{1}^{2}-\lambda_{2}^{2}+\lambda_{3}^{2}}{(u+1)(u-1)}
\right]
\text{, }\quad\lambda_{i}\in\mathbb{C}
$$
The $\lambda_i\in\mathbb{C}$ will eventually be fixed by enforcing the correct monodromy properties.
\subsubsection{Solving the ODE.}
We now solve this ODE and determine the $\lambda_{i}$ by transforming it to a hypergeometric equation. To do so,  set $\psi(u)=(u+1)^{(1-\lambda_{1})/2}(u-1)^{(1-\lambda_{2})/2}f(\frac{u+1}{2})$, and take $x=(u+1)/2$. The ODE then becomes:
$$
x(1-x)f''(x)+[c_{3}-(c_{1}+c_{2}+1)x]f'(x)-c_{1}c_{2}f(x)=0
$$
Where $c_{1}=(1-\lambda_1-\lambda_2+\lambda_3)/2$, $c_{2}=(1-\lambda_1-\lambda_2-\lambda_3)/2$, and $c_{3}=1-\lambda_{1}$. This is indeed the hypergeometric equation. We pick the following basis of solutions:
\begin{align*}
&
f_{1}(x)
=
(-x)^{-c_{1}}F\left(c_{1},c_{1}+1-c_{3},c_{1}+1-c_{2};\frac{1}{x}\right)
\\
&
f_2(x)
=
(1-x)^{c_{3}-c_{1}-c_{2}}
F(c_{3}-c_{1},c_{3}-c_{2},c_{3}+1-c_{1}-c_{2};1-x)
\end{align*}
The known asymptotic behaviour can now be used to fix the $\lambda_{i}$ and match these solutions with $a$ and $a_{D}$.

Firstly, as $x\rightarrow \infty$, $V(u)\sim -\frac{1-\lambda_{3}^{2}}{(2u)^{2}}$, resulting in a Cauchy-Euler equation for $\psi$. This is easy to solve:
$$
\psi(u)\sim
\begin{cases} 
      Au^{(1+\lambda_3)/2}+Bu^{(1-\lambda_{3})/2} & \lambda_{3}\neq 0\\
      Au^{1/2}+Bu^{1/2}\log(u) & \lambda_3=0\\
\end{cases}
\text{, }
\quad
A,B\in\mathbb{C}
$$
Only the $\lambda_3=0$ solution can match the known asymptotics \eqref{asym1} as $u\rightarrow\infty$, so we conclude that $\lambda_3=0$.

Next we consider what happens as $u\rightarrow 1$. In this limit the ODE is as follows:
$$
\psi''(u)
=
-
\frac{1-\lambda_{2}^{2}}{(u-1)^{2}}\psi(u)
+
\frac{1-\lambda_1^2-\lambda_2^2}{8(u-1)}\psi(u)
+
\bigO(1)
$$
Recalling that $a_{D}(u)\approx c_0(u-1)$ as $u\rightarrow 1$, $a_{D}(u)$ can only be a solution if as $u\rightarrow 1$ we have:
$$
0
=
-\frac{c_0(1-\lambda_2^2)}{u-1}
+
\frac{c_0}{8}(1-\lambda_1^2-\lambda_2^2)
+
\bigO(1)
$$
This is only possible for $\lambda_2=1$.

Finally, the fact that $u\rightarrow -u$ is a symmetry of $\M$ implies that $V$ should be an even function. This immediately yields $\lambda_1=1$.

In summary, $\lambda_{1}=\lambda_{2}=1$ and $\lambda_{3}=0$, so $c_{1}=c_{2}=-\frac{1}{2}$, and $c_{3}=0$. Furthermore $V(u)$ is now entirely fixed: $V(u)=-\frac{1}{4(u+1)(u-1)}$.
\subsubsection{The result.}
The solutions $\psi_{1,2}$ can still be scaled by constants to get the correct monodromies. It turns out that $a(u)=-2i\psi_{1}(u)$, and $a_{D}(u)=i\psi_{2}(u)$ are the correct choices as will be shown in Section $2.6$. Rewriting the associated hypergeometric functions in integral form gives:
\begin{align}
\begin{split}
a(u)
&=
\sqrt{2}\sqrt{u+1}F\left(-\frac{1}{2},\frac{1}{2},1;\frac{2}{u+1}\right)
=
\frac{\sqrt{2}}{\pi}\int_{-1}^{1}\frac{dx\sqrt{x-u}}{\sqrt{x^{2}-1}}
\\
a_{D}(u)
&=
\frac{i}{2}(u-1)F\left(\frac{1}{2},\frac{1}{2},2;\frac{1-u}{2}\right)
=
\frac{\sqrt{2}}{\pi}\int_{1}^{u}\frac{dx\sqrt{x-u}}{\sqrt{x^{2}-1}}
\end{split}
\label{result}
\end{align}
With $a(u)$ and $a_{D}(u)$ determined we have now implicitly determined the prepotential $\F$.
\subsection{The elliptic curve approach.}
We now present another way to obtain $\F$ from the monodromies. This method, due to Seiberg and Witten, relies on some facts from basic differential geometry and the theory of modular curves \cite{SW:1}. It is less direct than the differential equations approach but is more readily generalised to theories including matter multiplets \cite{SW:2}.
\subsubsection{Set up.}
The information we have is as follows: the metric on $\M$ is given by $ds^2=\Im(\tau)|da|^2$ where $\tau(u)=\frac{da_{D}/du}{da/du}$. This metric is positive definite: $\Im(\tau)>0$. Furthermore, $\M$ is the complex $u$-plane with a $\mathbb{Z}_{2}-$symmetry taking $u\to -u$ and singularities at $\{\pm1,\infty\}$. The coordinates $(a, a_D)$ on $\M$ are acted on by an $SL(2;\mathbb{Z})-$duality and have monodromies \eqref{mono} around the singularities of $\M$. 
\begin{remark}{2.5}
In the language of differential geometry this says that $(a,a_{D})$ forms a section of a flat $SL(2;\mathbb{Z})$ bundle over $\M$ with monodromies $M_{\pm 1}$ and $M_{\infty}$.
\end{remark}
\subsubsection{Identification of $\M$.}
The first thing to notice is that the three monodromies don't generate all of $SL(2;\mathbb{Z})$. In fact they generate the so-called principal congruence subgroup of level $2$:
$$
\langle M_{\pm1},M_{\infty}\rangle
=
\Gamma(2)
:=
\left\{
A\in SL(2;\mathbb{Z})|A\equiv\mathbbm{1}_{2\times 2}(\text{mod }2)
\right\}
\subseteq
SL(2;\mathbb{Z})
$$
There is an action of $\Gamma(2)$ on the upper half plane, $H$ and in fact $\M\cong H/\Gamma(2)$ \cite{Koblitz}. Next note that the space $H/\Gamma(2)$ parametrises the family of complex elliptic curves $E_{u}$ defined by $y^{2}=(x-1)(x+1)(x-u)$ so that each point $u\in\M$ can be associated to an elliptic curve $E_{u}$ \cite{SW:1}.
\begin{remark}{2.6}
For a sanity check, note that the equation defining $E_{u}$ is invariant under the symmetry group generated by $\{u\rightarrow -u,x\rightarrow -x,y\rightarrow \pm i y\}$, which is isomorphic to $\mathbb{Z}_{4}$. Of this symmetry, only the $\mathbb{Z}_{2}$ subgroup acts on $u$ (i.e. on all of $\M$). This is precisely the symmetry structure identified earlier.
\end{remark}
\begin{remark}{2.7}
The curve $y^{2}=(x-1)(x+1)(x-u)$ is known as the Seiberg-Witten curve for gauge group $SU(2)$.
\end{remark}
A given $E_{u}$ is essentially the surface on which $y(x)$ becomes a single valued function. Since the equation for $E_{u}$ is quadratic in $y$, encircling $\pm 1$, $u$, or $\infty$ in the $x-$plane takes $y\rightarrow -y$. For example if $x=1$ is encircled by a circle of small radius, then writing $x=1+\delta x$ and taking $(x-1)\rightarrow e^{2\pi i}(x-1)$, we have:
$$
y
=
\sqrt{(x-1)(x+1)(x-u)}
\sim
\sqrt{2\delta x(1-u)}
\rightarrow
\sqrt{2e^{2\pi i}\delta x(1-u)}
=
-\sqrt{2\delta x(1-u)}
=
-y+\bigO(\delta x)^{2}
$$
Thus the $x-$space $E_{u}$ should be a double cover of the complex plane with points at infinity added. Furthermore it should have square root branch points at $\pm 1$, $u$, and $\infty$, which we join pairwise by two cuts. Finally the two sheets are joined along these cuts, that is to say crossing a cut takes us from one sheet to the other. Thus we have that for generic $u$ (that is $u$ not a singularity of $\M$), this Riemann surface is a torus, as shown in Figure $2.2$.
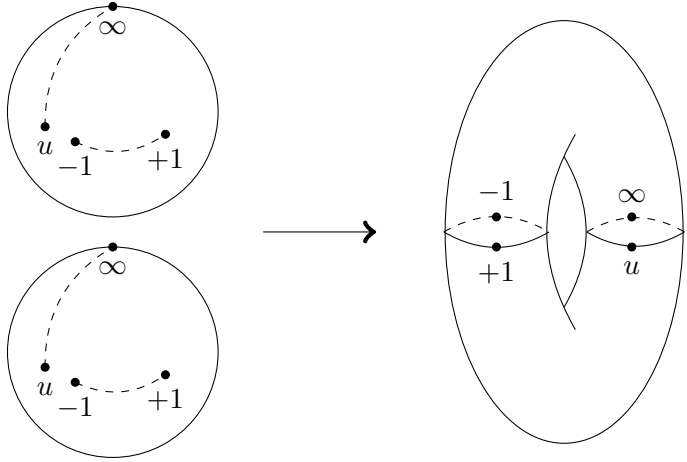
\begin{figure}
\centering
\begin{tikzpicture}
\draw (0,0) ellipse (45pt and 80pt);
\draw (0.15,-1.3) to[bend left] (0.15,1.3);
\draw (0.0,-1.0) to[bend right] (0.0,1.0);
\draw (-1.6,0) to[bend right] (-0.2,0);
\draw[dashed] (-1.6,0) to[bend left] (-0.2,0);
\draw (0.3,0) to[bend right] (1.6,0);
\draw[dashed] (0.3,0) to[bend left] (1.6,0);
\filldraw (-0.9,0.2) circle[radius=1.5pt];
\node[above=1pt of {(-0.9,0.2)}, outer sep=1pt] {$-1$};
\filldraw (-0.9,-0.2) circle[radius=1.5pt];
\node[below=1pt of {(-0.9,-0.2)}, outer sep=1pt] {$+1$};
\filldraw (0.9,0.2) circle[radius=1.5pt];
\node[above=1pt of {(0.9,0.2)}, outer sep=1pt] {$\infty$};
\filldraw (0.9,-0.2) circle[radius=1.5pt];
\node[below=1pt of {(0.9,-0.2)}, outer sep=1pt] {$u$};
\draw (-6,1.6) circle (1.4cm);
\draw[dashed] (-6.5,1.2) to[bend right] (-5.3,1.3);
\filldraw (-6.5,1.2) circle[radius=1.5pt];
\node[below=1pt of {(-6.5,1.2)}, outer sep=1pt] {$-1$};
\filldraw (-5.3,1.3) circle[radius=1.5pt];
\node[below=1pt of {(-5.3,1.3)}, outer sep=1pt] {$+1$};
\draw[dashed] (-6.9,1.4) to[bend left] (-6,3);
\filldraw (-6,3) circle[radius=1.5pt];
\node[below=1pt of {(-6,3)}, outer sep=1pt] {$\infty$};
\filldraw (-6.9,1.4) circle[radius=1.5pt];
\node[below=1pt of {(-6.9,1.4)}, outer sep=1pt] {$u$};
\draw (-6,-1.6) circle (1.4cm);
\draw[dashed] (-6.5,-2) to[bend right] (-5.3,-1.9);
\filldraw (-6.5,-2) circle[radius=1.5pt];
\node[below=1pt of {(-6.5,-2)}, outer sep=1pt] {$-1$};
\filldraw (-5.3,-1.9) circle[radius=1.5pt];
\node[below=1pt of {(-5.3,-1.9)}, outer sep=1pt] {$+1$};
\draw[dashed] (-6.9,-1.8) to[bend left] (-6,-0.2);
\filldraw (-6,-0.2) circle[radius=1.5pt];
\node[below=1pt of {(-6,-0.2)}, outer sep=1pt] {$\infty$};
\filldraw (-6.9,-1.8) circle[radius=1.5pt];
\node[below=1pt of {(-6.9,-1.8)}, outer sep=1pt] {$u$};
\draw[
        decoration={markings, mark=at position 1.00 with {\arrow[line width = 2pt]{>}}},
        postaction={decorate}
        ]
        (-4,0) -- (-2.5,0);
\end{tikzpicture}
\caption{The Riemann Surface $E_{u}$ as a torus}
\end{figure}
\subsubsection{The periods of $E_{u}$.}
We now seek to relate the family of tori $E_{u}$ to $a$ and $a_{D}$. To do so note that encircling the cut from $-1$ to $1$ corresponds to one of the basic cycles of the torus, while the other basic cycle corresponds to travelling from $1$ to $u$ on the first sheet, then back from $u$ to $1$ on the second sheet. This is clear from Figure $2.2$. We call the first cycle $\gamma_1$, and the second $\gamma_2$.

Since $E_{u}$ is generically a torus, the independent cycles $\gamma_{1},\gamma_{2}$ form a basis for the family of first homology groups $V_{u}:=H_{1}(E_{u};\mathbb{C})\cong \mathbb{Z}\oplus\mathbb{Z}$. De Rham's theorem then allows us to identify $\gamma_{1}$ and $\gamma_{2}$ with differential forms on $E_{u}$. In particular, it says that $V_{u}\cong H_{dR}^{1}(E_{u};\mathbb{C})$, that is the cycles on $E_{u}$ are in one to one correspondence with closed $1$-forms modulo exact $1$-forms. A basis for the first cohomology group is thus provided by the following $1$-forms:
$$
\lambda_{1}
=
\frac{dx}{y}
\quad
\lambda_{2}
=
\frac{xdx}{y}
$$
The periods of $E_{u}$ are $\Omega_{i}^{j}=\int_{\gamma_{i}}\lambda^{j}$. Letting $b_{i}=\Omega_{i}^{1}$ it is a well known fact of differential geometry that for a torus $\frac{b_1}{b_2}=\tau_{u}$ where $\tau_{u}$ is the modular parameter of said torus.
\subsubsection{Identification of the metric.}
In general, $\tau_{u}$ has the fundamental property that $\Im \tau_{u}>0$. Since this property is shared by the complexified coupling $\tau(u)$, we seek to identify these quantities.

Consider a general element $\lambda=a_1(u)\lambda_1+a_2(u)\lambda_2\in H_{dR}^{1}(E_{u};\mathbb{C})$. Define $a_{D}=\oint_{\gamma_1}\lambda$, and $a=\oint_{\gamma_2}\lambda$. We claim that $\tau(u)$ has $\Im(\tau(u))>0$ if and only if $\frac{d\lambda}{du}=f(u)\lambda_{1}$ for some function $f$.

For one implication, assume that $\frac{d\lambda}{du}=f(u)\lambda_{1}$ for some function $f$. In this case
$
\frac{da_D}{du}
=
\int_{\gamma_1}\frac{d\lambda}{du}
=
f(u)b_{1}
$
, and similarly 
$\frac{da}{du}=f(u)b_{2}$
so:
$$
\tau(u)=\frac{da_D/du}{da/du}=\frac{b_1}{b_2}=\tau_u
$$
So we have identified the modular parameter of $E_{u}$ with the complexified coupling $\tau(u)$, and thus also ensured that $\Im \tau >0$.

For the other implication, assume that $\Im\tau>0$ everywhere. Then for each $u\in\M$, $\tau(u)$ is the modular parameter of some elliptic curve. General considerations show that the family of curves thus determined have the same monodromies and singularities as those determined by $\tau_{u}$. It can then be shown that the two families coincide and thus $\tau(u)=\tau_{u}$ \cite{SW:1}.

The condition $\tau(u)=\tau_{u}$ implies that $\frac{da_{D}}{du}=\left(\frac{1}{b_{2}}\frac{da}{du}\right)b_{1}$ and $\frac{da}{du}=\left(\frac{1}{b_{1}}\frac{da_{D}}{du}\right)b_{2}$, so 
$$
\frac{1}{b_{2}}\frac{da}{du}=\frac{1}{b_{1}}\frac{da_{D}}{du}:=f(u)
$$
Thus $f(u)b_{i}=\oint_{\gamma_{i}}\frac{d\lambda}{du}$, and so $\oint_{\lambda_{i}}\left(f(u)\lambda_{1}-\frac{d\lambda}{du}\right)=0$. $\lambda_{1}$ is the unique holomorphic differential on $E_{u}$ so up to a redefinition of $f$ the integrand vanishes, as required.

It turns out that $f$ is entirely fixed by the asymptotic behaviour of $a$ and $a_{D}$ near the singularities. In fact we claim that $f(u)=-\frac{\sqrt{2}}{4\pi}$ is the unique choice with the correct properties. To verify that this is the case we need only to show that this choice of $f$ gives the correct monodromy properties. It is then clear that any other choice would have somewhere introduced extra poles or zeros. The monodromy properties will be verified in Section $2.6$.
\subsubsection{Identification of $a(u)$ and $a_{D}(u)$.}
To determine $a$ and $a_{D}$ explicitly, note that:
$$
\frac{d\lambda}{du}
=
f(u)\frac{dx}{y}
=
-
\frac{\sqrt{2}}{4\pi}\frac{dx}{\sqrt{(x-1)(x+1)(x-u)}}
$$
So upon integration $\lambda=\frac{\sqrt{2}}{2\pi}\sqrt{\frac{x-u}{x^{2}-1}}dx$. The constant of integration has been set to zero since constants are entire functions and so would not contribute to $a$ or $a_{D}$. Deforming the $\gamma_{i}$ to lie along the branch cuts then gives:
$$
a_{D}(u)
=
\oint_{\gamma_{1}}\frac{\sqrt{2}}{2\pi}\sqrt{\frac{x-u}{x^{2}-1}}dx
=
\frac{\sqrt{2}}{2\pi}
\left(
\int_{1}^{u}\sqrt{\frac{x-u}{x^{2}-1}}dx
-
\int_{u}^{1}\sqrt{\frac{x-u}{x^{2}-1}}dx
\right)
=
\frac{\sqrt{2}}{\pi}
\int_{1}^{u}\sqrt{\frac{x-u}{x^{2}-1}}dx
$$
In the above calculation the negative sign comes from traversing the cut in the opposite direction. Similarly we find that $a(u)=\frac{\sqrt{2}}{\pi}\int_{-1}^{1}\sqrt{\frac{x-u}{x^{2}-1}}dx$.

This is indeed the same result \eqref{result} obtained previously from the differential equation approach. All that is left now is to verify that these solutions have the correct monodromy properties.
\section{Verification of Monodromy Properties}
To conclude that the expressions obtained for $a(u)$ and $a_{D}(u)$ are indeed correct, it must be verified that they have the correct monodromy properties.
\subsection{The $u\rightarrow\infty$ limit.}
As $u\rightarrow \infty$, we have:
$$
a(u)
=
\frac{\sqrt{2}}{\pi}\int_{-1}^{1}\sqrt{\frac{x-u}{x^{2}-1}}dx
\approx
\frac{\sqrt{2}}{\pi}\int_{-1}^{1}\sqrt{\frac{u}{1-x^{2}}}dx
=\sqrt{2u}
$$
This result agrees with the assertion that $u=\frac{1}{2}a^{2}$ in the semiclassical region and has the correct monodromy. To examine $a_{D}(u)$ in this limit, set $x=uz$ and note that $u\rightarrow\infty$ implies $z\rightarrow 0$ for finite $x$:
$$
a_{D}(u)
=
\frac{\sqrt{2u}}{\pi}
\int_{1/u}^{1}\sqrt{\frac{z-1}{(z+u^{-1})(z-u^{-1})}}dz
\approx
\frac{\sqrt{2u}}{\pi}
\int_{1/u}^{1}\sqrt{\frac{z-1}{z^{2}}}dz
\approx
i\frac{\sqrt{2u}}{\pi}
\int_{1/u}^{1}\frac{dz}{z}
$$
So as $u\rightarrow\infty$, $a_{D}(u)\approx\frac{\sqrt{2u}}{\pi}i(\log(1)-\log(1/u))\approx \frac{\sqrt{2u}}{\pi}i\log(u)$ in accordance with our previous findings. It is easy to see that this form also has the correct monodromy.
\subsection{The $u\rightarrow \pm 1$ limits.}
For $u\rightarrow 1$, we again use that
$
a_{D}(u)
=
\frac{\sqrt{2u}}{\pi}
\int_{1/u}^{1}\sqrt{\frac{z-1}{(z+u^{-1})(z-u^{-1})}}dz
$. Note that as $u\rightarrow 1$ with $z\in(u^{-1},1)$, $z-1$ and $z-u^{-1}$ both vanish but $z+u^{-1}\rightarrow 2$, so:
$$
a_{D}(u)
=
\frac{\sqrt{2u}}{\pi}
\int_{1/u}^{1}\sqrt{\frac{z-1}{(z+u^{-1})(z-u^{-1})}}dz
\approx
\frac{\sqrt{2}}{\pi}
\int_{1/u}^{1}\sqrt{\frac{z-1}{2(z-u^{-1})}}dz
=
\frac{1}{\pi}
\int_{1/u}^{1}\sqrt{\frac{z-1}{z-u^{-1}}}dz
$$
This integral can be evaluated:
\begin{align*}
\frac{1}{\pi}
\int\sqrt{\frac{z-1}{z-u^{-1}}}dz
&=
\frac{1}{\pi}
\int\sqrt{\frac{z-1}{z-u^{-1}}}dz
\\
&=
\frac{-1}{\pi}
\int\frac{1}{x^{2}}\sqrt{x(u^{-1}-1)+1}dx
\text{ , where }x=\frac{1}{z-u^{-1}}
\\
&=
\frac{2(1-u^{-1})}{\pi}
\int\frac{y^{2}}{(y^{2}-1)^{2}}dy
\text{ , where }y=\sqrt{x(u^{-1}-1)+1}
\\
&=
\frac{1-u^{-1}}{2\pi}
\int
\left[
\frac{1}{y-1}
-
\frac{1}{y+1}
+
\frac{1}{(y-1)^{2}}
+
\frac{1}{(y+1)^{2}}
\right]
dy
\\
&=
\frac{1-u^{-1}}{2\pi}
\left[
-\log(\frac{y+1}{y-1})
-
\frac{2y}{y^{2}-1}
\right]
\\
&=
\frac{u^{-1}-1}{\pi}
\left[
\arctanh\left(\sqrt{\frac{u^{-1}-1}{z-u^{-1}}+1}\right)
+
\frac{z-u^{-1}}{u^{-1}-1}
\sqrt{\frac{u^{-1}-1}{z-u^{-1}}+1}
\right]
\end{align*}
So as $u\rightarrow 1$:
$$
a_{D}(u)
\approx
\frac{1-u^{-1}}{\pi}\lim_{z\rightarrow u^{-1}}\arctanh\left(i\sqrt{\frac{1-u^{-1}}{z-u^{-1}}-1}\right)
=
\frac{i}{2}(1-u^{-1})
\approx
\frac{i}{2}(u-1)
$$
This expression has the correct monodromy and furthermore gives that $c_{0}=\frac{i}{2}$.

Finally, we examine $a(u)$ as $u\rightarrow 1$. Unlike the other quantities, $a(1)$ can be computed exactly:
$$
a(1)
=
\frac{\sqrt{2}}{\pi}\int_{-1}^{1}\sqrt{\frac{x-1}{x^{2}-1}}dx
=
\frac{\sqrt{2}}{\pi}\int_{-1}^{1}\frac{dx}{\sqrt{x+1}}
=
\frac{4}{\pi}
$$
However, to determine the monodromy the leading order non-constant term is required. To obtain it, consider $a'(u)$ and integrate:
$$
a'(u)
=
-\frac{\sqrt{2}}{2\pi}\int_{-1}^{1}
\frac{dx}{\sqrt{(x+1)(x-1)(x-u)}}
$$
Near $u=1$ this integral develops a factor of $1/(x-1)$, leading to a logarithmic divergence at $x=1$:
$$
a'(u)
\approx
-\frac{1}{2\pi}
\int_{-1}^{1}
\frac{dx}{(x-1)(x-u)}
=
-\frac{1}{2\pi}\log(2x-1-u+2\sqrt{(x-1)(x-u)})\biggr|^{1}_{-1}
=
-\frac{1}{2\pi}\log(1-u)+\bigO(1)
$$
Upon integrating and using the expression for $a(1)$ we have:
$$
a(u)
=
-\frac{1}{2\pi}(u-1)\log(1-u)+C+\bigO(u)
=
\frac{4}{\pi}-\frac{1}{2\pi}(u-1)\log(1-u)+\bigO((u-1)^2\log(1-u))
$$
Thus $a(u)\approx\frac{4}{\pi}-\frac{u-1}{2\pi}\log(u-1)$, which yields the correct monodromy and also gives $a_{0}=\frac{4}{\pi}$.

Rather than analysing the $u\rightarrow -1$ limit, note that the $u\rightarrow -u$ symmetry fixes the  $u=-1$ monodromies. Thus all the monodromies have been verified and we are done!
\section{Re-expression as a Power Series}
We have now in principle determined $\F$, however it is expressed in a very indirect form. To obtain an explicit form, invert $a(u)$ to get $u(a)$, and substitute to get $a_D(a)=\frac{\del\F}{\del a}$. Integrating with respect to $a$ then yields $\F(a)$. These steps then result in an explicit power series for the prepotential. The result of this process is:
$$
\F_{\text{inst}}(a)
=
-\frac{1}{2}\frac{\Lambda^{4}}{a^{2}}
-
\frac{5}{64}
\frac{\Lambda^{8}}{a^6}
-
\frac{3}{64}
\frac{\Lambda^{12}}{a^{10}}
+
\bigO(\Lambda^{16})
$$
Deriving this result is a straightforward but tedious exercise in manipulating power series and so is omitted for brevity. For details see \cite{DHoker:1} and Appendix A of \cite{Tachikawa:1}.
\chapter{Localisation}
At this point we turn to the second technique for computing the prepotential of $\N=2$ SYM; localisation. Localisation involves showing that the partition function path integral only receives contributions from some subspace of the space of fields, that is to say it "localises". This reduces the path integral to a lower dimensional integral. For the case of $\N=2$ SYM the resulting integral is finite dimensional.

Localisation is a more direct approach than Seiberg-Witten theory. It gives $\F$ as a power series in $a$ from the get-go, and can be more easily generalised to other gauge groups. On the other hand it is much more complicated.
\section{Equivariant Cohomology}
As a warm-up for supersymmetric localisation we first introduce a "toy model"; the bosonic localisation of finite dimensional integrals with abelian symmetry. From here it is surprisingly only a mild generalisation to obtain localisation formulae for path integrals. The discussion in this section follows \cite{Cremonesi}.

To start, consider a $2l-$dimensional boundaryless Riemannian manifold $(M,g)$ with a symmetry group $G$. We want to reduce integrals over $M$ to integrals over the lower dimensional quotient space $M/G$.

In general $G$ may not act freely (indeed this is the case for most applications of interest), so this quotient may not be a manifold. In general it is an orbifold, a generalisation of a manifold to include singularities. This complication is dealt with by introducing the notions of equivariant differential forms and equivariant cohomology. These ideas generalise the usual cohomology of manifolds to include singularities.

In this discussion we take $G=U(1)$, although what follows can be generalised to the non-abelian case.

Let $V=V^{\mu}\partial_{\mu}$ be a Killing vector on $M$ and assume that it generates the $U(1)$ symmetry. By definition this means that $\Lag_{V}g=0$, where $\Lag_{V}$ is the Lie derivative along $V$. Equivalently we have that $\nabla_{\mu}V_{\nu}+\nabla_{\nu}V_{\mu}=0$, where $\nabla_{\mu}V_{\nu}=V_{\nu,\mu}-\Gamma^{\lambda}_{\mu\nu}V_{\lambda}$ is the covariant derivative with Levi-Civita connection.

Now let $\bigwedge^{n}M$ be the space of differential $n-$forms on $M$ and let $\bigwedge M=\bigoplus_{n=0}^{\infty}\bigwedge^{n}M$ be the space of polyforms on $M$. We define the $V-$equivariant differential, $d_{V}:\bigwedge M\rightarrow\bigwedge M\text{, }d_{V}=d-\iota_{V}$, where $d:\bigwedge^{n}M\rightarrow \bigwedge^{n+1}M$ is the usual exterior differential and $\iota_{V}:\bigwedge^{n}M\rightarrow\bigwedge^{n-1}M$ is contraction in the first slot with the vector $V$.

\begin{remark}{3.1}
Remarkably, $d_{V}^{2}=-\Lag_{V}$. This is easy to prove: $d^{2}=0$ by definition and $\iota_{V}^{2}=0$ by the antisymmetry of forms. The final step is to verify that $d\iota_{V}+\iota_{V}d=\Lag_{V}$, which is somewhat tedious but straightforward, see for example \cite{SG}.
\end{remark}

The space of equivariant polyforms is:
$$
\bigwedge\nolimits_{V}M=\{\alpha\in\bigwedge M|\Lag_{V}\alpha=0\}
$$
Remark $3.1$ shows that $(d_{V}|_{\bigwedge_{V}M})^{2}=0$, that is to say the equivariant differential acts as a coboundary operator on $\bigwedge_{V}M$. It thus makes sense to define equivariantly closed and equivariantly exact forms analogously to the usual case but with $d_{V}$ instead of just $d$. We then define the $n$'th $V$-equivariant de Rham cohomology group as the space of equivariantly closed $n-$forms modulo equivariantly exact $n-$forms:
$$
H_{V}^{n}(M)
:=
\ker\left(d_{V}|_{\wedge^{n}_{V}M}\right)
/
\im\left(d_{V}|_{\wedge^{n-1}_{V}M}\right)
$$
The point of this construction is that if the $U(1)$ action has no fixed points $H_{V}^{n}(M)=H_{dR}^{n}(M/U(1))$, while if it does, $H_{V}^{n}$ provides a well defined generalisation of the usual cohomology \cite{Cremonesi}.

Further motivation for introducing equivariant cohomology is provided by the behaviour of polyforms under integration. For a polyform $\alpha=\alpha_{2l}+\alpha_{2l-1}+\ldots+\alpha_{0}$, we define $\int_{M}\alpha=\int_{M}\alpha_{2l}$, where the right hand side is the usual integral.

For an equivariantly exact polyform $d_{V}\beta$, $d_{V}\beta=(d\beta_{2l-1})+(d\beta_{2l-2}-\iota_{V}\alpha_{2l})+\ldots+(d\alpha_{0}-\iota_{V}\alpha_{2})+(-\iota_{V}\alpha_{1})$ so the top-term is $d\beta_{2l-1}$ which is exact in the usual sense since $d^{2}=0$. Using Stokes' theorem and the fact that $M$ is boundaryless we have that $\int_{M}d_{V}\beta=\int_{M}d\beta_{2l-1}=0$, so for polyforms $\alpha$ and $\beta$:
$$
\int_{M}(\alpha+d_{V}\beta)
=
\int_{M}\alpha
$$
\begin{remark}{3.2}
We have shown that the integral is constant on members of an equivariant de Rham cohomology class.
\end{remark}
\section{Bosonic Localisation of Ordinary Integrals}
The zero locus of the Killing vector $V$ is the set of points fixed under the action of $V$:
$$
M_{V}
=
\{
x\in M|V|_{x}=0
\}
$$
Integrals of equivariantly closed polyforms localise to $M_{V}$. This will now be proven in two different ways.
\subsection{An indirect localisation argument.}
We claim that an equivariantly closed polyform $\alpha$ is equivariantly exact on $M\setminus M_{V}$. To prove this, define $\eta$ to be the $1-$form dual to $V$: $\eta=g(V,\cdot)=V_{\mu}dx^{\mu}$.

The 1-form $\eta$ is $V$-equivariant. To prove this fact let $Y$ be an arbitrary vector field, then:
$$
\Lag_{V}(Y)\eta
=
\Lag_{V}g(V,Y)
=
V(g(V,Y))
=
(\Lag_{V}g(V,\cdot))(Y)+g(V,\Lag_{V}Y)
=
(\Lag_{V}g(V,\cdot))(Y)+g(V,[V,Y])
$$
So 
Rewriting this expression in component form yields:
\begin{align*}
(\Lag_{V}\eta)(Y)
&=
V^{\rho}\del_{\rho}(g_{\mu\nu}V^{\mu}Y^{\nu})-g_{\mu\nu}V^{\mu}(V^{\rho}\del_{\rho}Y^{\nu}-Y^{\rho}\del_{\rho}V^{\nu})
\\
&=
V^{\rho}
\del_{\rho}(g_{\mu\nu}V^{\mu})Y^{\nu}
+
g_{\mu\nu}V^{\rho}V^{\mu}\del_{\rho}Y^{\nu}
-
g_{\mu\nu}V^{\mu}(V^{\rho}\del_{\rho}Y^{\nu}-Y^{\rho}\del_{\rho}V^{\nu})
\\
&=
(
V^{\nu}
\del_{\nu}(g_{\mu\rho}V^{\rho})
+
g_{\nu\rho}V^{\rho}\del_{\mu}V^{\nu}
)
Y^{\mu}
\\
&=
(
V^{\rho}
\del_{\rho}g_{\mu\nu}
+
g_{\mu\rho}\del_{\nu}V^{\rho}
+
g_{\rho\nu}\del_{\mu}V^{\rho}
)
V^{\nu}Y^{\mu}
\\
&=
(\Lag_{V}g)_{\mu\nu}V^{\nu}Y^{\mu}
\\
&=
0
\end{align*}
Since $Y$ is arbitrary, the above calculation shows that $\Lag_{V}\eta=0$ and so $\eta$ is equivariant.

It is easy to show that $d_{V}\eta=-|V|^{2}+d\eta=-|V|^{2}\left(1-\frac{d\eta}{|V|^{2}}\right)$. This expression can be inverted using the geometric series formula:
\begin{equation}
(d_{V}\eta)^{-1}
=
-\frac{1}{|V|^{2}}
\sum_{n=0}^{l}\left(\frac{d\eta}{|V|^{2}}\right)^{n}\label{geoser}
\end{equation}
The above series terminates since $d\eta$ is a $2-$form and $\dim M=2l$.

Equation $\eqref{geoser}$ holds only for $|V|\neq 0$, which is the case on $M\setminus M_{V}$. We now claim that $(d_{V}\eta)^{-1}$ is equivariantly closed on $M\setminus M_{V}$. Applying $d_{V}$ to $1=(d_{V}\eta)(d_{V}\eta)^{-1}$ gives:
$$
0
=
(d_{V}^{2}\eta)(d_{V}\eta)^{-1}
+
(d_{V}\eta)d_{V}(d_{V}\eta)^{-1}
=
0
-
(d\eta-|V|^{2})d_{V}(d_{V}\eta)^{-1}
$$
Since $|V|^{2}$ is a $0-$form and $d\eta$ is a $2-$form it must then be that $d_{V}(d_{V}\eta)^{-1}=0$, as claimed.

Finally we define the polyform $\Theta_{V}=\eta(d_{V}\eta)^{-1}$. Applying $d_{V}$ once gives
$$
d_{V}\Theta_{V}=(d_{V}\eta)(d_{V}\eta)^{-1}+\eta d_{V}(d_{V}\eta)^{-1}=1+0=1
$$
Applying it a second time gives $\Lag_{V}\Theta_{V}=-d_{V}^{2}\Theta_{V}=d_{V}(1)=0$, so $\Theta_{V}$ is equivariant on $M\setminus M_{V}$.

Let $\alpha$ be an equivariantly closed polyform on $M$. Then $\alpha=(d_{V}\Theta_{V})\alpha=d_{V}(\Theta_{V}\alpha)$, so $\alpha$ is equivariantly exact wherever $\Theta_{V}$ is defined, i.e. on $M\setminus M_{V}$. So since integrals of equivariantly exact forms vanish:
$$
\int_{M}\alpha
=
\int_{M\setminus M_{V}}\alpha
+\int_{M_{V}}\alpha
=
\int_{M\setminus M_{V}}d_{V}(\Theta_{V}\alpha)
+\int_{M_{V}}\alpha
=
0+
\int_{M_{V}}\alpha
=
\int_{M_{V}}\alpha
$$
So indeed the integral has localised to $M_{V}$.
\subsection{A direct localisation argument.}
Let $\alpha$ be an equivariantly closed polyform, and define $\alpha_{t}=\alpha e^{td_{V}\beta}$, where $\beta$ is a $V-$equivariant polyform. $\beta$ is arbitrary at this point, fixing it is called a choice of localisation scheme. Since $\alpha_{0}=\alpha$ and $\alpha_{t}$ is a continuous deformation of $\alpha$, $\alpha_{t}$ is cohomologous to $\alpha$, and:
$$
\int_{M}\alpha=\int_{M}\alpha_{t}
\text{ , }\quad\forall t\in\mathbb{R}
$$
We are free to take $t\rightarrow \infty$ as long as this limit exists. This will be the case if the $0$-form term of $d_{V}\beta$ is non-positive with maxima equal to $0$. A convenient choice  is $\beta=\eta$, yielding:
$$
\int_{M}\alpha=\lim_{t\rightarrow\infty}\int_{M}\alpha e^{td\eta}e^{-t|V|^{2}}
$$
To see why this process has localised the integral, note that $e^{td\eta}$ is a polynomial of degree $l$ in $t$, and $e^{-t|V|^{2}}$ is a Gaussian peaked at $M_{V}$. As $t\rightarrow \infty$, $e^{-t|V|^{2}}\rightarrow \delta(V)$, a delta function localised on $M_{V}$.
\section{The Bosonic Atiyah-Bott-Berline-Vergne Localisation Formula}
The second localisation argument can be extended to derive an explicit formula for the localised integral in the case in which $M_{V}$ consists of isolated points: $M_{V}=\{P_{k}\}$.

To derive the localisation formula we define local cartesian coordinates $(x_{i},y_{i})$ on $M$ near some $P_{k}\in M_{V}$, so the local metric is $ds^{2}\approx\sum_{i=1}^{l}(dx_{i}^{2}+dy_{i}^{2})$. In these coordinates the Killing vector is:
$$
V
\approx
\sum_{i=1}^{l}\omega_{P_{k},i}\left(-y_{i}\frac{\del}{\del x_{i}}+x_{i}\frac{\del}{\del y_{i}}\right)
=\sum_{i=1}^{l}\omega_{P_{k},i}\frac{\del}{\del\phi_{i}}
$$
To verify this, one can evaluate the local connection and show that $\Lag_{V}g=0$ locally.

By acting on $(x_{i},y_{i})^{T}$ and summing, one can show that the Killing vector generates the following transformation:
$$
e^{\omega_{P,i}\phi_{i}\del_{\phi_{i}}}
\begin{pmatrix}
x_{i}\\
y_{i}
\end{pmatrix}
=
\begin{pmatrix}
\cos(\omega_{P,i}\phi_{i})&-\sin(\omega_{P,i}\phi_{i})\\
\sin(\omega_{P,i}\phi_{i})&\cos(\omega_{P,i}\phi_{i})
\end{pmatrix}
\begin{pmatrix}
x_{i}\\
y_{i}
\end{pmatrix}
=:
R(\phi_{i})
\begin{pmatrix}
x_{i}\\
y_{i}
\end{pmatrix}
$$
This is an anticlockwise rotation by $\phi_{i}$ on the $i$'th eigenspace. The infinitesimal action of this rotation is specified by  the linear order expansion of the following equation:
$$
L_{V}\vec{x}
:=
\frac{\delta\vec{x}}{\phi_{i}}
=
\frac{1}{\phi_{i}}\left(R(\phi_{i})-1\right)\vec{x}
=
\begin{pmatrix}
-\omega_{P,i}y_{i}\\
\omega_{P,i}x_{i}
\end{pmatrix}
+\bigO(\phi_{i})
$$
The above equation is easily solved to yield
$
L_{V}
=
\begin{pmatrix}
0&-\omega_{P,i}\\
\omega_{P,i}&0
\end{pmatrix}
$
. 

In local coordinates $\eta$ and $d_{V}\eta$ are as follows:
$$
\eta
\approx
\sum_{i=1}^{l}\omega_{P,i}r_{i}^{2}d\phi_{i}
\text{, }
\quad
d_{V}\eta
\approx
\sum_{i=1}^{l}(\omega_{P,i}d(t_{i}^{2})\wedge d\phi_{i}-\omega_{P,i}^{2}r_{i}^{2})
$$
We can now evaluate the contribution to the integral from a neighbourhood $N_{P}$ of the point $P_{k}$:
\begin{align*}
\lim_{t\rightarrow \infty}\int_{N_{P}}\alpha e^{td_{V}\eta}
&=
\lim_{t\rightarrow \infty}\int_{N_{P}}\alpha e^{t\sum_{i=1}^{l}(\omega_{P,i}d(r_{i}^{2})\wedge d\phi_{i}-\omega_{P,i}^{2}r_{i}^{2})}
\\
&=
\lim_{t\rightarrow \infty}\int_{N_{P}}\alpha\prod_{i=1}^{l} e^{t(\omega_{P,i}d(r_{i}^{2})\wedge d\phi_{i}-\omega_{P,i}^{2}r_{i}^{2})}
\\
&=
\lim_{t\rightarrow \infty}\int_{N_{P}}(\alpha_{0}+\ldots+\alpha_{2l})\prod_{i=1}^{l}\biggr(1+t\omega_{P,i}d(r_{i}^{2})\wedge d\phi_{i}+\ldots
\\&\qquad\qquad\qquad\qquad\qquad\qquad\quad
+\frac{2}{l!}(t\omega_{P,i}d(r_{i}^{2})\wedge d\phi_{i})^{l}\biggr)e^{-t\omega_{P,i}^{2}r_{i}^{2}}
\\
&=
\lim_{t\rightarrow \infty}\alpha_{0}(P)t^{l}\prod_{i=1}^{l}\omega_{P,i}\int_{N_{P}}d(r_{i}^{2})\wedge d\phi_{i}e^{-t\omega_{P,i}^{2}r_{i}^{2}}
\\
&=
\lim_{t\rightarrow \infty}\alpha_{0}(P)t^{l}\prod_{i=1}^{l}\omega_{P,i}\int_{\mathbb{R}^{2}}d(r_{i}^{2})\wedge d\phi_{i}e^{-t\omega_{P,i}^{2}r_{i}^{2}}
\\
&=
\lim_{t\rightarrow \infty}\alpha_{0}(P)t^{l}\prod_{i=1}^{l}\omega_{P,i}\int_{0}^{\infty}d(r_{i}^{2})e^{-t\omega_{P,i}^{2}r_{i}^{2}}\int_{0}^{2\pi}d\phi_{i}
\\
&=
\lim_{t\rightarrow \infty}\alpha_{0}(P)\prod_{i=1}^{l}\frac{2\pi}{\omega_{P,i}}
\end{align*}
In the fourth equality we have noted that only top (i.e degree $2l$), terms contribute to the integral and have kept only the leading order in $t$ term since it is all that survives in the limit. In the fifth equality we have traded an integral over $N_{P}$ for one over all of $\mathbb{R}^{2}$ since in the limit the integrand vanishes off $N_{P}$.

The above expression can be rewritten so as to make its generalisation to the quantum case more straightforward. Recall the Pfaffian of a $2n\times 2n$ antisymmetric matrix $A$:
$$
Pf(A)
=
\frac{1}{2^{n}n!}\epsilon^{i_{1}\ldots i_{2n}}A_{i_{i}i_{2}}\ldots A_{i_{2n-1}i_{2n}}
$$
Note also that in general $Pf(A)^{2}=\det(A)$.

It is easy to verify that:
$$
\lim_{t\rightarrow \infty}\int_{N_{P}}\alpha e^{td_{V}\eta}
=
\frac{(2\pi)^{l}\alpha_{0}(P)}{Pf(-L_{V}(P))}
$$
So since the integral vanishes off $M_{V}$, we have by linearity:
\begin{equation}
\int_{M}\alpha=(2\pi)^{l}\sum_{x_{k}\in M_{V}}\frac{\alpha_{0}(x_{k})}{Pf(-L_{V}(x_{k}))}
\end{equation}
This is known as the Atiyah-Bott-Berline-Vergne localisation formula \cite{AB:1}. It has reduced an integral over a manifold with abelian symmetry to a sum of contributions from discrete points.
\section{Supersymmetric Localisation of Path Integrals}
The localisation arguments made in the previous section can be generalised to the case of supersymmetric path integrals. In fact, there is a direct correspondence between the various objects introduced in the equivariant case and those necessary in the SUSY case. This correpondence is presented in table $3.1$ \cite{Cremonesi}:
\begin{table}[h!]
  \begin{center}
    \caption{Equivariant and Supersymmetric Localisation Correspondence}
    \label{tab:table1}
    \begin{tabular}{l|c|r}
      \textbf{Equivariant Localisation} & \textbf{SUSY Localisation}\\
      \hline
      $d_{V}$ & $Q$ \\
      $d_{V}=-\Lag_{V}$ & $Q^{2}=B$\\
      Even polyforms/Odd polyforms & Bosons/Fermions\\
      $d_{V}\alpha=0$ & $Q\mathcal{O}=0$\\
      $\int_{M}\alpha=\int_{M}\alpha e^{t d_{V}\beta}$ with $\Lag_{V}\beta=0$ & $\int_{\mathfrak{F}}[\D X]\mathcal{O}e^{-S[X]}=\int_{\mathfrak{F}}[\D X]\mathcal{O}e^{-S[X]-tQ\mathcal{P}_{F}[X]}$ with $B\mathcal{P}_{F}[X]=0$\\
      $M_{V}$ & $\mathfrak{F}_{Q}$
    \end{tabular}
  \end{center}
\end{table}

In Table $3.1$, $Q$ is a fermionic supercharge which squares to some bosonic operator $B$, $\mathcal{O}$ is a BPS operator (an operator such that $Q\bigO=0$), and $\F$ is a space of fields. Finally, the localisation locus $M_{V}$ is swapped for the so-called "BPS-locus", $\mathfrak{F}_{Q}$ of supersymmetric field configurations.

We now present in very general terms some of the ideas of supersymmetric localisation before specialising to the case of $\N=2$ SYM in the next section.

Consider a SUSY gauge theory with supercharge $Q$ and action $S$. We wish to compute the expectation values of gauge invariant BPS observables:
\begin{equation}
\expval{\mathcal{O}_{BPS}}
=
\int_{\mathfrak{F}}[\D X]\mathcal{O}e^{iS[X]}
\end{equation}
\begin{remark}{3.3}
The partition function is given by the special case $\bigO_{BPS}=\mathbbm{1}$.
\end{remark}
As in the equivariant case, the expectation value of an operator depends only on its $Q-$cohomology class:
$$
\expval{Q\mathcal{O}}
:=
\int_{\mathfrak{F}}[\D X]Q(\mathcal{O})e^{iS[X]}
=
\int_{\mathfrak{F}}[\D X](Q(\mathcal{O}e^{iS[X]})+Qe^{iS[X]})
=
\int_{\mathfrak{F}}[\D X]Q(\mathcal{O}e^{iS[X]})
=
0
$$
In the above calculation we have used that the action of a SUSY theory is necessarily supersymmetric so that $QS[X]=0$, a supersymmetric generalisation of Stokes' theorem and the assumption that the fields decay sufficiently quickly at spatial infinity.

It is now clear that $\expval{\mathcal{O}_{BPS}+Q\mathcal{O}}=\expval{\mathcal{O}_{BPS}}$, that is the expectation value of a BPS operator in a supersymmetric theory depends only on the $Q$-cohomology class of said operator.
\subsection{SUSY path integrals localise.}
We now show that SUSY path integrals of BPS observables localise to the BPS locus. Again this can be done in two ways, by generalising either of the two arguments presented in Section $3.2$. Here we present briefly the generalisation of the first method, due to Witten \cite{Witten:1}. For details on generalising the second method see \cite{Cremonesi}. 

Assume there is a (super)group $G$, generated by a fermionic charge $Q$ and that $G$ acts freely on the field space $\mathfrak{F}$. For a $G$-invariant operator $\bigO$ we can reduce the path integral to the space $\mathfrak{F}/G$ by introducing $G$-collective coordinates, yielding $\expval{\bigO}=Vol(G)\int_{\mathfrak{F}/G}\D X\bigO e^{iS[X]}$.

However, since $G$ is generated by a fermionic variable and as $\int d\theta = 0$, we have that $\text{Vol}(G)=0$. This is a contradiction since if this is the case, even $\expval{\mathbbm{1}}$ would vanish meaning expectation values would be non-normalisable.

The way out is to conclude that $G$ does not act freely, instead it has fixed points forming the BPS locus $\mathfrak{F}_{Q}$ of $Q-$invariant field configurations. $G$ then acts freely on the complement $\mathfrak{F}\setminus\mathfrak{F}_{Q}$, so the path integral vanishes there. We thus see that $\expval{\bigO}$ localises to $\mathfrak{F}_{Q}$, as claimed.
\section{Topological Twist and Localisation of $\N=2$ SYM}
The goal of this section is to establish some more advanced facts about $\N=2$ SYM. In particular, we show that the action can be rewritten as $S=S_{\text{top}}+\bar{Q}(V_{YM}+V')$, where $\bar{Q}$ is a fermionic BRST, operator, $V_{YM}$ and $V'$ are potential terms, and $S_{\text{top}}=\frac{\Theta }{32\pi^{2}}g^{2}\int d^{4}x F_{\mu\nu}(\star{F})^{\mu\nu}$ is the topological part of the $\N=2$ SYM action \label{LYMexpanded}. Furthermore we identify the localisation locus of $\N=2$ SYM which will be explicitly constructed in the next chapter.
\subsection{Topological QFTs.}
In general a QFT is said to be topological if all correlation functions are independent of the metric; $\frac{\delta}{\delta g}\expval{\bigO_{\alpha _{1}}\ldots\bigO_{\alpha_{p}}}$. There are two general classes of QFTs which satisfy this condition \cite{Labastida:1}.

The first type are called Schwarz QFTs. A Schwarz QFT is a QFT in which each $\bigO_{\alpha_{i}}$ is individually metric independent. Such theories are clearly topological.

The second type are called Witten QFTs. A Witten type QFT has a nilpotent symmetry $\delta$ such that $\delta\bigO_{\alpha}(\phi_{i})=0$ where $\bigO_{\alpha_{i}}$ is a gauge invariant supersymmetric observable. Furthermore $T_{\mu\nu}=\delta G_{\mu\nu}$ for some tensor $G_{\mu\nu}$, where $T_{\mu\nu}=\frac{\delta S}{\delta g}$. We further assume that the variation of the $\bigO_{\alpha}$ with respect to the metric is $\delta$-exact, that is to say $\frac{\delta \bigO_{\alpha}(\phi_{i})}{\delta g_{\mu\nu}}=\delta \bigO_{\alpha}^{\mu\nu}(\phi_{i})$ for some functional $\bigO_{\alpha}^{\mu\nu}(\phi_{i})$.

Under the previous assumptions:
\begin{align*}
\frac{\delta}{\delta g_{\mu\nu}}
\expval{\bigO_{\alpha}}
&=
\frac{\delta}{\delta g_{\mu\nu}}
\int[\D\phi_{i}]\bigO_{\alpha}e^{-S[\phi_{i}]}
\\
&=
\int[\D\phi_{i}]
\left(
(\delta \bigO^{\mu\nu}_{\alpha})e^{-S[\phi_{i}]}
-
\bigO_{\alpha}T_{\mu\nu}e^{-S[\phi_{i}]}
\right)
\\
&=
\int[\D\phi_{i}]
\delta
\left(
(\bigO^{\mu\nu}_{\alpha})e^{-S[\phi_{i}]}
-
\bigO_{\alpha}G_{\mu\nu}e^{-S[\phi_{i}]}
\right)
\\
&=
0
\end{align*}
where the product rule for $\delta$ and the fact that $\delta\bigO_{\alpha}=\delta S=0$ have been used. We have also assumed that the measure $\D[\phi_{i}]$ is invariant under $\delta$.

So indeed Witten type QFTs are topological. We will show that $\N=2$ SYM is, after a topological twist, a Witten type QFT.
\subsection{The topological twist.}
The topological twist is performed by defining the twisted supercharges:
$$
\bar{Q}=\epsilon^{A\dot{\alpha}}\bar{Q}_{A,\dot{\alpha}}
\text{, }\quad
Q_{\mu}=\bar{\sigma}_{\mu}^{A\alpha}Q_{A,\alpha}
\text{, }\quad
\bar{Q}_{\mu\nu}=\bar{\sigma}_{\mu\nu}^{A\dot{\alpha}}\bar{Q}_{A,\dot{\alpha}}
$$
Introducing the twisted supercharges corresponds to redefining the spacetime symmetry group of $\N=2$ SYM. Initially the spacetime symmetry group is $SO(3,1)\cong SU(2)_{L}\times SU(2)_{R}$, and the global symmetry group is $SO(3,1)\times SU(2)_{\R}\times U(1)_{\R}\cong SU(2)_{L}\times SU(2)_{R}\times SU(2)_{\R}\times U(1)_{\R}$, where the proceeding isomorphisms are local (i.e. isomorphisms of the corresponding Lie algebras).

The topological twist corresponds to redefining this spacetime symmetry group to
\newline
$\diag\left(SU(2)_{L}\times SU(2)_{\R}\right)\times SU(2)_{R}$, where $\diag$ denotes the diagonal subgroup \cite{Labastida:1, Witten:2}.

Similarly we define the twisted fields:
$$
\psi^{\mu}=\bar{\sigma}^{\mu,A\alpha}\psi_{A,\alpha}
\text{, }
\bar{\psi}=\epsilon^{A\dot{\alpha}}\bar{\psi}_{A,\dot{\alpha}}
\text{, }
\bar{\psi}^{\mu\nu}=\bar{\sigma}^{\mu\nu}_{\dot{\alpha}A}\bar{\psi}^{A,\dot{\alpha}}
$$
\begin{remark}{3.4}
The field $\bar{\psi}^{\mu\nu}$ is anti-self-dual: $\bar{\psi}^{\mu\nu}=-i(\star\bar{\psi})^{\mu\nu}$, as is $\bar{Q}_{\mu\nu}$.
\end{remark}
The $\N=2$ SYM action can be rewritten using the twisted fields:
\begin{align*}
S
&=
\frac{1}{g^{2}}
\int d^{4}x
\text{Tr}
\biggr(
-\frac{1}{4}F_{\mu\nu}F^{\mu\nu}
+
\nabla_{\mu}H^{\dagger}\nabla^{\mu}H
-
\frac{1}{2}[H,H^{\dagger}]^{2}
+
\frac{i}{2}\psi^{\mu}\nabla_{\mu}\bar{\psi}
-
\frac{i}{2}(\nabla_{\mu}\psi_{nu}-\nabla_{\nu}\psi_{\mu})^{-}\bar{\psi}^{\mu\nu}
\\
&\qquad
+
\frac{i}{2\sqrt{2}}\psi_{\mu}[H^{\dagger},\psi^{\mu}]
-
\frac{i}{2\sqrt{2}}\bar{\psi}[H,\bar{\psi}]
-
\frac{i}{2\sqrt{2}}\bar{\psi}^{\mu\nu}[H,\bar{\psi}_{\mu\nu}]
\biggr)
+
\frac{\Theta}{32\pi^{2}}
\int d^{4}x \text{Tr} F_{\mu\nu}(\star{F})^{\mu\nu}
\end{align*}
Where in general $(X)^{\pm}=\frac{1}{2}(X\mp i\star X)$ refers to the (anti)self dual part of the field $X$.

The action of the twisted supercharges on the fields can also be calculated. This is a tedious but simple calculation, and so is deferred to Appendix B.

We can now establish three important facts.
\begin{itemize}
  \item $\bar{Q}$ is nilpotent up to a gauge transformation with parameter $-2\sqrt{2}H$. This can be shown by using the action of $\bar{Q}$.
  \item $S$ is $\bar{Q}$-exact up to a topological term:
$$
S
=
S_{\text{top}}
+
\Im
\left[
\bar{Q}
\left\{
\frac{\tau}{16\pi}
\int d^{4}x\text{Tr}
\left(
(F_{\mu\nu})^{-}\bar{\psi}^{\mu\nu}
-
i\sqrt{2}\psi^{\mu}\nabla_{\mu}\phi^{\dagger}
+
i\bar{\psi}[\phi,\phi^{\dagger}]
\right)
\right\}
\right]
$$
\item $S$ is $\bar{Q}-$cohomologous to $S_{top}$. This can be seen by using the equations of motion to show that $S_{\text{top}}$ is $\bar{Q}-$closed.
\end{itemize}
\subsection{Topological twist from gauge fixing.}
The twisted $\N=2$ SYM action can also be obtained from $S_{\text{top}}$ by a gauge fixing procedure known as BV quantisation. This process involves introducing  so-called ghost and antighost fields to eliminate unphysical degrees of freedom. A deep discussion of BV quantisation would lead us seriously off topic, but since we need the gauge fixed action to perform explicit computations, we will quote the important results. For more details on the general procedure as well as the case at hand, see \cite{BV:1, Shadchin}.

Essentially BV quantisation leads to the introduction of the fields shown in Table $3.2$. The action of $\bar{Q}$ on these fields is given in Appendix B. Using this action and the equations of motion it can be shown that the BRST operator $\bar{Q}$ coincides with the twisted supercharge $\bar{Q}$, and is nilpotent up to a gauge transformation \cite{Shadchin}.
\begin{table}[h!]
  \begin{center}
    \caption{BV ghost fields and their statistics}
    \label{tab:table2}
    \begin{tabular}{l|c|r|r|r|r|r|r|r|r|r|r}
      \hline
      Field & $b$ & $c$ & $\bar{c}$ & $\phi$ & $\eta$ & $\lambda$ & $A_{\mu}$ & $\psi_{\mu}$ & $H_{\mu\nu}$ & $\chi_{\mu\nu}$\\
      \hline
      Ghost number & $0$ & $+1$ & $-1$ & $+2$ & $-1$ & $-2$ & 0 & $+1$ & $0$ & $-1$\\
      \hline
      Statistics & B & F & F & B & F & B & B & F & B & F\\
      \hline
    \end{tabular}
  \end{center}
\end{table}

Furthermore BV quantisation imposes the following conditions: 
$$
\nabla^{\mu}A_{\mu}=(F_{\mu\nu})^{-}=\nabla^{\mu}\psi_{\mu}=0
$$
For later convenience, we  make the following field redefinitions:
$$
\phi=-2\sqrt{2}H
\text{, }
\quad
\lambda=-2\sqrt{2}H^{\dagger}
\text{, }
\quad
\chi_{\mu\nu}=\bar{\psi}_{\mu\nu}
\text{, }
\quad
\eta=-4\bar{\psi}
$$
The gauge fixed action is then given by $S_{top}+\bar{Q}\left(V_{YM}+V'\right)$, where \cite{Shadchin}:
\begin{align*}
&
V_{YM}
=
\frac{1}{g^{2}}\int d^{4}x
\text{Tr}
\left[
\frac{1}{2}\chi^{\mu\nu}
\left(
(F_{\mu\nu})^{-}
+
\frac{1}{4}H_{\mu\nu}
\right)
+
\frac{i}{8}\lambda\nabla_{\mu}\psi^{\mu}
+
\bar{c}
(
\nabla_{\mu}A^{\mu}
+
b
)
\right]
\\
&
V'
=
-
\frac{i}{128g^{2}}\int d^{4}x
\text{Tr}
\left(
\eta[\phi, \lambda]
\right)
\end{align*}
One can then show that indeed $S=S_{top}+\bar{Q}\left(V_{YM}+V'\right)$.
\subsection{The localisation locus of $\N=2$ SYM.}
We are now in a position to determine the localisation locus of $\N=2$ SYM. Since the twisted action is $\bar{Q}-$cohomologous to $S_{top}$, it can be deformed by a $\bar{Q}-$exact term. We choose:
\begin{equation}
\widetilde{V}
=
\int d^{4}x
\text{Tr}
\left[
-\chi^{\mu\nu}
\left(
t(F_{\mu\nu}^{-}-\frac{1}{4}H_{\mu\nu})
\right)
+
i\lambda\nabla_{\mu}\psi^{\mu}
\right]
\label{gaugepot}
\end{equation}
for some parameter $t$, then deform the twisted action to $S_{top}+\bar{Q}\widetilde{V}$.

The equation of motion for $H^{\mu\nu}$ will be required to proceed. We now determine the part of the $\widetilde{V}$ integrand which involves $H^{\mu\nu}$:
\begin{align*}
\bar{Q}
\left(
-t\chi^{\mu\nu}(F_{\mu\nu})^{-}
+\frac{1}{4}
\chi^{\mu\nu}H_{\mu\nu}
\right)
&=
-t\bar{Q}(\chi^{\mu\nu})(F_{\mu\nu})^{-}
+\frac{1}{4}
\bar{Q}(\chi^{\mu\nu}H_{\mu\nu})
\\
&=
-tH^{\mu\nu}(F_{\mu\nu})^{-}
+
\frac{1}{4}(H^{\mu\nu}-i\{c,\chi^{\mu\nu}\})H_{\mu\nu}
-
i\chi^{\mu\nu}[c,H_{\mu\nu}]
\\
&=
-tH^{\mu\nu}(F_{\mu\nu})^{-}
+
\frac{1}{4}H^{\mu\nu}H_{\mu\nu}
-
\frac{i}{4}\{c,\chi^{\mu\nu}\}H_{\mu\nu}
-
i\chi^{\mu\nu}[c,H_{\mu\nu}]
\end{align*}
where these equalities are up to terms not involving $H^{\mu\nu}$.

This results in the following equation of motion:
$$
H^{\mu\nu}
=
2t(F^{\mu\nu})^{-}
+
\frac{i}{2}\{c,\chi^{\mu\nu}\}
$$
Evaluating the action of $\bar{Q}$ and substituting in this equation of motion, the deformed action becomes:
$$
S
=
S_{top}
+
\int d^{4}x
\text{Tr}
\left(
-t^{2}(F_{\mu\nu})^{-}(F^{\mu\nu})^{-}
+
t\chi^{\mu\nu}(\nabla_{\mu}\psi_{\nu}-\nabla_{\nu}\psi_{\mu})^{-}
+
i\eta\nabla^{\mu}\psi_{\mu}
+
i\lambda\nabla^{\mu}\nabla_{\mu}\phi
\right)
$$
So as per the usual localisation procedure we have that upon taking $t\rightarrow \infty$ the action becomes large and negative leading to a vanishing exponential $e^{S}$ unless $(F_{\mu\nu})^{-}(F^{\mu\nu})^{-}=0$. Thus the localisation locus consists of field configurations for which the field tensor satisfies $(F^{\mu\nu})^{-}=0$, or equivalently $F^{\mu\nu}=i(\star{F})^{\mu\nu}$. This is called the self-dual equation, and such configurations are called self dual.

Satisfying the self-dual equation is in fact a sufficient condition for satisfying the Yang-Mills equation, so such configurations satisfy the classical equations of motion. They are called instanton configurations. The space of such configurations is called the instanton moduli space.

We have now shown that the partition function path integral localises to the instanton moduli space. This space turns out to be the direct sum of certain finite dimensional spaces, thus reducing a path integral to a sum of finite dimensional integrals.
\chapter{The ADHM construction}
The subject of this chapter is to explicitly construct the space of self dual field configurations of $\N=2$ SYM, thus providing a finite dimensional model of the instanton moduli space. The approach we will take to this problem is known as the ADHM construction, after Atiyah, Drinfeld, Hitchin and Manin \cite{ADHM}.

The ADHM construction proceeds by explicitly constructing a family of connections $A_{\mu}$, one for each $k$, then showing that the corresponding curvatures $F_{\mu\nu}$ are self dual. The connections constructed in this way then provide a family of instantons indexed by $k$. To show that this process does in fact yield all possible instantons, we will show that the solutions of the corresponding Dirac equation are in bijection with the constructed configurations.
\section{The Construction}
The ADHM construction is the only part of the localisation approach which is strongly gauge group dependent. For this thesis we will limit ourselves to the $SU(N)$ case, although the generalisation to other gauge groups is quite straightforward \cite{Ito, Shadchin}. Our discussion follows  \cite{CG:1} and \cite{Shadchin}.

Take $k\in\mathbb{Z}_{>0}$ and introduce $(N+2k)\times 2k$ matrices $\A=(\A_{s,rI})$ and $\B=(B_{s,rI})$ with $1\leq s\leq N+2k$, $1\leq r\leq k$, and $I\in\{1,2\}$.
\begin{remark}{4.1}
The matrices $\A$ and $\B$ can be seen as operators $\mathbb{C}^{N+2k}\rightarrow \mathbb{C}^{k}\otimes \mathbb{C}^{2}$.
\end{remark}
Next we introduce a complex structure on $\mathbb{R}^{4}$ by defining:
$$
x
=
(x_{IJ})
=
\sigma^{\mu}
x_{\mu}
=
\begin{pmatrix}
x_{0}-ix_{3}&-ix_{1}-x_{2}\\
-ix_{1}+x_{2}&x_{0}+ix_{3}
\end{pmatrix}
\text{ , }\quad
x_{\mu}\in\mathbb{R}^{4}
$$

Several more definitions are necessary. Firstly, define $\Delta(x)=\A+\B x$ which is assumed to have the maximal rank $2k$ for all $x$. Next introduce an $(N+2k)\times N$ matrix, $v(x)$ which consists of a basis for the null space of $\Delta^{\dagger}$: $\Delta^{\dagger}v(x)=0$. We normalise this basis in the following way: $v^{\dagger}v=\mathbbm{1}_{N}$.

Now define a candidate connection:
$$
A_{\mu}(x)
=
iv^{\dagger}(x)\del_{\mu}v(x)
$$
This connection is in fact Hermitian:
$$
A_{\mu}(x)^{\dagger}
=
-i(\del_{\mu}v^{\dagger}(x))v(x)
=
-i(\del_{\mu}(v^{\dagger}(x)v(x))-v(x)^{\dagger}\del_{\mu}v)
=
A_{\mu}(x)-i\del_{\mu}(\mathbbm{1}_{N})
=
A_{\mu}(x)
$$
For $A_{\mu}(x)$ to lead to a self dual curvature we require a further factorisation condition on $\Delta$:
\begin{equation}
(\Delta^{\dagger}\Delta)_{rI,sJ}=(\mathcal{R}^{-1}(x))_{rs}\delta_{IJ}
\label{factcond}
\end{equation}
Using this factorisation condition and the assumption that $\Delta$ is of maximal rank:
\begin{align*}
&
(\Delta^{\dagger}\Delta)_{rI,sJ}
=
(\mathcal{R}^{-1})_{rs}\delta_{IJ}
\\
\Rightarrow&\quad\ \ \;
\mathcal{R}\Delta^{\dagger}\Delta
=
\mathbbm{1}
\\
\Rightarrow&\quad\,
\Delta\mathcal{R}\Delta^{\dagger}\Delta
=
\Delta
\end{align*}
But $(\mathbbm{1}-vv^{\dagger})\Delta=\Delta-v(\Delta^{\dagger}v)^{\dagger}=\Delta$, since $\Delta^{\dagger}$ annihilate $v$, so:
$$
(\Delta\mathcal{R}\Delta^{\dagger}-(\mathbbm{1}-vv^{\dagger}))\Delta=0
$$
By the maximality of the rank of $\Delta$, a right inverse exists and thus:
$$
\Delta\mathcal{R}\Delta^{\dagger}=\mathbbm{1}-vv^{\dagger}
$$
We can use the above result to show that $F_{\mu\nu}$ is indeed self dual:
\begin{align*}
F_{\mu\nu}
&=
\del_{\mu}A_{\nu}
-
\del_{\nu}A_{\mu}
-i[A_{\mu},A_{\nu}]
\\
&=
i((\partial_{\mu}v^{\dagger})\partial_{\nu}v+v^{\dagger}(\del_{\mu}v)v^{\dagger}(\partial_{\nu}v))-(\mu \leftrightarrow \nu)
\\
&=
i(\partial_{\mu}v^{\dagger})(\mathbbm{1}-vv^{\dagger})(\partial_{\nu}v)+(\partial_{\mu}\mathbbm{1})-(\mu \leftrightarrow \nu)
\\
&=
i(\partial_{\mu}v^{\dagger})(\mathbbm{1}-vv^{\dagger})(\partial_{\nu}v)-(\mu \leftrightarrow \nu)
\\
&=
i(\partial_{\mu}v^{\dagger})\Delta\mathcal{R}\Delta^{\dagger}(\partial_{\nu}v)-(\mu \leftrightarrow \nu)
\\
&=
i(\partial_{\mu}((\Delta^{\dagger}v)^{\dagger})-v^{\dagger}(\partial_{\mu}\Delta))\mathcal{R}(\partial_{\nu}(\Delta^{\dagger}v)-\partial_{\nu}(\Delta^{\dagger})v)-(\mu \leftrightarrow \nu)
\\
&=
iv^{\dagger}(\partial_{\mu}\Delta)\mathcal{R}\partial_{\nu}(\Delta^{\dagger})v-(\mu \leftrightarrow \nu)
\end{align*}
So noting that $\del_{\mu}\Delta(x)=0+\B\sigma_{\nu}\partial_{\mu}(x^{\nu})=\B\sigma_{\mu}$, we have:
$$
F_{\mu\nu}
=
iv^{\dagger}\B(\sigma_{\mu}\R\sigma_{\nu}^{\dagger}-\sigma_{\nu}\R\sigma_{\mu}^{\dagger})\B^{\dagger}v
=
iv^{\dagger}\B\R(\sigma_{\mu}\sigma_{\nu}^{\dagger}-\sigma_{\nu}\sigma_{\mu}^{\dagger})\B^{\dagger}v
=
4iv^{\dagger}\B\R\sigma_{\mu\nu}\B^{\dagger}v
$$
where we have used that $\R=(\R_{rs})$ and $\sigma_{\mu}=(\sigma_{\mu,IJ})$
\begin{remark}{4.2}
The above condition says that as matrices, $\R$ and $\sigma_{\mu}$ have a block diagonal structure. 
\end{remark}
The quantity $\sigma_{\mu\nu}$ is easily shown to be self-dual, so $F_{\mu\nu}=i(\star{F})^{\mu\nu}$, as required. We have thus constructed a family of self-dual field configurations indexed by the non-negative integer $k$.
\section{Reformulation as Linear Operators}
Before showing that the previous construction is complete, we reformulate things in the language of linear operators. To do so, the factorisation condition \eqref{factcond} is reinterpreted to yield the so-called ADHM constraints. This reformulation will illuminate some residual freedom in the ADHM data which can then be used to determine a canonical form for the matrices $\A$ and $\B$, thus yielding the ADHM equations.
\subsection{The ADHM constraints.}
Noting that that $\delta^{\dot{\alpha}}_{\dot{\beta}}\delta^{\dot{\beta}}_{\dot{\alpha}}=2$, we have from \eqref{factcond}:
\begin{align*}
\R^{-1}(x)
&=
\frac{1}{2}\delta^{\dot{\beta}}_{\dot{\alpha}}\Delta^{\dagger\dot{\alpha}}\Delta_{\dot{\beta}}
\\
&=
\frac{1}{2}\delta^{\dot{\beta}}_{\dot{\alpha}}
\left(
\A^{\dagger,\dot{\alpha}}\A_{\dot{\beta}}
+
\A^{\dagger\dot{\alpha}}\B^{\beta}x_{\beta\dot{\beta}}
+
\bar{x}^{\alpha\dot{\alpha}}\B^{\dagger}_{\alpha}\A_{\dot{\beta}}
+
\bar{x}^{\alpha\dot{\alpha}}\B^{\dagger}_{\alpha}\B^{\beta}x_{\beta\dot{\beta}}
\right)
\\
&=
\frac{1}{2}
\left(
\A^{\dagger,\dot{\gamma}}\A_{\dot{\gamma}}
+
\A^{\dagger\dot{\alpha}}\B^{\alpha}x_{\alpha\dot{\alpha}}
+
\bar{x}^{\alpha\dot{\alpha}}\B^{\dagger}_{\alpha}\A_{\dot{\alpha}}
+
\bar{x}^{\alpha\dot{\beta}}\B^{\dagger}_{\alpha}\B^{\beta}x_{\beta\dot{\beta}}
\right)
\end{align*}
From the third line of the above expression we can easily obtain a Taylor series for $\R^{-1}(x)$ which terminates after three terms:
\begin{align*}
\R^{-1}(x)
&=
\frac{1}{2}\A^{\dagger,\dot{\gamma}}\A_{\dot{\gamma}}
+
\frac{1}{2}x^{\mu}
\left(
\A^{\dagger\dot{\alpha}}\B^{\alpha}\sigma_{\mu,\alpha\dot{\alpha}}
+
\bar{\sigma}_{\mu}^{\alpha\dot{\alpha}}\B^{\dagger}_{\alpha}\A_{\dot{\alpha}}
\right)
+
\frac{1}{4}
x^{\mu}x^{\nu}
\left(
\bar{\sigma}_{\mu}^{\alpha\dot{\beta}}\B^{\dagger}_{\alpha}\B^{\beta}\sigma_{\nu,\beta\dot{\beta}}
+
\bar{\sigma}_{\nu}^{\alpha\dot{\beta}}\B^{\dagger}_{\alpha}\B^{\beta}\sigma_{\mu\beta\dot{\beta}}
\right)
\\
&=
\frac{1}{2}\A^{\dagger,\dot{\gamma}}\A_{\dot{\gamma}}
+
\frac{1}{2}
\left(
\A^{\dagger\dot{\alpha}}\B^{\alpha}x_{\alpha\dot{\alpha}}
+
\bar{x}^{\alpha\dot{\alpha}}\B^{\dagger}_{\alpha}\A_{\dot{\alpha}}
\right)
+
\frac{1}{4}
\bar{x}^{\alpha\dot{\beta}}
\B^{\dagger}_{\alpha}\B^{\beta}
x_{\beta\dot{\beta}}
\end{align*}
where $(x_{\alpha \dot{\alpha}})^{\dagger}=x_{\mu}\bar{\sigma}^{\mu,\alpha\dot{\alpha}}:=\bar{x}^{\alpha\dot{\alpha}}=x^{\dagger,\alpha\dot{\alpha}}$.

Using the above Taylor series as well as \eqref{factcond}, we have:
\begin{align*}
\A^{\dagger,\dot{\alpha}}\A_{\dot{\beta}}
+
\A^{\dagger\dot{\alpha}}\B^{\beta}x_{\beta\dot{\beta}}
+
&
\bar{x}^{\alpha\dot{\alpha}}\B^{\dagger}_{\alpha}\A_{\dot{\beta}}
+
\bar{x}^{\alpha\dot{\alpha}}\B^{\dagger}_{\alpha}\B^{\beta}x_{\beta\dot{\beta}}
\\
&=
\delta^{\dot{\alpha}}_{\dot{\beta}}
\left(
\frac{1}{2}\A^{\dagger,\dot{\gamma}}\A_{\dot{\gamma}}
+
\frac{1}{2}
\biggr(
\A^{\dagger\dot{\gamma}}\B^{\alpha}x_{\alpha\dot{\gamma}}
+
\bar{x}^{\alpha\dot{\gamma}}\B^{\dagger}_{\alpha}\A_{\dot{\gamma}}
\right)
+
\frac{1}{4}
\bar{x}^{\alpha\dot{\gamma}}
\B^{\dagger}_{\alpha}\B^{\beta}
x_{\beta\dot{\gamma}}
\biggr)
\end{align*}
Equating coefficients in the above expression then gives the following three conditions:
\begin{align}
\begin{split}
\B^{\dagger}_{\alpha}\B^{\beta}
=
\frac{1}{2}\delta^{\beta}_{\alpha}\B^{\dagger}_{\gamma}\B^{\gamma}
\text{ , }\quad
\B^{\dagger}_{\alpha}\A_{\dot{\alpha}}
=
\A^{\dagger}_{\dot{\alpha}}\B_{\alpha}
\text{ , }\quad
\A^{\dagger\dot{\alpha}}\A_{\dot{\beta}}
=
\frac{1}{2}\delta^{\dot{\alpha}}_{\dot{\beta}}\A^{\dagger\dot{\gamma}}\A_{\dot{\gamma}}
\end{split}
\label{ADHMconds}
\end{align}
For example, the first condition is obtained by equating:
$$
\bar{x}^{\alpha\dot{\alpha}}\B^{\dagger}_{\alpha}\B^{\beta}x_{\beta\dot{\beta}}
=
\frac{1}{2}\delta^{\dot{\alpha}}_{\dot{\beta}}
\bar{x}^{\alpha\dot{\gamma}}
\B^{\dagger}_{\alpha}\B^{\beta}
x_{\beta\dot{\gamma}}
$$
which holds for all $x$ and thus for $x=\bar{x}=\mathbbm{1}_{2}$, so:
$$
\delta^{\alpha\dot{\alpha}}\B^{\dagger}_{\alpha}\B^{\beta}\delta_{\beta\dot{\beta}}
=
\frac{1}{2}\delta^{\dot{\alpha}}_{\dot{\beta}}
\delta^{\alpha\dot{\gamma}}
\B^{\dagger}_{\alpha}\B^{\beta}
\delta_{\beta\dot{\gamma}}
=
\frac{1}{2}\delta^{\dot{\alpha}}_{\dot{\beta}}
\delta^{\alpha}_{\beta}
\B^{\dagger}_{\alpha}\B^{\beta}
=
\frac{1}{2}\delta^{\dot{\alpha}}_{\dot{\beta}}
\B^{\dagger}_{\gamma}\B^{\gamma}
$$
Multiplying through by $\delta^{\beta}_{\alpha}$ then yields:
\begin{align*}
&
\delta^{\beta}_{\alpha}\delta^{\alpha\dot{\alpha}}\B^{\dagger}_{\alpha}\B^{\beta}\delta_{\beta\dot{\beta}}
=
\frac{1}{2}\delta^{\dot{\alpha}}_{\dot{\beta}}
\delta^{\beta}_{\alpha}
\B^{\dagger}_{\gamma}\B^{\gamma}
\\
\Rightarrow&\quad\ \ 
\delta^{\dot{\alpha}}_{\dot{\beta}}
\left(
\B^{\dagger}_{\alpha}\B^{\beta}
\right)
=
\delta^{\dot{\alpha}}_{\dot{\beta}}
\left(
\frac{1}{2}
\delta^{\beta}_{\alpha}
\B^{\dagger}_{\gamma}\B^{\gamma}
\right)
\end{align*}
The result is then immediately obtained by multiplying through with $\delta^{\dot{\beta}}_{\dot{\alpha}}$.

The three position independent conditions \eqref{ADHMconds} are called the ADHM constraints. They are a set of coupled quadratic conditions on $\A$ and $\B$ which must be obeyed to yield a self-dual connection.
\subsection{Residual freedom.}
The input data for the construction consists of the two conditions defining $v$ and the factorisation condition on $\Delta$ (or equivalently the three ADHM constraints). We now want to see what freedom we have with this data. In particular we perform the transformations; $\Delta\mapsto U\Delta M$, $v\mapsto Uv$ and $\R\mapsto M^{\dagger}\R M$ with $U$ unitary and $M$ invertible. It is easy to show that this does not effect any of the ADHM conditions \eqref{ADHMconds}.
\subsection{Canonical forms.}
The above set of symmetries allows the freedom to choose a nice canonical form for $\A$ and $\B$. Firstly since $\B$ is assumed to be of rank $2k$ we can take $\Delta\mapsto U\Delta M$ to set 
$
\B
=
\left(0|\B'\right)^{T}
$
, with $\B'$ $2k\times 2k$. This is essentially a partial singular value decomposition.

The first ADHM condition says that $\B^{\dagger}\B'\propto\mathbbm{1}$, so $\B'^{\dagger}\B'$ is real and symmetric. So from basic linear algebra there exist matrices $O$ and $\mu$ with $O$ real orthogonal and $\mu$ real diagonal such that $\B'^{\dagger}\B=O\mu O^{T}$.

Using the symmetry transformation again with $U=\mathbbm{1}$ and $M=O\mu^{-\frac{1}{2}}$, $\B'\mapsto \B' O \mu^{-\frac{1}{2}}$. Thus:
$$\B'^{\dagger}\B'\mapsto \mu^{-\frac{1}{2}}O^{T}\B'^{\dagger}\B'O\mu^{-\frac{1}{2}}=\mu^{-\frac{1}{2}}O^{\dagger}O\mu O^{T}O\mu^{-\frac{1}{2}}=\mathbbm{1}
$$
This means that the matrix 
$
\begin{pmatrix}
\begin{array}{c|c}
\mathbbm{1}&0
\end{array}
\\
\hline
\B'^{\dagger}
\end{pmatrix}
$
 is unitary, allowing us to perform the transformation one more time with $M=\mathbbm{1}$ and
$
U
=
\begin{pmatrix}
\begin{array}{c|c}
\mathbbm{1}&0
\end{array}
\\
\hline
\B'^{\dagger}
\end{pmatrix}
$ as follows:
\begin{equation}
\B
\mapsto
\begin{pmatrix}
\begin{array}{c|c}
0&0
\end{array}
\\
\hline
\B'^{\dagger}\B'
\end{pmatrix}
=
\begin{pmatrix}
0\\
\hline
\mathbbm{1}_{2k}
\end{pmatrix}
=
\begin{pmatrix}
0\\
\hline
\mathbbm{1}_{k}\otimes\mathbbm{1}_{2}
\end{pmatrix}\label{form1}
\end{equation}
In this form the third ADHM constraint is manifestly obeyed. We then partition $\A$ and $v$ according to the form of $\B$:
\begin{equation}
\A
=
\begin{pmatrix}
S_{\dot{\alpha}}\\
\hline
X^{\mu}\otimes\sigma_{\mu}
\end{pmatrix}
\text{ , }\quad
v
=
\begin{pmatrix}
T\\
\hline
Q_{\alpha}
\end{pmatrix}\label{form2}
\end{equation}
Where $S_{\dot{1}}$ and $S_{\dot{2}}$ are $N\times k$, $X^{\mu}$ is $k\times k$, $T$ is $N\times N$ and $Q_{1}$ and $Q_{2}$ are $k\times N$.

The redundancy in this description can be further reduced by transforming with elements of the unitary subgroups of $U(N+2k)$ and $GL(2k)$ since such transformations preserve the ADHM conditions \eqref{ADHMconds}. This amounts to the following component transformations:
\begin{equation}
S_{\dot{\alpha}}\mapsto U_{N}S_{\dot{\alpha}}U_{k}^{\dagger}
\text{ ,}\quad
X^{\mu}\mapsto U_{k}X^{\mu}U_{k}^{-1}
\text{ ,}\quad
T\mapsto U_{N}T
\text{ ,}\quad
Q_{\alpha}
\mapsto
U_{k}Q_{\alpha}
\text{ ,}\quad
\R\mapsto (U_{k}^{\dagger}\otimes\mathbbm{1}_{2})R(U_{k}\otimes\mathbbm{1}_{2})
\label{resid}
\end{equation}
\begin{remark}{4.3}
This residual freedom arising from the transformations \eqref{resid} defines an action of $U(k)$ on the ADHM fields. For different gauge groups $G$, the residual transformations define actions of different groups. The resulting group is called the dual group $G_{D}$. We have just shown that for $G=SU(N)$, the $k$-instanton dual group is $U(k)$.
\end{remark}
\subsection{The ADHM equations.}
The first ADHM constraint is automatically satisfied with $\B$ in canonical form, but what about the other two? It is easy to see that the second constraint leads to $X^{\mu}=(X^{\mu})^{\dagger}$, that is the matrix $X^{\mu}$ is necessarily hermitian. The third condition is slightly harder to interpret. First note that for any Pauli matrix $\tau_{i}$ we have upon contraction:
$$
\mu^{i}
:=
(\A^{\dagger\dot{\alpha}})(\tau_{i})_{\dot{\alpha}}^{\dot{\beta}}\A_{\dot{\beta}}=\frac{1}{2}\text{Tr}(\tau_{i})\A^{\dagger\dot{\gamma}}\A_{\dot{\gamma}}=0
$$
Since $\{\mathbbm{1}_{2},\tau_{i}\}$ is a basis for the set of $2\times 2$ hermitian matrices and as $\A^{\dagger}\A$ is hermitian (and $2\times 2$ if considered to be matrix valued), this condition is enough to conclude that $\A^{\dagger}\A\propto\mathbbm{1}$, where the constant of proportionality is obtained by taking the trace. This shows that the third ADHM condition is obeyed if and only if $\mu^{i}=0$ for $i=1,2,3$. These three non-linear conditions are a first form of the so-called ADHM equations.

In summary, for a self-dual curvature, $X^{\mu}$ must be hermitian and each $\mu^{i}=0$ for $i=1,2,3$.

To bring the ADHM equations into their most well known form, we define the following quantities:
$$
J
=
S_{\dot{1}}
\text{, }\quad
J
=
S_{\dot{2}}
\text{, }\quad
B_{1}
=
X^{0}-iX^{3}
\text{, }\quad
B_{2}
=
-iX^{1}+X^{2}
$$
The columns of $X^{\mu}\otimes \sigma_{\mu}$ can be reordered using \eqref{resid} to obtain:
$$
X^{\mu}\otimes\sigma_{\mu}
=
\begin{pmatrix}
B_{1}&-B_{2}^{\dagger}\\
B_{2}&B_{1}^{\dagger}
\end{pmatrix}
$$
Then $\A_{\dot{1}}=(S_{\dot{1}},B_{1},B_{2})^{T}$ and $\B_{\dot{2}}=(S_{\dot{2}},-B_{2}^{\dagger},B_{1}^{\dagger})^{T}$, which yields:
\begin{align}
\begin{split}
&
\mu^{1}
=
J^{\dagger}I^{\dagger}+[B_{2}^{\dagger},B_{1}^{\dagger}]+IJ+[B_{1},B_{2}]
\\
&
\mu^{2}
=
i
\left(
-J^{\dagger}I^{\dagger}-[B_{2}^{\dagger},B_{1}^{\dagger}]+IJ+[B_{1},B_{2}]
\right)
\\
&
\mu^{3}
=
-II^{\dagger}+J^{\dagger}J-[B_{1},B_{1}^{\dagger}]-[B_{2},B_{2}^{\dagger}]
\end{split}
\end{align}
Next note that $\mu^{1}$ and $\mu^{2}$ are purely hermitian and antihermitian respectively, so defining
$
\mu^{\mathbb{R}}=-\mu^{3}
$ and $\mu^{\mathbb{C}}=\frac{1}{2}(\mu^{1}-i\mu^{2})$, we have the final form of the ADHM equations:
\begin{equation}
\mu^{\mathbb{R}}=\mu^{\mathbb{C}}=0\label{ADHMeqns}
\end{equation}
The matrices $I,J,B_{1}$, and $B_{2}$ can now be interpreted as linear operators acting on vector spaces $\mathcal{V}\cong \mathbb{C}^{k}$ and $\mathcal{W} \cong \mathbb{C}^{N}$, which obey the ADHM equations:
\begin{equation}
I:\mathcal{W}\rightarrow\mathcal{V}
\quad
J:\mathcal{V}\rightarrow\mathcal{W}
\quad
B_{1,2}:\mathcal{V}\rightarrow\mathcal{V}\label{ADHMoperators}
\end{equation}
We call the space of such operators (modulo the dual group transformations \eqref{resid}, $\mathfrak{m}_{k}$, and claim that this is the $k$-instanton moduli space. The last step to verifying this claim is to show that the ADHM construction does indeed yield all instantons. From a given $A_{\mu}$ yielding a self-dual $F_{\mu\nu}$ we must be able to recover matrices $\A$ and $\B$ satisfying the ADHM equations and hermiticity properties derived above. This can be done using the relevant massless Dirac equation.
\section{Completeness of the Construction}
Consider the massless Dirac equation in the instanton background, that is for $A_{\mu}$ a connection corresponding to a self-dual field strength. We first study its solutions in the formalism of ADHM, then later show how knowledge of said solutions allows inversion of the ADHM construction.
\subsection{Recovering $\A$ and $\B$.}
The massless Dirac equation is given by $\gamma^{\mu}D_{\mu}\Psi=0$, where $D_{\mu}=\partial_{\mu}+A_{\mu}$, $\Psi=(\psi^{+},\psi^{-})^{T}$, and 
$
\gamma^{\mu}
=
\begin{pmatrix}
0&\sigma^{\mu}\\
\bar{\sigma}^{\mu}&0
\end{pmatrix}
$.
This gives two equations for the two component Weyl spinors $\psi^{\pm}$:
$$
\sigma^{\mu}D_{\mu}\psi^{+}=\bar{\sigma}^{\mu}D_{\mu}\psi^{-}=0
$$
It can be shown that there are no solutions to the positive chirality equation, and exactly $k$ linearly independent solutions to the negative chirality one, where $k$ is the instanton number associated with the connection $A_{\mu}$, \cite{CG:1, Osborn:1}. We thus rename $\psi^{-}$ to $\psi$ and arrange these solutions in an $N\times 2k$ matrix which turns out to be \cite{Osborn:1}:
\begin{equation}
\psi^{\alpha}
=
v^{\dagger}\B^{\alpha}\R
\end{equation}
Where $\A$, $\B$, $\R$ and $v$ are the matrices determined in the previous section \cite{CG:1}. In particular $\B$ has its canonical form and $\R$ is defined by the equation:
$$
2\R^{-1}(x)
=
\A^{\dagger,\dot{\gamma}}\A_{\dot{\gamma}}
+
\left(
\A^{\dagger\dot{\alpha}}\B^{\alpha}x_{\alpha\dot{\alpha}}
+
\bar{x}^{\alpha\dot{\alpha}}\B^{\dagger}_{\alpha}\A_{\dot{\alpha}}
\right)
+
\bar{x}^{\alpha\dot{\beta}}
\B^{\dagger}_{\alpha}\B^{\beta}
x_{\beta\dot{\beta}}
=
\A^{\dagger}\A
+
2X^{\mu}x_{\mu}
+
x^{2}\mathbbm{1}_{k}
$$
Certain moments of these solutions can then be calculated, as can some relevant asymptotic behaviour. Together this is enough to recover that $X^{\mu}$ is hermitian as well as the ADHM constraints, thus showing that the ADHM construction is complete.

The moments are calculated via the following formula: $\bar{\psi}_{\alpha}\psi^{\alpha}=-\frac{1}{4}\del_{\mu}\del^{\mu}\R$ \cite{Osborn:1}. This formula and the definition of $\R$ give:
$$
\int d^{4}x\bar{\psi}_{\alpha}\psi^{\alpha}
=
\pi^{2}\mathbbm{1}_{k}
\text{ , }\quad
\int d^{4}x\bar{\psi}_{\alpha}\psi^{\alpha}x^{\mu}
=
-\pi^{2}X^{\mu}
$$
The first equation says that the solutions to the Dirac equation are orthonormal up to a constant, and the second says that part of $\A$ can be recovered from these solutions. In particular if we can find a complete orthonormal set of solutions to the massless Dirac equation, we can determine part of the matrix $\A$ by calculating the first moment of $\bar{\psi}_{\alpha}\psi^{\alpha}$.

To recover $S_{\dot{\alpha}}$, we use that \cite{CG:1}:
$$
\psi^{\alpha}x_{\alpha\dot{\alpha}}
\rightarrow-\frac{1}{x^{2}}S_{\dot{\alpha}}
\text{ , as }
x\rightarrow \infty
\text{.}
$$
Given this asymptotic behaviour as a definition, an orthonormal set of solutions to the Dirac equation gives $S_{\dot{\alpha}}$ by simply taking a limit.
\subsection{Inverting the ADHM construction.}
Given a connection $A_{\mu}$ corresponding to a self-dual curvature, the ADHM construction can be inverted as follows: we first solve the corresponding massless Dirac equation to obtain a complete set of orthonormal solutions $\psi^{\alpha}$. The corresponding $X^{\mu}$ and $S_{\dot{\alpha}}$ are then determined by doing a moment integral and taking a limit respectively.

With these quantities at hand, we define a derived $\A$, $\B$ and thus $\Delta$ by the canonical forms \eqref{form1} and \eqref{form2}. The corresponding $v$ is then obtained by simply finding the null space of $\Delta^{\dagger}$. We have thus recovered all the initial data.

To conclude that the ADHM construction is complete it must also be shown that the derived quantities satisfy the required hermiticty properties as well as the ADHM equations.

The hermiticity of $X^{\mu}$ and the ADHM equations were both derived from the assumption that $\Delta^{\dagger}\Delta=\R^{-1}\mathbbm{1}_{2}$, which is itself derived from the assumption that $\Delta^{\dagger}\Delta$ commutes with the quaternions, so it is sufficient to show that this holds for the derived $\Delta$  \cite{CG:1}. A lengthy but relatively straightforward calculation involving Green's functions establishes the result. The calculation will be omitted for brevity, but the details can be found in \cite{CG:1}.

We have now explicitly constructed the $k$-instanton moduli space $\mathfrak{m}_{k}$ of $\N=2$ SYM with gauge group $SU(N)$. It consists of linear operators $I,J,B_{1,2}$ \eqref{ADHMoperators} satisfying the ADHM equations \eqref{ADHMeqns}, modulo some dual group transformations \eqref{resid}. The full instanton moduli space is then:
\begin{equation}
\M_{\text{inst}}:=\bigoplus_{k=1}^{\infty}\mathfrak{m}_{k}
\label{Instmodspace}
\end{equation}

\begin{remark}{4.4}
For large $k$ the ADHM equations become very difficult so the integration must be performed without introducing local coordinates on the $k-$instanton moduli space. This can be achieved using equivariant integration.
\end{remark}
\section{Application to $\N=2$ SYM}
In section $3.5.4$, the path integral of $\N=2$ SYM was shown to localise to the instanton moduli space $\M_{\text{inst}}$. In section $4.3.2$ an explicit construction of $\M_{\text{inst}}$ was obtained. With these two pieces of information we can now deduce localisation formulae for expectation values in terms of finite dimensional integrals. It is at this point we perform a Wick rotation, taking our spacetime from $\mathbb{R}^{1,3}$ to $\mathbb{R}^{4}$ and introducing a factor of $i$ in \eqref{Z}.

For a BRST closed, gauge invariant operator $\bigO$, we have that 
$
\expval{\mathcal{O}}
=
\int_{\M_{\text{inst}}}[\D X]\mathcal{O}e^{-S[X]}
$.
Recalling the localisation argument of Section $3.4$, a $\bar{Q}-$exact term can be added to the action without changing the value of this integral. Since $S$ is $\bar{Q}-$cohomologous to $S_{\text{top}}$ and recalling the gauge fixing potential $\widetilde{V}$ \eqref{gaugepot}, we have:
$$
\expval{\mathcal{O}}
=
\int_{\M_{\text{inst}}}[\D X]\mathcal{O}e^{-(S_{\text{top}}[X]+\bar{Q}\widetilde{V}(x,t))}
$$
Taking $t\rightarrow\infty$ then sets $F_{\mu\nu}=i(\star{F})^{\mu\nu}$ everywhere that the integral is non-vanishing. Substituting this condition into the action leads to following contribution from $\mathfrak{m}_{k}$:
\begin{align*}
S|_{\mathfrak{m}_{k}}
&=
\left(
S_{\text{top}}
-
\frac{1}{4g^{2}}
\int d^{4}x\text{Tr}F_{\mu\nu}F^{\mu\nu}
+
0
\right)
\biggr|_{\mathfrak{m}_{k}}
\\
&=
\left(
k\Theta
-
\frac{i}{4g^{2}}
\int d^{4}x\text{Tr}F_{\mu\nu}(\star{F})^{\mu\nu}
+
0
\right)
\biggr|_{\mathfrak{m}_{k}}
\\
&=
-\Theta k
-
i\frac{8\pi^{2}k}{g^{2}}
\\
&=
i(2\pi i k \tau)
\end{align*}
The expectation value of $\bigO$ is then:
\begin{equation}
\expval{\mathcal{O}}
=
\int_{\M_{\text{inst}}}[\D X]\mathcal{O}e^{-i(S_{\text{top}}[X]+\bar{Q}\widetilde{V}(x,t))}
=
\sum_{k=0}^{\infty}
e^{2\pi ik\tau}
\int_{\mathfrak{m}_{k}}\widetilde{\bigO}_{k}
\end{equation}
where $\widetilde{\bigO}_{k}$ is $\bar{Q}$-cohomologous to $\bigO$. The expectation value of a general BRST closed and gauge invariant observable has thus been reduced to a sum of finite dimensional integrals over known spaces!
\chapter{Lorentz Deformation, the Omega Background and the Prepotential}
Using the results of the previous chapter, we can calculate the partition function of $\N=2$ SYM with gauge group $SU(N)$. However, to access the prepotential a little more work will be necessary. This will lead to a second localisation argument which gives the partition function in terms of the prepotential. We can then equate the two expressions for the partition function to obtain the prepotential in terms of known quantities.

The idea is to compactify a six dimensional $\N=1$ SYM theory in a non-trivial spacetime to give a deformed four dimensional $\N=2$ theory. This is called a Lorentz deformation and the resulting spacetime is called the $\Omega$-background. It turns out that the partition function of this theory is easily expressed in terms of the prepotential. We then take the limit in which the deformation of the $\Omega$-background vanishes and demand that the two expressions for the partition function agree, thus yielding the prepotential.
\section{Lorentz Deformation}
This section follows \cite{NO:1, Shadchin}, which also contain additional details. To carry out the Lorentz deformation we exploit the dimensional reduction construction of $\N=2$ SYM from Section $1.6$. This construction used the six dimensional flat metric:
$$
ds^{2}
=
g_{\mu\nu}dx^{\mu}dx^{\nu}-(dx^{4})^{2}-(dx^{5})^{2}
$$
To accomplish the Lorentz deformation, let the torus $\mathbbm{T}^{2}$ act on $\mathbb{R}^{3,1}$ by Lorentz rotations. This is achieved by introducing two independent Lorentz rotations corresponding to the $x^4$ and $x^5$ directions, packaged as follows:
$$
V_{a}^{\mu}
=
(\Omega_{a})^{\mu}_{\nu}x^{\nu}
\text{, }\quad
a=4,5\text{.}
$$
The deformed metric is:
\begin{equation}
ds^{2}
:=
g_{\mu\nu}(dx^{\mu}+V_{a}^{\mu}dx^{a})(dx^{\nu}+V_{b}^{\nu}dx^{b})-(dx^{4})^{2}-(dx^{5})^{2}
=:
G_{IJ}dx^{I}dx^{J}
\end{equation}
The components of the deformed metric and its inverse can easily be determined.
\begin{align}
\begin{split}
&
G_{\mu\nu}=g_{\mu\nu}
\text{ , }\qquad\qquad\quad\  \,
G^{\mu\nu}=g^{\mu\nu}-V_{a}^{\mu}V_{a}^{\nu}
\\
&
G_{a,\mu}=V_{a\mu}
\text{ , }\qquad\qquad\quad\,\,
G^{a\mu}=V_{a}^{\mu}
\\
&
G_{ab}=-\delta_{ab}+V_{a}^{\mu}V_{b,\mu}
\text{ , }\quad
G^{ab}=-\delta^{ab}
\end{split}
\end{align}
A more tedious but straightforward computation shows that $G:=\det G_{IJ}=-1$, and that the curvature tensor vanishes if $\Omega_{4}$ and $\Omega_{5}$ commute. Since we want a flat metric, this will from now on be assumed to be the case.

To facilitate computations, introduce a vielbein:
$$
ds^{2}
=
g_{\mu\nu}e_{I}^{(\mu)}e_{J}^{(\nu)}dx^{I}dx^{J}-e_{I}^{(a)}e_{J}^{(a)}dx^{I}dx^{J}
$$
The components can be calculated by expanding the metric:
\begin{align}
\begin{split}
&
e_{\nu}^{(\mu)}=\delta^{\mu}_{\nu}
\text{, }\quad
e_{a}^{(\mu)}=V^{\mu}_{a}
\text{, }\quad
e_{\mu}^{(a)}=0
\text{, }\quad\ \ \ \ \,
e_{b}^{(a)}=\delta^{a}_{b}
\\
&
e^{\mu}_{(\nu)}=\delta^{\mu}_{\nu}
\text{, }\quad
e^{a}_{(\mu)}=0
\text{, }\quad\ \ \;
e^{\mu}_{(a)}=-V^{\mu}_{a}
\text{, }\quad
e^{a}_{(b)}=\delta^{a}_{b}
\end{split}
\end{align}
We now rewrite the $\N=1$, $d=6$ SYM action \eqref{SYMd6} in this background and once again compactify in the $x^4$ and $x^5$ directions. This process will yield $\N=2$ SYM in the $\Omega$-background.

We first rewrite the gauge kinetic term in flat vielbein indices:
\begin{align*}
F_{IJ}F^{IJ}
&=
(e_{I}^{(N)}e_{J}^{(M)}F_{(N)(M)})(e_{K}^{(O)}e_{L}^{(P)}F_{(O)(P)})G^{IK}G^{JL}
\\
&=
F_{(N)(M)}F_{(O)(P)}G^{(N)(O)}G^{(M)(P)}
\\
&=
F_{(I)(J)}F^{(I)(J)}
\end{align*}
Using that $F_{(I)(J)}=e^{K}_{(I)}e^{L}_{(J)}F_{KL}$, the components of the curvature can be obtained:
\begin{align}
\begin{split}
&
F_{(\mu)(\nu)}=F_{\mu\nu}
\text{, }\quad
F_{(a)(\mu)}=F_{a\mu}-V^{\rho}_{a}F_{\rho\mu}
\text{, }\quad
F_{(a)(b)}=V^{\mu}_{a}V^{\nu}_{b}F_{\mu\nu}-F_{a\nu}V^{\nu}_{b}-V^{\mu}_{a}F_{\mu b}+F_{ab}
\\
&
F^{(\mu)(\nu)}=V^{\mu}_{a}V^{\nu}_{b}F^{ab}+F^{\mu a}V^{\nu}_{a}+V^{\mu}_{a}F^{a\nu}+F^{\mu\nu}
\text{, }\quad
F^{(a)(\mu)}=F^{a\mu}+V_{b}^{\mu}F^{ab}
\text{, }\quad
F^{(a)(b)}=F^{ab}
\end{split}
\end{align}
Recalling the definition of $H$ and defining $V^{\mu}$ and $\Omega_{\mu\nu}$ analogously, the gauge term can be rewritten as:
\begin{align*}
-\frac{1}{4}F_{IJ}F^{IJ}
=
-
&
\frac{1}{4}F_{\mu\nu}F^{\mu\nu}
+
(\nabla_{\mu}H+V^{\nu}F_{\nu\mu})(\nabla^{\mu}H^{\dagger}+\bar{V}^{\nu}F_{\nu}^{\mu})
\\
&-
\frac{1}{2}
\left(
[H,H^{\dagger}]-i\bar{V}^{\mu}V^{\nu}F_{\mu\nu}-i(V^{\mu}\nabla_{\mu}H^{\dagger}-\bar{V}^{\nu}\nabla_{\nu}H)
\right)^{2}
\end{align*}
This resembles the undeformed expression with some shifts having been introduced. In fact this observation can be strengthened by defining a deformed version of $H$:
$$
\HH
:=
H-iV^{\mu}\nabla_{\mu}
,\quad
\HH^{\dagger}
:=
H^{\dagger}-i\bar{V}^{\mu}\nabla_{\mu}
$$
Recalling that $[\nabla_{\mu},\nabla_{\nu}]=-iF_{\mu\nu}$, we have:
$$
[\HH,\HH^{\dagger}]
=
[H,H^{\dagger}]
-
i\bar{V}^{\mu}V^{\nu}F_{\mu\nu}
-
i
(
V^{\mu}\nabla_{\mu}H^{\dagger}
-
\bar{V}^{\mu}\nabla_{\mu}H
)
-
i
(
H\bar{V}^{\mu}
-
H^{\dagger}V^{\mu}
)
\nabla_{\mu}
$$
However for commuting $\Omega_{a}$'s the last term vanishes, yielding:
$$
-\frac{1}{4}F_{IJ}F^{IJ}
=
-
\frac{1}{4}F_{\mu\nu}F^{\mu\nu}
+
(\nabla_{\mu}H+V^{\nu}F_{\nu\mu})(\nabla^{\mu}H^{\dagger}+\bar{V}^{\nu}F_{\nu}^{\mu})
-
\frac{1}{2}
[\HH,\HH^{\dagger}]^{2}
$$
Proceeding similarly for the fermionic term (and adding a spin  operator term to the definition of $\HH$, as detailed in \cite{Shadchin}), and rewriting a little it turns out that the only difference between the undeformed action \eqref{LYMexpanded} and the deformed action is the swapping of $H$ for $\HH$ and the addition of the following term:
$$
-
\frac{1}{4}\bar{\Omega}_{\rho\mu}F^{\rho\mu}\HH
-
\frac{1}{4}\Omega_{\rho\mu}F^{\rho\mu}\HH^{\dagger}
-
\frac{1}{2\sqrt{2}}\Omega_{\mu\nu}\bar{\psi}^{A}\bar{\sigma}^{\mu\nu}\bar{\psi}_{A}
-
\frac{1}{2\sqrt{2}}\bar{\Omega}_{\mu\nu}\psi^{A}\sigma^{\mu\nu}\psi_{A}
$$
In the $\N=1$ superspace formalism this corresponds to making the coupling constant coordinate dependent:
$$
\tau\mapsto\tau(x,\theta):=\tau-\frac{1}{\sqrt{2}}(\bar{\Omega}_{\mu\nu})^{+}\theta^{\mu}\theta^{\nu}
$$
\section{Localisation in the $\Omega-$background}
We wish to use localisation to evaluate the partition function of $\N=2$ SYM in the $\Omega-$background. This requires a nilpotent (to define a cohomology), BRST operator with respect to which the Lorentz deformed action is still exact. This is the case with the deformed BRST operator:
$$
\bar{Q}_{\Omega}
=
\bar{Q}
+
\frac{1}{2\sqrt{2}}\Omega^{\mu}_{\nu}x^{\nu}Q_{\mu}
$$
provided we make another shift of the coupling constant \cite{Shadchin}:
\begin{equation}
\tau(x,\theta)
:=
\tau-\frac{1}{\sqrt{2}}(\bar{\Omega}_{\mu\nu})^{+}\theta^{\mu}\theta^{\nu}
+
\frac{1}{2\sqrt{2}}\bar{\Omega}_{\mu\nu}\Omega^{\mu}_{\rho}x^{\rho}x^{\nu}
\end{equation}
Since this is a real shift it only changes the topological term of the action.

From the transformation properties, $\{\bar{Q},Q_{\mu}\}=4i\nabla_{\mu}$, and $\bar{Q}^{2}=Q_{\mu}^{2}=0$. Thus we have that up to a gauge transformation:
$$
\bar{Q}_{\Omega}^{2}
=
\frac{1}{2\sqrt{2}}\Omega^{\mu}_{\nu}\bar{Q}(x^{\nu})Q_{\mu}
+
\frac{1}{2\sqrt{2}}\Omega^{\mu}_{\nu}x^{\nu}\{\bar{Q},Q_{\mu}\}
+
\frac{1}{8}\Omega^{\mu}_{\nu}\Omega^{\rho}_{\sigma}x^{\nu}Q_{\mu}(x^{\sigma}Q_{\rho})
=
\sqrt{2}i\Omega^{\mu}_{\nu}x^{\nu}\nabla_{\mu}+\text{C}
=
\sqrt{2}i\Omega^{\mu}_{\nu}x^{\nu}\del_{\mu}
$$
where we have used that constants are gauge equivalent to zero, and that the covariant derivative is gauge equivalent to the ordinary derivative.
 
Recalling the antisymmetry of $\Omega$ we have that:
$$
\Omega^{\mu}_{\nu}x^{\nu}\del_{\mu}f
=
-\Omega^{\nu}_{\mu}x^{\nu}\del_{\mu}f
=
-\Omega^{\nu}_{\mu}(\del_{\mu}(x^{\nu}f)-\delta^{\nu}_{\mu}f)
=
-\del_{\mu}(\Omega^{\nu}_{\mu}x^{\nu}f)-\text{Tr}(\Omega)f
=
\del_{\mu}(-\Omega^{\nu}_{\mu}x^{\nu}f)
$$
So up to gauge transformations and total derivatives, we have that $\bar{Q}_{\Omega}^{2}=0$. Thus $\bar{Q}_{\Omega}$ is nilpotent and a valid BRST operator.

Since the action is $\bar{Q}_{\Omega}$-exact, we can once again localise the path integral defining the partition function. Recall that the action is 
$
\frac{1}{8\pi}\Im \int d^{4}x d^{4}\theta \tau \text{Tr}\Psi^{2}
$
, 
so the path integral localises to the zero modes of $\Psi$. At low energies the effective action \eqref{NRaction} is valid, so \cite{Shadchin}:
\begin{equation}
Z_{\Omega}(a)
=
\int_{\text{Zero modes}}\D Xe^{S[X]}
=
\exp
\left\{
\Im
\left(
\frac{1}{8\pi i}\int d^{4}x d^{4}\theta \F\left(-\frac{1}{2\sqrt{2}}a;\Lambda(x,\theta)\right)
\right)
\right\}\label{Zomega}
\end{equation}
where we have used that zero modes correspond to all non-scalar fields vanishing and the Higgs field being given by the constant $a$ as discussed in Section $2.1$.
\section{The Prepotential}
To access the prepotential, the superspace integral in \eqref{Zomega} must be evaluated. To this end we define another operator:
$$
\R_{\Omega}
:=
\theta^{\mu}\del_{\mu}
+
\frac{1}{2\sqrt{2}}\Omega^{\mu}_{\nu}x^{\nu}\frac{\del}{\del \theta^{\mu}}
$$
The key observation is that $\tau(x,\theta)$ is annihilated by this operator:
\begin{align*}
-4\R_{\Omega}\tau(x,\theta)
=
\Omega^{\mu}_{\nu}x^{\nu}\frac{\del}{\del\theta^{\mu}}
(
(\bar{\Omega}_{\rho\sigma})^{+}\theta^{\rho}\theta^{\sigma}
)
-
\theta^{\sigma}\del_{\sigma}
(
\bar{\Omega}_{\mu\nu}\Omega^{\mu}_{\rho}x^{\rho}x^{\nu}
)
=
0
\end{align*}
Recall that the renormalization group equation relates the complex coupling constant to the dynamically generated scale: $\Lambda=e^{\frac{2\pi i}{\beta}(\tau-\tau_{0})}$. So in the $\Omega$-background, $\Lambda$ is effectively superspace dependent and $\R_{\Omega}\Lambda(x,\theta)=0$. Furthermore $\R_{\Omega}$ is nilpotent up to total derivatives and gauge transformations \cite{Shadchin}.

Since $a$ is a constant and $\Lambda(x,\theta)$ is $\R_{\Omega}$-closed, we see that the integrand on the right hand side of \eqref{Zomega} is $\R_{\Omega}$-closed. A slight generalization of the localisation formula derived in Section $3.3$ can thus be applied. To do so, we fix a system of coordinates where $\Omega$ has the following canonical form:
\begin{equation}
\Omega
=
\frac{1}{\sqrt{2}}
\begin{pmatrix}
0&0&0&-\epsilon_{1}\\
0&0&-\epsilon_{2}&0\\
0&\epsilon_{2}&0&0\\
\epsilon_{1}&0&0&0\\
\end{pmatrix}
\end{equation}
In the above expression, the terms$\epsilon_{1}$ and $\epsilon_{2}$ measure the deformation of spacetime and will eventually be sent to zero to recover $\N=2$ SYM on undeformed spacetime.

The infinitesimal torus action is given by $\Omega$, yielding in the language of Section $3.3$, $Pf(-L_{V}(0))=\frac{1}{2}\epsilon_{1}\epsilon_{2}$. The only fixed point is $x=\theta=0$, so:
\begin{equation}
\Im
\left(\frac{1}{8\pi i}\int d^{4}x d^{4}\theta \F\left(-\frac{1}{2\sqrt{2}}a;\Lambda(x,\theta)\right)
\right)
=
\frac{1}{\epsilon_{1}\epsilon_{2}}\F(a,\Lambda|_{0};\epsilon)
\end{equation}
Where we have used that $\F$ is homogeneous of degree 2, and where $\F(a,\Lambda|_{0};\epsilon)$ is the Lorentz deformed prepotential.
\begin{remark}{5.1}
By assumption, the Lorentz deformed prepotential is related to the usual prepotential by:
\begin{equation}
\F(a,\Lambda)=\lim_{\epsilon_{1,2}\rightarrow 0}\F(a,\Lambda|_{0};\epsilon)
\end{equation}
\end{remark}
The prepotential can now be evaluated! We equate the Lorentz deformed and undeformed partition functions, use the ADHM construction to evaluate the undeformed partition function, then take the $\epsilon_{1,2}\to 0$ limit:
\begin{equation}
\lim_{\epsilon_{1,2}\rightarrow 0}e^{\frac{1}{\epsilon_{1}\epsilon_{2}}\F(a,\Lambda|_{0};\epsilon)}
=
\sum_{k=0}^{\infty}e^{2\pi ik\tau}\int_{\mathfrak{m}_{k}}\mathbbm{1}
=
\sum_{k=0}^{\infty}e^{2\pi ik\tau}\text{Vol}(\mathfrak{m}_{k})
\label{prepot}
\end{equation}
For later convenience, define $Z_{k}(a)=\text{Vol}(\mathfrak{m}_{k})$.
\chapter{Explicit Calculation of the Partition Function}
With the ADHM construction at hand, the integrals \eqref{prepot} over the instanton moduli space can be explicitly evaluated. This process will yield the right hand side of \eqref{prepot}, and thus the prepotential. Directly evaluating the prepotential in this way gives a power series which will agree order by order with that derived from Seiberg Witten theory.
\section{Transformation Properties}
Recall that the $k$-instanton moduli space $\mathfrak{m}_{k}$ is the space of linear operators $I,J,B_{1,2}$ \eqref{ADHMoperators} satisfying the ADHM equations \eqref{ADHMeqns}, modulo the dual group transformations \eqref{resid}. To deal with the fact that the operators $I,J,B_{1,2}$ must satisfy the ADHM equations we introduce two supplementary multiplets which act as Lagrange multiplers:
$$
(\chi_{\mathbb{R}},H_{\mathbb{R}}),\quad (\chi_{\mathbb{C}},H_{\mathbb{C}})
$$
The transformation properties of the ADHM fields under the torus action on $\mathbb{R}^{3,1}$ will be required. They are obtained from the transformation properties of the position vector $x^{\mu}$. Recalling that $\Omega=\frac{1}{\sqrt{2}}(\Omega_{4}+i\Omega_{5})$: 
$$
x^{0}\mapsto -\epsilon_{1}x^{3}
\quad
x^{1}\mapsto -\epsilon_{2}x^{2}
\quad
x^{2}\mapsto \epsilon_{2}x^{1}
\quad
x^{3}\mapsto \epsilon_{1}x^{0}
$$
yielding for example:
$$
B_{2}=-iX^{1}+X^{2}\mapsto i\epsilon_{2}X^{2}+\epsilon_{2}X^{1}=i\epsilon_{2}B_{2}
$$
so that under finite rotations, $B_{2}\mapsto e^{i\epsilon_{2}}B_{2}$.

For the spinorial quantities $I$ and $J$, recall that spinors transform under infinitesimal rotations as 
$
\psi \mapsto e^{-\frac{i}{2}\theta}\psi
$
. So since $\mathbbm{T}^{2}\cong U(1)\times U(1)$ is abelian,
$
\psi \mapsto e^{-\frac{i}{2}(\epsilon_{1}+\epsilon_{2})}\psi
$
.

The end result is:
\begin{align}
\begin{split}
B_{1}\mapsto e^{i\epsilon_{1}}B_{1},\quad B_{2}\mapsto e^{i\epsilon_{2}}B_{2}
,\quad 
I\mapsto e^{-i\epsilon_{+}}I,\quad J\mapsto e^{-i\epsilon_{+}}J
,\quad 
\mu^{\mathbb{R}}\mapsto\mu^{\mathbb{R}},\quad\mu^{\mathbb{C}}\mapsto e^{i\epsilon}\mu^{\mathbb{C}}
\end{split}
\end{align}
where $\epsilon:=\epsilon_{1}+\epsilon_{2}$, and $\epsilon_{+}:=\frac{1}{2}\epsilon$.

The BRST transformation properties of the ADHM fields and the supplementary multiplets will also be needed. A detailed derivation would be quite off topic, so we present only a brief overview. More details can be found in \cite{Shadchin}. The transformation properties are:
\begin{align}
\begin{split}
&
\bar{Q}_{\Omega}B_{1,2}=\psi_{1,2}, \quad \bar{Q}_{\Omega}\psi_{1,2}=[\phi,B_{1,2}]+i\epsilon_{1,2}B_{1,2}
\\
&
\bar{Q}_{\Omega}I=\psi_{I}, \qquad\ \ \,\, \bar{Q}_{\Omega}\psi_{I}=\phi I-Ia-i\epsilon_{+}I
\\
&
\bar{Q}_{\Omega}J=\psi_{J}, \qquad\ \ \, \bar{Q}_{\Omega}\psi_{J}=-J\phi+aJ-i\epsilon_{+}J
\\
&
\bar{Q}_{\Omega}\chi_{\mathbb{R}}=H_{\mathbb{R}}, \qquad\, \bar{Q}_{\Omega}H_{\mathbb{R}}=[\phi,\chi_{\mathbb{R}}]
\\
&
\bar{Q}_{\Omega}\chi_{\mathbb{C}}=H_{\mathbb{C}}, \qquad\, \bar{Q}_{\Omega}H_{\mathbb{C}}=[\phi,\chi_{\mathbb{C}}]+i\epsilon\chi_{\mathbb{C}}
\\
&
\bar{Q}_{\Omega}\lambda = \eta, \qquad\quad\ \; \bar{Q}_{\Omega}\eta = [\lambda,\psi]
\end{split}
\end{align}
Where $\psi_{1,2,I,J}$ are the basis of "1-forms" corresponding to the "coordinates" $B_{1,2},I,J$, on $\mathfrak{m}_{k}$.

The $H$ and $\psi$ expressions come from considering the weights of $\bar{Q}_{\Omega}^{2}$, and that $\bar{Q}_{\Omega}$ squares to the sum of a gauge transformation with parameter $a$, and a dual group transformation with parameter $\phi$. The $\epsilon$ terms come from the torus transformations derived above.
\section{The Calculation}
In this section we finally reduce the instanton partition function to a sum of contour integrals.
\subsection{Set up.}
The terms $Z_{k}(a;\epsilon)$ of the partition function sum \eqref{prepot} are:
\begin{equation}
Z_{k}(a;\epsilon)
=
\int_{\mathfrak{m}_{k}}\mathbbm{1}
=
\frac{1}{\text{Vol}(G_{D})}\int \D\phi\D\eta\D\lambda\D H\D\chi\D B_{1}\D B_{2}\D I\D J\D\psi e^{i\bar{Q}(\chi\cdot\mu+t\chi\cdot H+\psi\cdot V(\lambda))}
\end{equation}
where we have converted an integral over the quotient space $\mathfrak{m}_{k}\cong \mu^{-1}(0)/G_{D}$ to an integral over the full space by imposing the constraints defining $\mathfrak{m}_{k}$ via Lagrange multipliers \cite{Shadchin}.

In the above expression the dotted terms are:
\begin{align*}
&
\chi\cdot\mu
=
\text{Tr}
\left(
\chi_{\mathbb{R}}\mu_{\mathbb{R}}+\frac{1}{2}(\chi_{\mathbb{C}}^{\dagger}\mu_{\mathbb{C}}+\chi_{\mathbb{C}}\mu_{\mathbb{C}}^{\dagger})
\right)
\\
&
\chi\cdot H
=
\text{Tr}
\left(
\chi_{\mathbb{R}}H_{\mathbb{R}}+\frac{1}{2}(\chi_{\mathbb{C}}^{\dagger}H_{\mathbb{C}}+\chi_{\mathbb{C}}H_{\mathbb{C}}^{\dagger})
\right)
\end{align*}
and $V(\psi)$ is the dual group flow:
$$
\psi\cdot V(\lambda)
=
\text{Tr}
\left(
\psi_{1}[\lambda,B_{1}^{\dagger}]
+
\psi_{2}[\lambda,B_{2}^{\dagger}]
+
\bar{\psi}_{1}[\lambda,B_{1}]
+
\bar{\psi}_{2}[\lambda,B_{2}]
+
\psi_{I}\lambda I
-
I^{\dagger}\lambda\bar{\psi}_{I}
-
J\lambda\bar{\psi}_{J}
+
\psi_{J}\lambda J^{\dagger}
\right)
$$
To make the computation easier, we make a choice of localisation scheme by adding another $\bar{Q}$-exact term to the action:
$$
\bar{Q}
\left(
it'\text{Tr}
(
\chi_{\mathbb{R}}\lambda
)
-
\frac{1}{2}t''\text{Tr}
\left\{
\sum_{s=1}^{2}(B_{s}^{\dagger}\psi_{s}-\bar{\psi_{s}}B_{s})
-
I^{\dagger}\psi_{I}
+
\bar{\psi}_{I}I
-
J^{\dagger}\psi_{J}
+
\bar{\psi}_{J}J
\right\}
\right)
$$

The idea is to now take $t,t',t''\rightarrow\infty$ which will integrate out various fields, eventually resulting in a sum of contour integrals.
\subsection{The $t'\rightarrow \infty$ limit.}
We first take $t'\rightarrow\infty$. To see what this does, note that
$
\bar{Q}
\text{Tr}
(
\chi_{\mathbb{R}}\lambda
)
=
\text{Tr}(H_{\mathbb{R}}\lambda-\chi_{\mathbb{R}}\eta)
$, so the exponent contains the term 
$
-t'\text{Tr}(H_{\mathbb{R}}\lambda-\chi_{\mathbb{R}}\eta)
$. We can thus choose to first localise $H_{\mathbb{R}}$ to zero, giving a delta function at $t'\text{Tr}(\lambda)$ and allowing us to integrate out $\lambda$:
\begin{align*}
Z_{k}(a;\epsilon)
&=
\frac{1}{\text{Vol}(G_{D})}\int \D\lambda\D(\ldots)\delta(t'\text{Tr}(\lambda))e^{i\bar{Q}(\chi\cdot\mu+t\chi\cdot H+\psi\cdot V(\lambda))+\ldots}
\\
&=
\frac{1}{\text{Vol}(G_{D})}\int \D\lambda\D(\ldots)\delta(\text{Tr}(\lambda))t'^{-\dim \D\lambda}e^{i\bar{Q}(\chi\cdot\mu+t\chi\cdot H+\psi\cdot V(\lambda))+\ldots}
\end{align*}
Where $t'^{-\dim \lambda}=\det(t'\lambda)^{-1}$ is the Jacobian of the bosonic field redefintion; $\lambda\mapsto t'\lambda\Rightarrow \D\lambda\mapsto\det(t'\lambda)^{-1}\D\lambda'$.

Now, $\dim\lambda=\frac{1}{2}\dim(\text{adjrep}(G_{D}))=\frac{1}{2}k^{2}$ (the factor of $1/2$ is because $\lambda$ is real), so the integral over $\lambda$ sets $\lambda=0$ and introduces a factor of $t'^{-\frac{1}{2}k^{2}}$.

In this limit the $\chi_{\mathbb{R}}$ integral localises to a delta function at $t'\text{Tr}(\eta)$, allowing us to integrate out the fermionic field $\eta$. To do so, redefine the field; $\eta\mapsto t'\eta$. Since $\eta$ is fermionic, this changes the integration measure by a factor of $\det(t')=t'^{\frac{1}{2}k^{2}}$ (as is familiar from Grassmann integrals). This leads to cancellation of the determinant factors. 

The partition function now becomes:
$$
Z_{k}(a;\epsilon)
=
\frac{1}{\text{Vol}(G_{D})}\int \D\phi\D H_{\mathbb{C}}\D\chi_{\mathbb{C}}\D B_{1}\D B_{2}\D I\D J\D\psi e^{i\bar{Q}(\chi\cdot\mu+t\chi\cdot H+\psi\cdot V(\lambda))+\ldots}\biggr|_{\lambda=\eta=H_{\mathbb{R}}=\chi_{\mathbb{R}}=0}
$$
\subsection{The $\phi$ integration measure.}
We now discuss the $\D\phi$ part of the integration measure. This is where the non-trivial part of the resulting integral comes from. Since $\phi\in \text{Lie}(G_{D})$, this integral can be reduced from the whole Lie algebra to its maximal torus by using the Weyl integral formula \cite{Hall}.

The maximal torus of the Lie group $G_{D}=U(k)$ can be parametrised as $T=\{e^{i\theta_{1}},\ldots,e^{i\theta_{k}}|\theta_{j}\in\mathbb{R}\}$, so an element of the corresponding Cartan subalgebra can be given by $\phi=\diag\{i\phi_{1},\ldots,i\phi_{k}\}$. The Weyl integral formula then reduces the integration measure:
\begin{equation}
\D\phi
\mapsto
\frac{1}{k!}\prod_{i=1}^{k}\frac{d\phi_{i}}{2\pi i}\prod_{i<j\leq k}(\phi_{i}-\phi_{j})
\end{equation}
\subsection{The $t\rightarrow \infty$ limit.}
We now take $t\rightarrow\infty$ and expand the $\bar{Q}(\chi\cdot H)$ term of the exponent to see which terms can be integrated out:
\begin{align*}
\bar{Q}(\chi\cdot H)
=
\text{Tr}
\left(
H_{\mathbb{R}}^{2}+H_{\mathbb{C}}H_{\mathbb{C}}^{\dagger}+\chi_{\mathbb{R}}[\phi,\chi_{\mathbb{R}}]+\chi_{\mathbb{C}}^{\dagger}([\phi,\chi_{\mathbb{C}}]+i\epsilon\chi_{\mathbb{C}})
\right)
=
\text{Tr}
\left(
H_{\mathbb{C}}H_{\mathbb{C}}^{\dagger}+\chi_{\mathbb{C}}^{\dagger}([\phi,\chi_{\mathbb{C}}]+i\epsilon\chi_{\mathbb{C}})
\right)
\end{align*}
The only part of the integral containing $H_{\mathbb{C}}$ is a Gaussian factor $\int\D H_{\mathbb{C}}e^{itH_{\mathbb{C}}H_{\mathbb{C}}^{\dagger}}$, which can be evaluated:
$$
\int\D H_{\mathbb{C}}\D H_{\mathbb{C}}^{\dagger}e^{itH_{\mathbb{C}}H_{\mathbb{C}}^{\dagger}}
=
e^{-\text{Tr}(\log(t\mathbbm{1}_{k^{2}}))}
=
t^{-k^{2}}
$$
Where we have used that $H_{\mathbb{C}}$ is $k^{2}\times k^{2}$ since it is in the adjoint representation of the dual group, $U(k)$. This factor will later be eliminated by integrating out the fermionic $\chi_{\mathbb{C}}$.

\begin{remark}{6.1}
The cancellation of $\det$ and $\frac{1}{\det}$ terms arising from bosonic and fermionic fields is the famous bosonic/fermionic cancellation of supersymmetry.
\end{remark}

The fields $\chi_{\mathbb{C}}$ and $\chi_{\mathbb{C}}^{\dagger}$ can now be integrated out. To do so, note that $\phi_{ij}=\sqrt{-1}\phi_{i}\delta_{ij}$ so:
\begin{align*}
it\text{Tr}
\left(
\chi_{\mathbb{C}}^{\dagger}
(
[\phi,\chi_\mathbb{C}]
+
i\epsilon\chi_{\mathbb{C}}
)
\right)
&=
it\sum_{i,j=1}^{k}
(\chi_{\mathbb{C}}^{\dagger})_{ij}
(
[\phi,\chi_\mathbb{C}]
+
i\epsilon\chi_{\mathbb{C}}
)_{ji}
\\
&=
-t
\left(
\sum_{i,j=1}^{k}(\phi_{i}-\phi_{j})\chi^{\dagger}_{ji}\chi_{ij}
+
\epsilon\sum_{i,j=1}^{k}\chi_{ji}^{\dagger}\chi_{ij}
\right)
\\
&=
-t
\left(
\sum_{i,j=1}^{k}(\phi_{i}-\phi_{j})\bar{\chi_{ij}}\chi_{ij}
+
\epsilon\sum_{i,j=1}^{k}\bar{\chi_{ij}}\chi_{ij}
\right)
\\
&=
-t
\left(
\sum_{i<j\leq k}
\left(
(\phi_{i}-\phi_{j}+\epsilon)|\chi_{ij}|^{2}
+
(\phi_{j}-\phi_{i}+\epsilon)|\chi_{ji}|^{2}
\right)
+
\epsilon\sum_{i=1}^{k}|\chi_{ii}|^{2}
\right)
\end{align*}
Since the corresponding integral is fermionic, integrating out $\chi_{ij}$ and $\chi_{ij}^{\dagger}$ gives a factor $t(\phi_{i}-\phi_{j}+\epsilon)$ for $i\neq j$, or a factor $t\epsilon$, for $i=j$. The case $i=j$ occurs $k$ times, so we have:
$$
\int\D\chi_{\mathbb{C}}^{\dagger}\chi_{\mathbb{C}}e^{it\text{Tr}
\left(
\chi_{\mathbb{C}}^{\dagger}
(
[\phi,\chi_\mathbb{C}]
+
i\epsilon\chi_{\mathbb{C}}
)
\right)}
=
t^{k^{2}}\epsilon^{k}\prod_{i<j\leq k}(\phi_{i}-\phi_{j}+\epsilon)(\phi_{j}-\phi_{i}+\epsilon)
=
t^{k^{2}}\epsilon^{k}\prod_{i<j\leq k}((\phi_{i}-\phi_{j})^{2}+\epsilon^{2})
$$

The current result is as follows:
\begin{align*}
Z_{k}(a;\epsilon)
=&
\frac{\epsilon^{k}}{k!\text{Vol}(G_{D})}\int\prod_{i=1}^{k}\frac{d\phi_{i}}{2\pi i}\D B_{1}\D B_{2}\D I\D J\D\psi
\prod_{i<j\leq k}(\phi_{i}-\phi_{j})((\phi_{i}-\phi_{j})^{2}+\epsilon^{2})
\\&
e^
{
\bar{Q}
\left(
-
\frac{1}{2}t''\text{Tr}
\left\{
\sum_{s=1}^{2}(B_{s}^{\dagger}\psi_{s}-\bar{\psi_{s}}B_{s})
-
I^{\dagger}\psi_{I}
+
\bar{\psi}_{I}I
-
J^{\dagger}\psi_{J}
+
\bar{\psi}_{J}J
\right\}
\right)
}
\end{align*}
\subsection{The $t''\rightarrow \infty$ limit.}
We now send $t''\rightarrow\infty$ to integrate out $B_{1,2},I,J$ and the corresponding fermionic fields. To do so, first note that the exponent is:
\begin{align*}
-t''
\biggr(
\bar{\psi}_{1}\psi_{1}
+
\bar{\psi}_{2}\psi_{2}
+
\bar{\psi}_{I}\psi_{I}
+
\bar{\psi}_{J}\psi_{J}
+&
\sum_{s=1}^{2}
B_{s}^{\dagger}([\phi,B_{s}]+i\epsilon_{s}B_{s})
\\
&
-
I^{\dagger}(\phi I-Ia-i\epsilon_{+}I)
-
J^{\dagger}(-J\phi+aJ-i\epsilon_{+}J)
\biggr)
\end{align*}
The $B_{s}$ terms above have the exact same form as the $\chi_{\mathbb{C}}$ terms discussed previously, so by identical workings and noting that the $B_{s}$ are complex bosonic fields, we find that the corresponding integrals yield:
$$
t''^{-2k^{2}}\prod_{s=1}^{2}\frac{1}{\epsilon_{s}^{k}}\prod_{i<j\leq k}\frac{1}{(\phi_{i}-\phi_{j})^{2}-\epsilon_{s}^{2}}
$$
The Gaussian integrals over the $\psi_{s}$ then eliminate the $t''^{-2k^{2}}$ factor.

The process for $I$ and $J$ is similar. The only difference is that this time rather than a commutator with $\phi$, an $a$ appears in the $\bar{Q}_{\Omega}$ action. Noting that $a\in\text{Lie}(SU(N))$ is a gauge parameter at infinity, we can through similar reasoning to the $\phi$ case write $a=\diag(ia_{1},\ldots,ia_{N})$. Proceeding as above, the resulting factor (after also integrating out $\psi_{I,J}$), is:
$$
\prod_{i=1}^{k}\prod_{j=1}^{N}\frac{1}{(\phi_{i}-a_{j})-\epsilon_{+}^{2}}
$$
Only the $\phi_{i}$ integrals remain!
\subsection{The result.}
To write the final result nicely, we introduce some special polynomials:
$$
\Delta_{\pm}(x)=\prod_{i<j\leq k}\left((\phi_{i}\pm\phi_{j})^{2}-x^{2}\right)
\text{, }
\quad
\mathcal{P}(x)=\prod_{i=1}^{N}(x-a_{i})
$$
The $k$'th contribution to the partition function can now be written as follows:
\begin{equation}
Z_{k}(a;\epsilon)=\int\prod_{i=1}^{k}\frac{d\phi_{i}}{2\pi i}\mathfrak{z}_{k}(a,\phi;\epsilon)
\text{ , }\ 
\mathfrak{z}_{k}(a,\phi;\epsilon)
=
\frac{1}{k!}\frac{\epsilon^{k}}{\epsilon_{1}^{k}\epsilon_{2}^{k}}\frac{\Delta_{-}(0)\Delta_{-}(\epsilon)}{\Delta_{-}(\epsilon_{1})\Delta_{-}(\epsilon_{2})}\prod_{i=1}^{k}\frac{1}{\mathcal{P}(\phi_{i}+\epsilon_{+})\mathcal{P}(\phi_{i}-\epsilon_{+})}\label{Zk}
\end{equation}
In principle, the partition function and prepotential of $\N=2$ SYM with gauge group $SU(N)$ have now been determined:
\begin{equation}
Z(a;\epsilon)=Z_{pert}(a;\epsilon)\sum_{k=0}^{\infty}e^{2\pi i k \tau}Z_{k}(a;\epsilon)=\lim_{\epsilon_{1},\epsilon_{2}\rightarrow\infty}e^{\frac{1}{\epsilon_{1},\epsilon_{2}}\F(a,\Lambda)}
\text{ , }\  
Z_{k}(a;\epsilon)=\lim_{\epsilon_{1},\epsilon_{2}\rightarrow\infty}\int\prod_{i=1}^{k}\frac{d\phi_{i}}{2\pi i}\mathfrak{z}_{k}(a,\phi;\epsilon)
\end{equation}
\begin{remark}{6.2}
The factor $\frac{1}{k!\text{Vol}(G_{D})}$ has been removed in the above by normalisation.
\end{remark}
\section{Instanton Corrections for Gauge Group $SU(2)$}
The problem of calculating the prepotential has been reduced to calculating certain contour integrals and limits. In this section we consider in detail the $SU(2)$ case. In particular, the $k=1$ contribution will be calculated exactly and the procedure for calculating higher terms will be clarified.
\subsection{Calculating $Z_{1}$.}
The integrals $Z_{k}(a;\epsilon)$ must be calculated explicitly. This is done by the standard complex analytic trick of integrating around a large semicircular contour in the complex plane and taking the radius to infinity. Naively it appears that the $\phi_{i}$ integrals pass through the poles of the $\mathfrak{z}_{k}(a,\phi;\epsilon)$, which would would lead to non-convergence. To cure this problem we introduce a small imaginary shift. Schematically; $\epsilon_{s}\rightarrow \epsilon_{s} +i0$.
\begin{remark}{6.3}
The shift $\epsilon_{s}\rightarrow \epsilon_{s} +i0$ can be derived rigorously by considering a modified contour which excises the poles using small semicircular contours. 
\end{remark}
For $k=1$ the integrand is:
$$
\mathfrak{z}_{1}(a,\phi;\epsilon)
=
\frac{\epsilon}{\epsilon_{1}\epsilon_{2}}\frac{1}{(\phi_{1}+\epsilon_{+}-a_{1})(\phi_{1}+\epsilon_{+}-a_{2})(\phi_{1}-\epsilon_{+}-a_{1})(\phi_{1}-\epsilon_{+}-a_{2})}
$$
The contour of integration is shown in Figure $6.1$ in which the poles of $\mathfrak{z}_{1}(a,\phi;\epsilon)$ are denoted by black dots. Noting that the $\mathfrak{z}_{k}(a,\phi;\epsilon)$ decay sufficiently fast at infinity, the residue theorem yields:
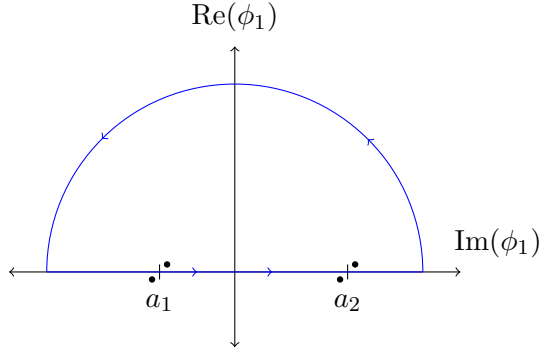
\begin{figure}
\centering
\begin{tikzpicture}
\draw[<->] (-3,0) -- (3,0);
\draw[<->] (0,-1) -- (0,3);
\draw [blue,->] (2.5,0) arc[x radius=2.5, y radius=2.5, start angle=0, end angle=45];
\draw [blue] (1.7677,1.7677) arc[x radius=2.5, y radius=2.5, start angle=45, end angle=90];
\draw [blue,->] (0,2.5) arc[x radius=2.5, y radius=2.5, start angle=90, end angle=135];
\draw [blue] (-1.7677,1.7677) arc[x radius=2.5, y radius=2.5, start angle=135, end angle=180];
\draw[blue,->] (-2.5,0) -- (-0.5,0);
\draw[blue,->] (-0.5,0) -- (0.5,0);
\draw[blue] (0.5,0) -- (2.5,0);
\node[above=1pt of {(0,3)}, outer sep=1pt] {$\Re(\phi_{1})$};
\node[above=1pt of {(3.5,0)}, outer sep=1pt] {$\Im(\phi_{1})$};
\draw (1.5,0.1) -- (1.5,-0.1);
\node[below=1pt of {(1.5,-0.1)}, outer sep=1pt] {$a_{2}$};
\draw (-1,0.1) -- (-1,-0.1);
\node[below=1pt of {(-1,-0.1)}, outer sep=1pt] {$a_{1}$};
\filldraw (1.6,0.1) circle[radius=1pt];
\filldraw (1.4,-0.1) circle[radius=1pt];
\filldraw (-0.9,0.1) circle[radius=1pt];
\filldraw (-1.1,-0.1) circle[radius=1pt];
\end{tikzpicture}
\caption{The contour of integration for $Z_{k}(a;\epsilon)$}
\end{figure}
$$
Z_{1}(a;\epsilon)
=
\frac{\epsilon}{\epsilon_{1}\epsilon_{2}}
\left(
\frac{1}{2\epsilon_{+}(a_{1}-a_{2}+2\epsilon_{+})(a_{1}-a_{2})}
+
\frac{1}{2\epsilon_{+}(a_{2}-a_{1}+2\epsilon_{+})(a_{2}-a_{1})}
\right)
=
-
\frac{1}{2\epsilon_{1}\epsilon_{2}}
\frac{1}{a^{2}-\epsilon_{+}^{2}}
$$
where $2a:=a_{1}-a_{2}$.
\subsection{Calculating $\F_{1}$.}
To calculate the corresponding term $\F_{1}(a)$ of the prepotential note that $\F_{\text{inst}}(a)=\lim_{\epsilon_{1,2}\rightarrow 0}(\epsilon_{1}\epsilon_{2}Z(a;\epsilon))$. Setting $q:=\Lambda^{2N}$ then gives:
$$
\sum_{k=1}^{\infty}\F_{k}(a)q^{k}=\lim_{\epsilon_{1,2}\rightarrow 0}\left(\epsilon_{1}\epsilon_{2}\sum_{k=0}^{\infty}q^{k}Z_{k}(a;\epsilon)\right)
$$
Viewing $q$ as a generating parameter and setting $Z_{0}(a;\epsilon)=1$, we can calculate $\F_{k}(a)$ by taking $k$ derivatives and setting $q=0$. Explicitly for $k=1$:
$$
\F_{1}(a)
+
\bigO(q)
=
\lim_{\epsilon_{1,2}\rightarrow 0}
\left(
\frac{\epsilon_{1}\epsilon_{2}}{1+\bigO(q)}(Z_{1}(a;\epsilon)+\bigO(q))
\right)
$$
So that upon setting $q=0$ we find $\F_{1}(a)=-\frac{1}{2}\lim_{\epsilon_{1,2}\rightarrow 0}\left(\frac{1}{a^{2}-\epsilon_{+}^{2}}\right)=-\frac{1}{2}\frac{1}{a^{2}}$.
\subsection{The instanton series.}
The above procedure is easily generalised to larger values of $k$ at the cost of increasingly complicated contour integrals. The general solution will be discussed in Chapter $8$. Proceeding in this way, the first three terms are \cite{Shadchin}:
$$
\F_{\text{inst}}(a)
=
-\frac{1}{2}\frac{\Lambda^{4}}{a^{2}}
-
\frac{5}{64}
\frac{\Lambda^{8}}{a^6}
-
\frac{3}{64}
\frac{\Lambda^{12}}{a^{10}}
+
\bigO(\Lambda^{16})
$$
This result is in agreement with the result of Chapter $2$.
\chapter{The Many Instanton Limit}
In the previous chapter, the prepotential was determined as a power series with terms indexed by $k$. In this form its analytic properties are not very clear. To this end we now discuss the large $k$ limit which will lead to two important results.

Firstly, the perturbative part of the prepotential (which has so far been neglected), will be recovered. Secondly, the Seiberg-Witten geometry will emerge. This will verify that the localisation approach is consistent with the Seiberg-Witten approach. In particular it will guarantee that the power series arising from the two methods agree with one another to arbitrarily high order.
\section{The $k\rightarrow\infty$ Limit}
The main contribution to $Z(a,\Lambda;\epsilon)$ comes from the region $k\sim\frac{1}{\epsilon_{1}\epsilon_{2}}$ \cite{Shadchin}. Since we take $\epsilon_{1},\epsilon_{2}\rightarrow 0$, it follows that the dominant contribution comes from the region where $k\rightarrow \infty$.

The large $k$ limit is equivalent to taking some sort of semiclassical limit,  effectively converting a problem in field theory to one in quantum mechanics. Each $\phi_{i}$ is interpreted as a particle, converting the many contour integrals to a single quantum mechanical path integral. To achieve this in practice, we must derive the Hamiltonian of the corresponding quantum mechanical system.
\section{The Hamiltonian from the Equivariant Index}
The integrand appearing in a quantum mechanical path integral generally has the form $e^{-\frac{1}{\epsilon_{1}\epsilon_{2}}H}$, so the integrand $\mathfrak{z}_{k}$ of $Z_{k}$ should be converted to a similar form. From general considerations, it turns out that the relevant Hamiltonian is given by  \cite{Shadchin}:
\begin{equation}
H
=
-\lim_{\epsilon_{1}\epsilon_{2}\rightarrow 0}\sum_{\alpha}\epsilon_{\alpha}\log\biggr|\frac{w_{\alpha}}{\Lambda}\biggr|
\end{equation}
where the $w_{\alpha}$ are the weights of a certain torus action, and the $\epsilon_{\alpha}$ are $\pm 1$ depending on whether the coordinates from which the corresponding weights come are bosonic or fermionic.
\subsection{Weights from $\text{Ind}_{q}$.}
The weights $w_{\alpha}$ are obtained from the equivariant index of the Dirac operator:
$$
\text{Ind}_{q}=\sum_{\alpha}\epsilon_{\alpha}e^{w_{\alpha}}
$$
For the case of $SU(N)$ in the adjoint representation, this index is:
\begin{align}
\begin{split}
\text{Ind}_{q}
&=
\frac{1}{(e^{i\epsilon_{1}}-1)(e^{i\epsilon_{2}}-1)}
\left(
N+\sum_{l\neq m}^{N}e^{ia_{l}-ia_{m}}
\right)
\\
&
\qquad
-
\sum_{i=1}^{k}\sum_{l=1}^{N}(e^{i\phi_{i}-i\epsilon_{+}-ia_{l}}+e^{-i\phi_{i}+ia_{l}-i\epsilon_{+}})
+
k(1-e^{-i\epsilon_{1}})(1-e^{-i\epsilon_{2}})
\\
&
\qquad
+
\sum_{i\neq j}^{k}(e^{i\phi_{i}-i\phi_{j}}+e^{i\phi_{i}-i\phi_{j}-i\epsilon_{1}-i\epsilon_{2}}-e^{i\phi_{i}-i\phi_{j}-i\epsilon_{1}}-e^{i\phi_{i}-i\phi_{j}-i\epsilon_{2}})\label{indq}
\end{split}
\end{align}
The above result is derived using the equivariant Atiyah-Singer index theorem and known results about Chern classes \cite{Nekrasov:1, Shadchin}.

The weights can be extracted by using the following integral transform to convert $e^{w_{\alpha}}$ to $\log(w_{\alpha})$:
\begin{equation}
f(x)
\mapsto
\frac{d}{ds}\biggr|_{s=0}\frac{\Lambda^{s}}{\Gamma(s)}\int_{0}^{\infty}\frac{dt}{t}t^{s}f(itx)
\end{equation}
which is closely related to the Mellin transform. In the case at hand we have:
$$
e^{iw_{\alpha}}
\mapsto
\frac{d}{ds}\biggr|_{s=0}\frac{\Lambda^{s}}{\Gamma(s)}\int_{0}^{\infty}\frac{dt}{t}t^{s}e^{-tw_{\alpha}}
=
\log
\left|
\frac{w_{\alpha}}{\Lambda}
\right|
$$
Approximating this integral in the small $\epsilon$ limit then allows the weights $w_{\alpha}$ to be extracted.
\subsection{Extracting the weights.}
To determine the contribution to $H$ from the first line of \eqref{indq}, define:
$$
\gamma_{\epsilon_{1}\epsilon_{2}}(a_{l}-a_{m},\Lambda)
:=
\frac{d}{ds}\biggr|_{s=0}\frac{\Lambda^{s}}{\Gamma(s)}\int_{0}^{\infty}\frac{dt}{t}t^{s}
\frac{e^{-t(a_{l}-a_{m})}}{(e^{-t\epsilon_{1}}-1)(e^{-t\epsilon_{2}}-1)}
$$
And expand to $\bigO(\frac{1}{\epsilon_{1}\epsilon_{2}})$:
\begin{equation}
\begin{aligned}[b]
\gamma_{\epsilon_{1}\epsilon_{2}}(a_{l}-a_{m},\Lambda)
&=
\frac{1}{\epsilon_{1}\epsilon_{2}}\frac{d}{ds}\biggr|_{s=0}\frac{\Lambda^{s}}{\Gamma(s)}\int_{0}^{\infty}t^{s-3}e^{-t(a_{l}-a_{m})}dt
+\bigO(1)
\\
&=
\frac{1}{\epsilon_{1}\epsilon_{2}}\frac{d}{ds}\biggr|_{s=0}\frac{\Lambda^{s}}{\Gamma(s)}(a_{l}-a_{m})^{2-s}\Gamma(s-2)
+\bigO(1)
\\
&=
\frac{1}{\epsilon_{1}\epsilon_{2}}\frac{d}{ds}\biggr|_{s=0}\frac{\Lambda^{s}}{(s-1)(s-2)}(a_{l}-a_{m})^{2-s}
+\bigO(1)
\\
\Rightarrow
\gamma_{\epsilon_{1}\epsilon_{2}}(a_{l}-a_{m},\Lambda)
&=
\frac{1}{\epsilon_{1}\epsilon_{2}}
\frac{(a_{l}-a_{m})^{2}}{2}
\left(
\log\biggr|\frac{\Lambda}{a_{l}-a_{m}}\biggr|+\frac{3}{2}
\right)
+\bigO(1)
\end{aligned}
\end{equation}
The first line of \eqref{indq} thus gives a contribution of $\sum_{l\neq m }\frac{1}{2}(a_{l}-a_{m})^{2}\left(\log\biggr|\frac{a_{l}-a_{m}}{\Lambda}\biggr|-\frac{3}{2}\right)$ to the Hamiltonian.
\begin{remark}{7.1}
This contribution to $H$ is the $SU(N)$ generalisation of the perturbative part of the prepotential as determined using Seiberg-Witten theory in Section $2.4.1$
\end{remark}
The contribution to $H$ from each term in the last line of \eqref{indq} is $-\epsilon_{1}\epsilon_{2}\frac{1}{(\phi_{i}-\phi_{j})^{2}}+\ldots$, as is easily verified using the following identity:
$$
f(0)+f(\epsilon_{1}+\epsilon_{2})-f(\epsilon_{1})-f(\epsilon_{2})=-\epsilon_{1}\epsilon_{2}f''(0)+\ldots
$$
Similarly the term $k(1-e^{-i\epsilon_{1}})(1-e^{-i\epsilon_{2}})$ from the second line of \eqref{indq} gives a contribution $-k\epsilon_{1}\epsilon_{2}$ to $H$.

The final term of \eqref{indq} is rather simple as well. Since it is already in an exponential form we can read off the contribution right away:
$$
\sum_{i=1}^{k}\sum_{l=1}^{N}
\left(
\log
\left(
\frac{(\phi_{i}-a_{l})-\epsilon_{+}}{\Lambda}
\right)
+
\log
\left(
\frac{-(\phi_{i}-a_{l})-\epsilon_{+}}{\Lambda}
\right)
\right)
=
2\sum_{l=1}^{N}
\log
\left(
\biggr|\frac{\mathcal{P}(\phi_{i})}{\Lambda^{N}}\biggr|
\right)
$$
where for the last equality we have taken $\epsilon_{+}\rightarrow 0$.

The full Hamiltonian has now been determined:
$$
H
=
-
\sum_{l\neq m }\frac{1}{2}(a_{l}-a_{m})^{2}\left(\log\biggr|\frac{a_{l}-a_{m}}{\Lambda}\biggr|-\frac{3}{2}\right)
+
2\epsilon_{1}\epsilon_{2}
\sum_{i=1}^{k}
\log
\left(
\biggr|\frac{\mathcal{P}(\phi_{i})}{\Lambda^{N}}\biggr|
\right)
+
(\epsilon_{1}\epsilon_{2})^{2}
\sum_{i\neq j}\frac{1}{(\phi_{i}-\phi_{j})^{2}}
$$
\subsection{Particle densities and the profile function.}
As $k\rightarrow \infty$ the number of $\phi_{i}$ fields becomes infinite and it makes sense to instead work with particle densities. To this end we introduce the density function:
$$
\rho(x)
=
\epsilon_{1}\epsilon_{2}
\sum_{i=1}^{k}\delta(x-\phi_{i})
$$
The Hamiltonian can now be rewritten:
$$
H
=
-
\sum_{l\neq m }\frac{1}{2}(a_{l}-a_{m})^{2}\left(\log\biggr|\frac{a_{l}-a_{m}}{\Lambda}\biggr|-\frac{3}{2}\right)
+
2\sum_{l=1}^{N}
\int dx \rho(x)
\log
\left(
\biggr|\frac{x-a_{l}}{\Lambda}\biggr|
\right)
+
\fint_{x\neq y}dxdy
\frac{\rho(x)\rho(y)}{(x-y)^{2}}
$$
where the bar denotes a principal value integral \cite{NO:1}.

We now introduce the so-called profile function $f(x)$:
$$
f(x)
=
-2\rho(x)
+\sum_{l=1}^{N}|x-a_{l}|
$$
\begin{remark}{7.2}
The profile function will later be the key to connecting the partition function to integer partitions and Young diagrams.
\end{remark}
Upon integration by parts, one can show that the Hamiltonian can be written in a very nice form:
\begin{equation}
H[f]
=
-\frac{1}{4}\int dxdy
f''(x)f''(y)k_{\Lambda}(x-y)
\end{equation}
where $k_{\Lambda}(x):=\epsilon_{1}\epsilon_{2}\gamma_{\epsilon_{1}\epsilon_{2}}(x,\Lambda)$. This expression is tough to dream up, but straightforward to verify.

The deformed partition function in the many instanton limit can now be expressed as follows:
\begin{equation}
Z(a,\Lambda;\epsilon)
\sim
\int\D f
e^{-\frac{1}{\epsilon_{1}\epsilon_{2}}H[f]}\label{pathint}
\end{equation}
\section{Lagrange Multipliers and the Space of  Profile Functions}
To analyse the path integral \eqref{pathint} we nee to understand the space of profile functions $f$. The profile function was defined in terms of $\rho$ which is itself dependent on the particles $\phi_{i}$. The first goal of this section is to derive some easier to analyse conditions which also fix $f$. We will then introduce Lagrange multipliers to enforce these conditions, thus allowing \eqref{pathint} to be replaced with an unconstrained path integral.

For future convenience we make several remarks on the parameters $\{a_{1},\ldots,a_{\text{rank}(G)}\}$. These parameters are Cartan subalgebra elements of $\text{Lie}(G)$, so for the $SU(N)$ case there are $N-1$ of them which sum to some non-zero matrix. If an extra parameter $a_{N}:=-(a_{1}+\ldots+a_{N-1})$ is introduced, then only the differences $a_{1}-a_{2},\ldots,a_{N-1}-a_{N}$ matter and $\sum_{l=1}^{N}a_{l}=0$. From now on ths will be assumed to be the case.

Following \cite{NO:1, Shadchin}, note that by definition $\rho(x)$ is only supported on the compact set $\{\phi_{i}\}_{i=1}^{k}$, so $f(x)\sim N|x|$ as $x\rightarrow \infty $. This implies that $f$ satisfies $f(+\infty)=f(-\infty)$. Also note that $f''(x)=2\sum_{l=1}^{N}\delta(x-a_{l})-2\rho''(x)$, so $\int_{-\infty}^{\infty} f''(x)dx=2N$ since the $a_{l}$ are distinct.

In a similar way, the first two moments of $f''(x)$ can be determined. Explicitly $\int_{-\infty}^{\infty}xf''(x)dx=2\sum_{l=1}^{N}a_{l}=0$, and $\int_{-\infty}^{\infty}x^{2}f''(x)dx=2\sum_{l=1}^{N}a_{l}^{2}-4\epsilon_{1}\epsilon_{2}k$. In fact, since the $a_{l}$ are distinct and $f''(x)$ has compact support, there exist intervals $[\alpha_{l}^{-},\alpha_{l}^{+}]$ containing a single $a_{l}$, so that:
\begin{equation}
\int_{\alpha_{l}^{-}}^{\alpha_{l}^{+}}xf''(x)dx=2a_{l}
\end{equation}
It turns out that it is sufficient to only enforce this condition in the path integral \eqref{pathint} \cite{Shadchin}.

We now introduce the Lagrange multipliers $\xi_{l}$ and consider the following modified Lagrangian:
\begin{equation}
L[f,\xi]
:=
H[f]+\sum_{l=1}^{N}\xi_{l}
\left(
\frac{1}{2}\int_{\alpha_{l}^{-}}^{\alpha_{l}^{+}}xf''(x)dx-a_{l}
\right)
=
S[f,\xi]-\sum_{l=1}^{N}\xi_{l}a_{l}
\end{equation}
where $S[f,\xi]:=H[f]+\frac{1}{2}\sum_{l=1}^{N}\xi_{l}\int_{\alpha_{l}^{-}}^{\alpha_{l}^{+}}xf''(x)dx$.
\begin{remark}{7.3}
Physically, the Lagrange multipliers $\xi_{l}$ can be interpreted as dual charges to the $a_{l}$.
\end{remark}
\section{Extremising the Action}
One can show that in the $\epsilon_{1,2}\rightarrow 0$ limit the functional $L[f,\xi]$ has a unique extremum \cite{NO:1}. The extremising $f$ corresponds to the limiting partition profile shape, and is closely related to the Seiberg-Witten curve. This extremiser will be constructed explicitly.
\subsection{The surface tension function.}
Given the unique extremiser $f_{*}(x)$ of $L[f,\xi]$, we will also need the stationary points with respect to the $\xi_{i}$. Solving $\frac{\del L[f_{*},\xi]}{\del \xi_{i}}=0$ easily gives the following condition for the stationary points:
$$
\frac{\del S[f_{*},\xi_{l}]}{\del \xi_{l}}=a_{l}
$$
The idea now is to solve this system for the $\xi_{l}$ in terms of the $a_{l}$ and substitute back into $H$ to get the value of the Hamiltonian at the extremiser. There is however, an issue. We need $N$ independent equations to solve for the $a_{l}$ but since $\sum_{l=1}^{N}a_{l}=0$, at most $N-1$ of these equations are independent. This issue can be resolved by introducing the surface tension function $\sigma$.

Recalling that $f(x)\sim N|x|$ as $x\rightarrow \infty$, we have that $f'(+\infty)=N$ and $f'(-\infty)=-N$. Similarly since $\int_{\alpha_{l}^{-}}^{\alpha_{l}^{+}}f''(x)=2$, we have that $f'(\alpha_{l}^{+})=2+f'(\alpha_{l}^{-})$. It can be assumed without loss of generality that $\alpha_{0}^{-}=-\infty$, $\alpha_{N}^{+}=+\infty$, and $\alpha_{l+1}^{-}=\alpha_{l}^{+}$ (while possibly deleting some isolated points such as the $\phi_{i}$). With this in mind it is clear that for $x\in [\alpha_{l}^{-},\alpha_{l}^{+}]$, $f'(x)\in[-N+2(l-1),-N+2l]$. This observation motivates the following definition of the surface tension function.

The surface tension, $\sigma:[-N,N]\rightarrow \mathbb{R}$ is defined to be the unique continuous piecewise linear function such that:
\begin{align}
\begin{split}
&
\sigma'(t)=\xi_{l}\text{, for }t\in[-N+2(l-1),-d+2l]
\\
&
\sigma(-N)+\sigma(N)=0
\end{split}
\end{align}
Upon integration the first condition gives several linear functions at various vertical positions. Demanding continuity yields a system of $N-1$ equations for $N$ unknowns and the final condition gives the $N$'th equation needed to solve the system. For example, if $(\xi_1,\xi_2,\xi_3,\xi_4)=(4,1,-2,-3)$, these conditions are easily solved to yield:
\begin{align*}
\sigma(t)=4t+16\text{, on }[-4,-2]
\text{ , }\quad
\sigma(t)=t+10\text{, on }[-2,0]
\\
\sigma(t)=-2t+10\text{, on }[0,2]
\text{ , }\quad
\sigma(t)=-3t+12\text{, on }[2,4]
\end{align*}
The second term of the action functional $S[f,\xi]$ can now be rewritten in a nicer way. Working backwards we have:
\begin{align*}
\fint_{-\infty}^{\infty}\sigma(f'(x))dx
&=
\fint_{-\infty}^{\infty}1\sigma(f'(x))dx
\\
&=
\lim_{a\rightarrow \infty}
\left[
x\sigma(f(x))
\right]\biggr|_{-a}^{a}
-
\int_{-\infty}^{\infty}x\sigma'(f'(x))f''(x)dx
\\
&=
\lim_{a\rightarrow \infty}
\left(
a(\sigma(N)+\sigma(-N))
\right)
-
\sum_{l=1}^{N}
\int_{\alpha_{l}^{-}}^{\alpha_{l}^{+}}\sigma'(f'(x))xf''(x)dx
\\
&=
-
\sum_{l=1}^{N}
\xi_{l}\int_{\alpha_{l}^{-}}^{\alpha_{l}^{+}}xf''(x)dx
\end{align*}
We thus have:
$$
\frac{1}{2}\sum_{l=1}^{N}
\xi_{l}\int_{\alpha_{l}^{-}}^{\alpha_{l}^{+}}xf''(x)dx
=
-\frac{1}{2}\fint_{-\infty}^{\infty}\sigma(f'(x))dx
$$
This implies that $\sum_{l=1}^{N}\xi_{l}=0$, and thus the $\xi_{l}$ are in principle all fixed \cite{Shadchin}.

We seek to extremise the Lagrange functional $L[f,\xi]$. To do so we extremise the action functional:
\begin{equation}
S[f,\xi]
=
-\frac{1}{4}\int dxdy
f''(x)f''(y)k_{\Lambda}(x-y)
-\frac{1}{2}\fint_{-\infty}^{\infty}\sigma(f'(x))dx
\end{equation}
Since $\sigma$ is not smooth there will be two cases. One for points of continuity of $\sigma '$ and one for its discontinuities.
\subsection{Variation at points of continuity.}
At a point of continuity of $\sigma'$ we vary $S[f,\xi]$ with respect to $f'(x)$ to obtain the corresponding Euler Lagrange equation:
\begin{align*}
S[f',\xi]
&\rightarrow
-\frac{1}{4}\int dxdy
(f''(x)+\del_{y}\delta f'(x))(f''(y)+\del_{y}\delta f'(y))k_{\Lambda}(x-y)
-\frac{1}{2}\fint_{-\infty}^{\infty}\sigma(f'(x)+\delta f'(x))dx
\\
&=
-\frac{1}{4}\int dxdy
(f''(x)f''(y)+2f''(y)\del_{x}\delta f'(x))k_{\Lambda}(x-y)
-\frac{1}{2}\fint_{-\infty}^{\infty}\sigma(f'(x))+\sigma'(f'(x))\delta f'(x)dx
\\
&=
S[f',\xi]
-\frac{1}{2}
\int dx
\left(
\int \del_{x}(\delta f'(x))f''(y)k_{\Lambda}(x-y)dy
+
\sigma'(f'(x))\delta f'(x)
\right)
\\
&=
S[f',\xi]
+\frac{1}{2}
\int dx
\left(
\int f''(y)k_{\Lambda}'(x-y)dy
-
\sigma'(f'(x))
\right)\delta f'(x)
\end{align*} 
so that:
\begin{equation}
2\frac{\delta S[f(x),\xi]}{\delta f'(x)}
=
\int f''(y)k_{\Lambda}'(x-y)dy
-
\sigma'(f'(x))
\end{equation}
Recalling the definition of $k_{\Lambda}$, we see that a function $f(x)$ extremises $S$ at a point of continuity if and only if
$
\int f''(y)(x-y)\left(\log|\frac{x-y}{\Lambda}|-1\right)dy
=
\sigma'(f'(x))
$. For later convenience, define the following integral transform:
$$
[Xf](x)
=
\int f''(y)(x-y)\left(\log\biggr|\frac{x-y}{\Lambda}\biggr|-1\right)dy
$$
Then $f(x)$ extremises $S$ at a point of continuity if and only if $[Xf](x)=\sigma '(f'(x))$.
\subsection{Variation at the discontinuities.}
The discontinuities of $\sigma'$ occur when $\sigma(f(x))=\xi_{l}$, that is when $f'(x)\in\{-N+2l|l\in\mathbb{Z}_{[1,N-1]}\}$. At these points the left hand side of the previous calculation is unchanged, but the right hand side does change.

For $f'(x)=-N+2l$, $\sigma(-N+2l+\delta f'(x))$ cannot be determined since it depends on the sign of the variation. All that can be said is that in this case: 
$$
\sigma(-N+2l+\delta f'(x))\in \sigma(-N+2l)+\{\sigma(-N+2l-0)\delta f'(x),\sigma(-N+2l+0)\delta f'(x)\}
$$
leading to the requirement that for $f'(x)=-N+2l$, $\xi_{l}>[Xf](x)>\xi_{l+1}$.
\subsection{Equivalent conditions for a critical point.}
In summary, a function $f_{*}(x)$ is a critical point of $S[f,\xi]$ if and only if the following two conditions are met for $l=0,1,\ldots,N$:
\begin{align}
\begin{split}
&
(i).\quad [Xf_{*}](x)=\xi_{l}\text{, for }-N+2(l-1)<f_{*}'(x)<-N+2l
\\
\\
&
(ii).\quad \xi_{l}>[Xf_{*}](x)>\xi_{l+1}\text{, for }f_{*}'(x)=-N+2l
\end{split}
\label{conditions}
\end{align}
where we define $\xi_{0}=-\infty$, $\xi_{N+1}=+\infty$ and without loss of generality order $\xi_{0}<\xi_{1}<\ldots<\xi_{N+1}$.

The conditions \eqref{conditions} have a geometric interpretation which allows them to be solved quite easily. To this end, define the following function:
$$
\phi(x)
=
f_{*}'(x)
+
\frac{1}{i\pi}[Xf_{*}]'(x)
$$
We claim that $f_{*}(x)$ obeying \eqref{conditions} is equivalent to the condition that $f_{*}(x)$ maps the real line to the boundary of the slitted strip domain $\Delta$:
$$
\Delta
=
\{z||\Re(z)|<N,\Im(z)>0\}\setminus\{z|\Re(z)=-N+2l,\Im(z)\in[0,\eta_{l}]\text{, }l=1,2,\ldots,N-1\}
$$
for some $\eta_{1},\ldots,\eta_{N-1}\in\mathbb{R}$.
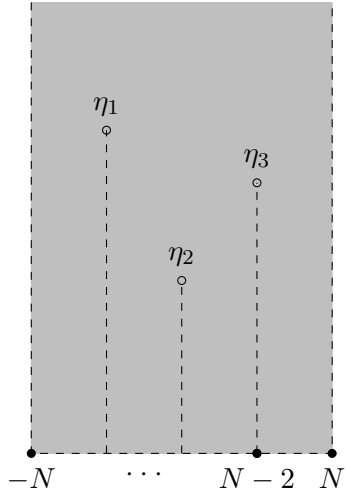
\begin{figure}
\centering
\begin{tikzpicture}
\filldraw[lightgray]  (-2,6) -- (-2,0) -- (2,0) -- (2,6);
\draw[dashed] (-2,6) -- (-2,0) -- (2,0) -- (2,6);
\filldraw (-2,0) circle[radius=1.5pt];
\node[below=1pt of {(-2,0)}, outer sep=1pt] {$-N$};
\draw[dashed] (-1,0) -- (-1,4.3);
\draw (-1,4.3) circle[radius=1.5pt];
\node[above=1pt of {(-1,4.3)}, outer sep=1pt] {$\eta_{1}$};
\draw[dashed] (0,0) -- (0,2.3);
\node[below=1pt of {(-0.5,0)}, outer sep=1pt] {$\cdot\cdot\cdot$};
\draw (0,2.3) circle[radius=1.5pt];
\node[above=1pt of {(0,2.3)}, outer sep=1pt] {$\eta_{2}$};
\draw[dashed] (1,0) -- (1,3.6);
\filldraw (1,0) circle[radius=1.5pt];
\node[below=1pt of {(1,0)}, outer sep=1pt] {$N-2$};
\draw (1,3.6) circle[radius=1.5pt];
\node[above=1pt of {(1,3.6)}, outer sep=1pt] {$\eta_{3}$};
\filldraw (1,0) circle[radius=1.5pt];
\node[below=1pt of {(2,0)}, outer sep=1pt] {$N$};
\filldraw (2,0) circle[radius=1.5pt];
\end{tikzpicture}
\caption{The domain $\Delta$ for $N=4$.}
\end{figure}

To see this, first note that in case $(i)$, $\phi(x)=f_{*}'(x)+0$, so $-N+2(l-1)<\phi(x)<-N+2l$. Thus $\phi$ maps the corresponding parts of $\mathbb{R}$ to the "gaps" between the slits of $\Delta$.

In case $(ii)$, $f_{*}'(x)=-N+2l$, so $\Re\phi(x)$ is positioned on a slit. Then $\Im \phi(x)=\frac{1}{\pi}[Xf_{*}]'(x)\geq 0$ (Since the $\xi$'s are in increasing order and $\xi_{l}>[Xf_{*}](x)>\xi_{l+1}$), and thus $\phi(x)\in \{-N+2l+i[0,\eta_{l}]|l=1,2,\ldots,N-1\}$ for some $\eta_{l}\in\mathbb{R}_{>0}$. So $\phi$ maps the corresponding sections of $\mathbb{R}$ to the slits of $\Delta$.

We have now shown that indeed $\phi:\mathbb{R}\rightarrow\del\Delta$. Each statement above is trivially reversed to show that this is equivalent to $f_{*}(x)$ obeying the conditions \eqref{conditions}. 
\begin{remark}{7.4}
Since $\xi_{0}=-\infty$ and $\xi_{N+1}=+\infty$, $\eta_{0}=\eta_{N+1}=\infty$.
\end{remark}
\section{Construction of the Extremising Profile Function and the Seiberg-Witten Differential}
Using the geometric interpretation developed in the previous section, the extremiser $f_{*}(x)$ can now be constructed explicitly. Doing so will also lead to the Seiberg-Witten differential, which will also help to show that this construction is complete.
\subsection{Construction of $f_{*}(x)$.}
We first construct $\phi(x)$ by defining a conformal map
\newline
$\Phi(z;\eta_{1},\ldots,\eta_{N-1}):\mathbb{H}\rightarrow \Delta$ and restricting back to the real line. The map $\Phi$ is guaranteed to exist by the Riemann Mapping Theorem, where the slitted strip is interpreted along with the point at infinity as a polygon on the Riemann sphere. In principle, this map can be recovered from the Schwarz-Christoffel formula of complex analysis, although the details are somewhat messy.

One can show that up to normalisation and the addition of a constant  the following map is the unique choice \cite{NO:1}:
$$
\Phi(z)=N+\frac{2}{i\pi}\log(w)
$$
where $P_{N}(z)$ is defined to be a monic polynomial of degree $N$ such that all roots of
\begin{equation}
P_{N}(z)^{2}-4\Lambda^{2N}=\prod_{l=1}^{N}(z-\alpha_{l}^{+})(z-\alpha_{l}^{-})\label{SW curve}
\end{equation}
are real, and $w$ is the smaller root of
$$
P_{N}(z)=\Lambda^{N}\left(w+\frac{1}{w}\right)
$$
We now have a map from the upper half plane to $\Delta$, but desire a map between the real line and $\del\Delta$. To this end, $\phi$ is defined as follows:
$$
\phi(x):=\Phi(x+i\epsilon)
\text{, }\quad
f_{*}'(x):=\Re\phi(x)
$$
It remains to show that the constructed $f_{*}(x)$ obeys \eqref{conditions} and that for a given set of $\xi_{l}$'s there exists a set of $\eta_{l}$'s such that $f_{*}'(x):=\Re\phi(x)$ extremises $S[f,\xi]$. To do so, we first make the connection to Seiberg-Witten theory.
\subsection{The Seiberg-Witten differential.}
Note that \ref{SW curve} defines a genus $N-1$ hyperelliptic curve, $\mathcal{C}_{u}$ which is in particular a Riemann surface. This curve will be identified with the Seiberg-Witten curve.

We introduce basic cycles $\bf{a}_{l}$ and $\bf{b}_{l}$ with unit intersection number on $\mathcal{C}_{u}$. These cycles are illusrated in Figure $7.2$ for the case $N=3$ . 
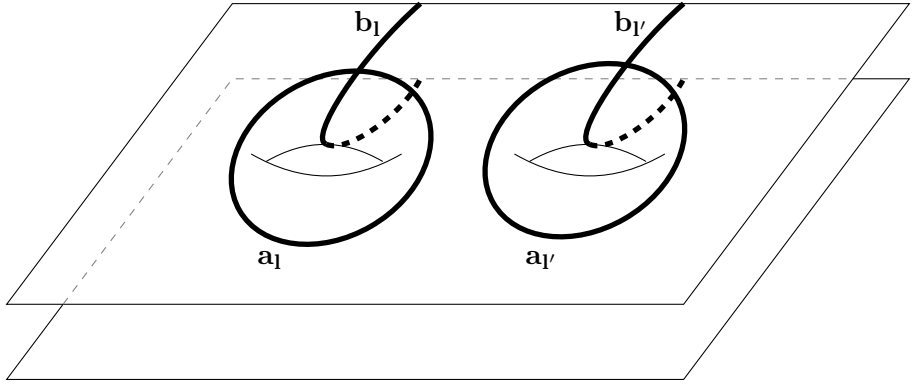
\begin{figure}
\centering
\begin{tikzpicture}
\draw (-6,-2) -- (-3,2) -- (6,2) -- (3,-2) -- cycle;
\draw (-5.25,-2) -- (-6,-3) -- (3,-3) -- (6,1) -- (5.25,1);
\draw[dashed, gray] (-5.25,-2) -- (-3,1) -- (5.25,1);
\draw[line width = 2pt, rotate around={30:(-1.6,0)}] (-1.7,0) ellipse (40pt and 30pt);
\node[below=1pt of {(-2.5,-1.1)}, outer sep=1pt] {$\bold{a_{l}}$};
\draw[line width = 2pt, rotate around={30:(1.6,0)}] (1.7,0) ellipse (40pt and 30pt);
\node[below=1pt of {(1.11,-1.1)}, outer sep=1pt] {$\bold{a_{l'}}$};
\draw (-2.55,-0.1) to[bend left] (-1.0,-0.1);
\draw (-2.75,0.0) to[bend right] (-0.75,0.0);
\draw (0.95,-0.1) to[bend left] (2.5,-0.1);
\draw (0.75,0.0) to[bend right] (2.75,0.0);
\draw[line width = 2pt] (-1.75,0.13) .. controls (-2,0.2) and (-1.3,1.3) .. (-0.5,2);
\draw[dashed, line width = 2pt] (-1.75,0.13) .. controls (-1.5,0.0) and (-0.75,0.5) .. (-0.5,1);
\node[below=1pt of {(-1.2,2.1)}, outer sep=1pt] {$\bold{b_{l}}$};
\draw[line width = 2pt] (1.75,0.13) .. controls (1.5,0.2) and (2.2,1.3) .. (3,2);
\draw[dashed, line width = 2pt] (1.75,0.13) .. controls (2,0.0) and (2.75,0.5) .. (3,1);
\node[below=1pt of {(2.3,2.1)}, outer sep=1pt] {$\bold{b_{l'}}$};
\end{tikzpicture}
\caption{The hyperelliptic curve $\mathcal{C}_{u}$ and its cycles.}
\end{figure}

Using the Schwarz reflection principle, it can be shown that \cite{NO:1, SG}:
\begin{equation}
\Phi'(z)
=
\frac{1}{i\pi}\int_{-\infty}^{\infty}\frac{f_{*}''(x)}{x-z}dx
=:
-\frac{2}{i\pi}R_{f_{*}}(z)
\end{equation}
Let $\bold{a_{l}}$ be a circular contour encircling $a_{l}$ and no other $a_{k}$ for $k\neq l$, then:
\begin{equation}
\begin{aligned}[b]
2
\oint_{\bold{a_{l}}}
zR_{f_{*}}(z)dz
=
\oint_{\bold{a_{l}}}dz\int_{\mathbb{R}}dx\frac{zf_{*}''(x)}{z-x}
=
\int_{\mathbb{R}}dxf''_{*}(x)
\left\{
    \begin{array}{ll}
      2\pi i x & \text{, }x\in\bold{a_{l}}\\ 
      0 & \text{, }x\notin\bold{a_{l}} \\
    \end{array}
  \right.
=
\pi i
\int_{\alpha_{l}^{-}}^{\alpha_{l}^{+}}dxxf''_{*}(x)
=
4\pi i a_{l}
\end{aligned}
\end{equation}
where the circular contour has been deformed to the interval in which $a_{l}$ lies.

Noting the following identity:
$$
\frac{dw}{w}
=
\frac{1}{w}\frac{dw}{dz}dz
=
\left(\frac{d}{dz}\log(w)\right)dz
=\frac{i\pi}{2}\Phi'(z)dz
$$
the arbitrary $N$ Seiberg-Witten differential $dS=\frac{1}{2\pi i}z\frac{dw}{w}$ can be rewritten as  $dS=\frac{1}{2\pi i}zR_{f_{*}}(z)$, and thus:
\begin{equation}
a_{l}=\oint_{\bold{a_{l}}}dS
\end{equation}
in agreement with the Seiberg-Witten approach for $N=2$, and now generalised to arbitrary $N$.

To complete the connection to Seiberg-Witten theory we also need the dual quantities $a_{D,l}=\frac{\del\F}{\del a_{l}}$, which will allow us to fix $\F$. The $a_{D,l}$ were originally defined as the Legendre transform of the $a_{l}$'s. Inverting the Legendre transform, we find that as expected \cite{Shadchin}:
\begin{equation}
a_{D,l}=\frac{\del\F}{\del a_{l}}=2\pi i\oint_{\bold{b_{l}}}dS=\xi_{l}
\end{equation}
\subsection{Verification of the extremising property of $f_{*}(x)$.}
The map $\phi$ is easily shown to have the following properties: $\Im\phi(x)\geq 0$, $\Im\phi(x)=0$ on the "bands", $[\alpha_{l}^{-},\alpha_{l}^{+}]$, $\Re\phi'(z)\geq 0$ and $\Re\phi(x)=0$ on the "gaps", $(\alpha_{l}^{+},\alpha_{l+1}^{-})$.

Since $\Im\phi(z)=-\frac{1}{\pi}[Xf_{*}]'(x)$, $[Xf_{*}](x)$ is monotonically decreasing everywhere and is constant on the bands. We thus identify $\xi_{l}$ with the value of $[Xf_{*}](x)$ on the $l$'th band.

Since $\Re\phi(z)$ is constant on the gaps, on the $l$'th gap its value is 
$
\Re\phi(\alpha_{l}^{+})
$
.
Here $w=-1$, so we see that $\Re\phi(z)=-N+2l$. 

Taking into account the monotonicity properties, \eqref{conditions} have now been verified and thus $f_{*}(x)$ does indeed extremise $S[f,\xi]$.
\subsection{Completeness of the construction.}
It only remains to show that this construction is complete, that is for generic $\xi_{l}$ there must exist corresponding slit lengths $\eta_{l}$. To this end we must first recover the $\xi_{l}$.

Recalling that $[Xf_{*}]'(x)=-\pi\Im\phi(x)$, and integrating along the $l$'th gap gives:
$$
\xi_{l+1}-\xi_{l}
=
-\pi\int_{\alpha_{l}^{+}}^{\alpha_{l+1}^{-}}\Im\phi(x)dx
=
-\pi[x\Im\phi(x)]\biggr|_{\alpha_{l}^{+}}^{\alpha_{l+1}^{-}}
+
\pi\int_{\alpha_{l}^{+}}^{\alpha_{l+1}^{-}}xd\Im\phi(x)
=
0
+
\pi\int_{\alpha_{l}^{+}}^{\alpha_{l+1}^{-}}xd\Im\phi(x)
$$
Then $\Im\phi(x)=-\frac{1}{\pi}[Xf_{*}]'(x)=i(\phi(x)-f_{*}'(x))$, so as $f_{*}'(x)$ is constant on the gaps, $d\Im\phi(x)=-id\phi(x)$:
$$
\xi_{l+1}-\xi_{l}
=
-i\pi\int_{\alpha_{l}^{+}}^{\alpha_{l+1}^{-}}xd\phi(x)
=
4i\pi\int_{\alpha_{l}^{+}}^{\alpha_{l+1}^{-}}dS
=
2i\pi\oint_{\bold{b_{l}}-\bold{b_{l+1}}}dS
$$
Where the last equality is clear upon considering Figure $7.2$.

So for a choice of slit lengths we can calculate the Seiberg-Witten differential, integrate along the gaps and (since $\sum_{l=1}\xi_{l}=0$), solve the resulting system of equations to recover the corresponding $\xi_l$'s.

With the $\xi_{l}$ in hand we define the period map \cite{NO:1}:
$$
(\eta_{1},\ldots,\eta_{N-1})
\mapsto
(\xi_{1}>\ldots>\xi_{N})
$$
This map is a continuous map between open sets, and since the extremiser $f_{*}$ is unique, it is injective. Surjectivity follows from the general fact that a continuous map which maps boundaries to boundaries is surjective. This map is thus invertible, showing that for any choice of dual charges $\xi_{l}$, there exists a corresponding set of slit lengths $\eta_{l}$.
\chapter{Charged Partitions and Young Diagrams}
So far the $Z_{k}$ contour integrals have only been evaluated to low order (as in Chapter $6$), or evaluated in the large $k$ limit (as in Chapter $7$). However not only can these integrals be evaluated to arbitrary order, doing so provides a connection to integer partitions. Making this connection will also provide a nice interpretation of some of the quantities introduced in the previous chapter. In this chapter we follow \cite{Nekrasov:1}.
\section{Charged and Coloured Partitions}
To make the connection to integer partitions we must first introduce some notation. A coloured integer partition of $k\in\mathbb{Z}_{\geq 0}$ is an $N$-tuple $\vec{k}=(\bm{k}_{1},\ldots,\bm{k}_{N})$, where each $\bm{k}_{l}$ is itself an integer partition of some $k_{l}:=|\bm{k}_{l}|<k$, and $\sum_{l}k_{l}=k$. That is to say $\bm{k}_{l}=(k_{l,1}\geq k_{l,2}\geq \ldots\geq k_{l,n_{l}}>k_{l,n_{l}+1}=0=\ldots)$, such that $|\vec{k}|:=k=\sum_{l,i}k_{l,i}$.

More visually, a coloured partition consists of $N$ integer partitions. Each one can be drawn as a (possibly empty), Young diagram in the usual way, resulting in $N$ Young diagrams with $k$ boxes shared between them.

A charged partition of $k\in\mathbb{Z}_{\geq 0}$ is a set $\{k_{i}'=k_{i}+a|i\in\mathbb{Z}_{>0}\}$ of non-increasing integers such that $(k_{1}\geq k_{2}\geq\ldots\geq k_{n}>k_{n+1}=0=\ldots)$ is a partition of $k$. The integer $a$ is  called the charge.

For a given partition $\bm{k}$, the corresponding dual partition is obtained by swapping the rows and columns of the Young diagram of $\bm{k}$. It is denoted by $\widetilde{\bm{k}}:=(\nu^{1}\geq\ldots\geq\nu^{k_{1}}>0)$.
\section{The Arbitrary $k$ Residue Formula}
Recall the expression for the integrands of the partition function instanton series \eqref{Zk}:
\begin{equation}
\mathfrak{z}_{k}(a,\phi;\epsilon)
=
\frac{1}{k!}\frac{\epsilon^{k}}{\epsilon_{1}^{k}\epsilon_{2}^{k}}\prod_{i=1}^{k}\frac{1}{\mathcal{P}(\Phi_{i})\mathcal{P}(\Phi_{i}+\epsilon)}
\prod_{1\leq i<j\leq k}
\frac{\Phi_{ij}^{2}(\Phi_{ij}^{2}-\epsilon^{2})}{(\Phi_{ij}^{2}-\epsilon_{1}^{2})(\Phi_{ij}^{2}-\epsilon_{2}^{2})}
\end{equation}
where we have defined $\Phi_{i}=\phi_{i}-\epsilon_{+}$, and $\Phi_{ij}:=\Phi_{i}-\Phi_{j}$.

We seek to classify the singularities of $\mathfrak{z}_{k}$ and find their residues, thus allowing the calculation of the corresponding contour integrals.

Each singularity of $\mathfrak{z}_{k}$ is a simple pole and thus must have $\Phi_{IJ}\neq 0$, otherwise the numerator would vanish. Each such singularity comes from some $\Phi_{i}$ taking on the value $\Phi_{i}=a_{l}+\epsilon_{1}(\alpha-1)+\epsilon_{2}(\beta-1)$, for some $\alpha,\beta\in\mathbb{Z}_{\geq 0}$. This is easy to see for the $\mathcal{P}$ factors and can be seen recursively for the remaining product by noting that its poles occur at $\Phi_{i}=\Phi_{j}\pm\epsilon_{1,2}$, so each successive application of the residue formula replaces one $\Phi_{i}$ with some term of the form $a_{l}+\epsilon_{1}(\alpha-1)+\epsilon_{2}(\beta-1)$.

In fact it can be shown that the poles of $\Phi_{i}$ are in one to one correspondence with the points $a_{l}+\epsilon_{1}(\alpha-1)+\epsilon_{2}(\beta-1)$, where $0\leq\alpha\leq\nu^{l,\beta}$, and $0\leq\beta\leq k_{l,\alpha}$, i.e. box $(\alpha,\beta)$ of the $l$'th Young tableau  \cite{Nekrasov:1}. Schematically:
$$
\vec{k}
\leftrightarrow
a_{l}+\epsilon_{1}(\alpha-1)+\epsilon_{2}(\beta-1)
$$
So coloured partitions of $k$ are in one to one correspondence with the poles of $\mathfrak{z}_{k}$.

The residue corresponding to an arbitrary $\vec{k}$ is \cite{Nekrasov:1}:
\begin{align}
\begin{split}
&
\frac{1}{(\epsilon_{1}\epsilon_{2})^{k}}
\prod_{l}\prod_{\alpha=1}^{\nu^{l,1}}\prod_{\beta=1}^{k_{l,\alpha}}
\frac{\mathcal{S}(\epsilon_{1}(\alpha-1)+\epsilon_{2}(\beta-1))}{(\epsilon(\ell(s)+1)-\epsilon_{2}h(s))(\epsilon_{2}h(s)-\epsilon\ell(s))}
\\
&\quad
\times
\prod_{l<m}\prod_{\alpha=1}^{\nu^{l,1}}\prod_{\beta=1}^{k_{m,1}}
\left(
\frac
{(a_{lm}+\epsilon_{1}(\alpha-\nu^{m,\beta})+\epsilon_{2}(1-\beta))(a_{lm}+\epsilon_{1}\alpha+\epsilon_{2}(k_{l,\alpha}+1-\beta))}
{(a_{lm}+\epsilon_{1}\alpha+\epsilon_{2}(1-\beta))(a_{lm}+\epsilon_{1}(\alpha-\nu^{m,\beta})+\epsilon_{2}(k_{l,\alpha}+1-\beta))}
\right)^{2}
\end{split}
\end{align}
where we define:
$$
\mathcal{S}(x)
=
\left(
\prod_{m\neq l}(x+a_{lm})(x+\epsilon+a_{lm})
\right)^{-1}
\text{ , }\quad
\ell(s)=k_{l,\alpha}-\beta
\text{ , }\quad
h(s)=k_{l,\alpha}+\nu^{l,\beta}-\alpha-\beta+1
$$
After some simplification, we then have the following contribution from this coloured partition:
\begin{equation}
Z_{\vec{k}}(a;\epsilon_{1},\epsilon_{2})
:=
\frac{1}{\epsilon_{2}^{2N|\vec{k}|}}
\prod_{(l,i)\neq(n,j)}
\frac
{\Gamma(k_{li}-k_{nj}+\nu(j-i+1)+b_{ln})\Gamma(\nu(j-i)+b_{ln})}
{\Gamma(k_{li}-k_{nj}+\nu(j-i)+b_{ln})\Gamma(\nu(j-i+1)+b_{ln})}
\end{equation}
where $b_{ln}=\frac{a_{l}-a_{n}}{\epsilon_{2}}$, and $\nu=-\frac{\epsilon_{1}}{\epsilon_{2}}$.

Since clearly $Z_{k}=\sum_{\vec{k},|\vec{k}|=k}Z_{\vec{k}}$, we can now relabel the partition function sum to a sum over partitions $\vec{k}$, rather than instanton number $k$:
\begin{equation}
Z(a,\Lambda;\epsilon_{1},\epsilon_{2})
=
Z^{pert}(a,\Lambda;\epsilon_{1},\epsilon_{2})\sum_{k=0}^{\infty}\Lambda^{2Nk}Z_{k}(a;\epsilon_{1},\epsilon_{2})
=
Z^{pert}(a,\Lambda;\epsilon_{1},\epsilon_{2})\sum_{\vec{k}}\Lambda^{2N|\vec{k}|}Z_{\vec{k}}(a;\epsilon_{1},\epsilon_{2})
\end{equation}
\begin{remark}{8.1}
Physically we have reinterpreted the partition function as a sum over an ensemble of random coloured partitions.
\end{remark}
\section{Partitions and the Many Instanton Limit}
Relating the reformulation of the partition function in terms of Young Diagrams to the many instanton limit exposes a new interpretation of the profile function $f$, and in particular the extremiser $f_{*}$.
\subsection{Partitions and their profiles.}
We draw Young diagrams in the so-called Russian convention, that is rotated 90 degrees and arranged right to left as in Figure $8.1$.
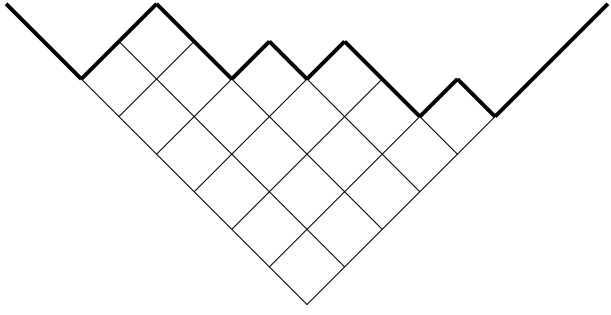
\begin{figure}
\centering
\begin{tikzpicture}
\draw (-4,4) -- (0,0) -- (4,4);
\draw(-0.5,0.5) -- (2,3);  
\draw(-1,1) -- (1,3);  
\draw(-1.5,1.5) -- (0.5,3.5);  
\draw(-2,2) -- (-0.5,3.5);  
\draw(-2.5,2.5) -- (-1.5,3.5);  
\draw(-3,3) -- (-2,4); 
\draw(-2.5,3.5) -- (0.5,0.5);  
\draw(-2,4) -- (1,1);  
\draw(-0.5,3.5) -- (1.5,1.5);  
\draw(0.5,3.5) -- (2,2);  
\draw(2,3) -- (2.5,2.5);  
\draw(3,3) -- (3,3); 
\draw [line width=1.5pt] (-4,4) -- (-3,3);
\draw [line width=1.5pt] (-3,3) -- (-2,4);
\draw [line width=1.5pt] (-2,4) -- (-1,3);
\draw [line width=1.5pt] (-1,3) -- (-0.5,3.5);
\draw [line width=1.5pt] (-0.5,3.5) -- (0,3);
\draw [line width=1.5pt] (0,3) -- (0.5,3.5);
\draw [line width=1.5pt] (0.5,3.5) -- (1.5,2.5);
\draw [line width=1.5pt] (1.5,2.5) -- (2,3);
\draw [line width=1.5pt] (2,3) -- (2.5,2.5);
\draw [line width=1.5pt] (2.5,2.5) -- (4,4);
\end{tikzpicture}
\caption{The partition $(5,4,4,3,2,2)$ in the Russian convention.}
\end{figure}

The profile $f_{\bold{k}}$ of a partition $\bold{k}$ is defined to be the piecewise linear curve forming the upper boundary of the Young Diagram corresponding to $\bold{k}$. For example, in Figure $8.1$ $f_{\bold{k}}$ is  shown in bold.

In the Russian convention the profile has a convenient expression:
$$
f_{\bold{k}}(x)
=
|x|
+
\sum_{i=1}^{\infty}
\left(
|x-k_{i}+i-1|
-
|x-k_{i}+i|
+
|x+i|
-
|x+i-1|
\right)
$$
We also define the profile of a squeezed Young diagram. A squeezed Young diagram is one for which the two axes have been scaled by constants $\epsilon_{1}$, and $\epsilon_{2}$. Explicitly:
$$
f_{\bold{k}}(x|\epsilon_{1},\epsilon_{2})
=
|x|
+
\sum_{i=1}^{\infty}
\left(
|x-\epsilon_{2}k_{i}+\epsilon_{1}(i-1)|
-
|x-\epsilon_{2}k_{i}+\epsilon_{1}i|
+
|x+\epsilon_{1}i|
-
|x+\epsilon_{1}(i-1)|
\right)
$$
It is straightforward to verify that the profile of a squeezed Young Diagram satisfies the following properties:
\begin{equation}
f_{\bold{k}}'(x|\epsilon_{1},\epsilon_{2})=\pm 1\text{, }
\quad
f_{\bold{k}}(x|\epsilon_{1},\epsilon_{2})\geq |x|\text{, }
\quad
f_{\bold{k}}(x|\epsilon_{1},\epsilon_{2})\sim |x|\text{, as } x\rightarrow\infty\label{disccons}
\end{equation}
The profile of a charged partition of charge $a$ is defined to be $f_{a;\bold{k}}(x|\epsilon_{1},\epsilon_{2}):=f_{\bold{k}}(x-a|\epsilon_{1},\epsilon_{2})$.

Finally, the profile of a general charged and coloured partition is defined as the sum of the constituent profiles:
\begin{equation}
f_{\bold{a};\vec{k}}(x|\epsilon_{1},\epsilon_{2})
=
\sum_{l=1}^{\infty}
f_{a_{l};\bold{k_{l}}}(x|\epsilon_{1},\epsilon_{2})\label{sum}
\end{equation}
Where $\vec{k}$ is a coloured partition and $\bold{a}$ is some vector of charges.
\subsection{The $\epsilon_{1,2}\rightarrow 0$ limit.}
The connection between profiles of partitions and the profile function $f(x)$ is established via the $\epsilon_{1,2}\rightarrow 0$ limit. To understand this limit, note that for a charged partition \cite{NO:1}:
\begin{equation}
|\bold{k}|
=
\frac{a^{2}}{2\epsilon_{1}\epsilon_{2}}
-
\frac{1}{4\epsilon_{1}\epsilon_{2}}\int dx f''_{a;\bold{k}}(x;\epsilon_{1},\epsilon_{2})
\end{equation}
So for $\epsilon_{1,2}\rightarrow 0$ the size of a typical partition goes as $\frac{1}{\epsilon_{1}\epsilon_{2}}$, a very large number! Thus we seek to convert the sum \eqref{sum} over discrete Young diagrams to an integral over continuous Young diagrams.

A continuous Young diagram is a continuous function $f$ satisfying the following conditions:
\begin{equation}
|f(x)-f(y)|\leq|x-y|\text{, }
\quad
\fint_{\mathbb{R}}dxf'(x)=0\text{, }
\quad
\int_{\mathbb{R}}dx(f(x)-|x|)<\infty\text{, }
\quad
f(x)\sim|x|\text{, as }x\rightarrow\infty
\end{equation}
\begin{remark}{8.2}
The above conditions are a weaker version of the conditions \eqref{disccons} for the profile of a discrete Young diagram.
\end{remark}
Associating profiles with the corresponding partitions gives a measure on the space of continuous Young Diagrams. In the $\epsilon_{1,2}\rightarrow 0$ kimit this measure concentrates to a delta measure at a single function. This function is the limiting profile shape of the random partition, it ends up being the extremised profile function $f_{*}$ derived earlier \cite{NO:1}.
\subsection{The partition function in terms of Young Diagrams.}
It can be shown from the representation theory of the symmetric group and the Plancherel measure on partitions that \cite{NO:1}:
$$
Z_{f}(\epsilon_{1},\epsilon_{2},\Lambda)
:=
Z_{\vec{\bold{k}}}(\bold{a};\epsilon_{1},\epsilon_{2},\Lambda)
=
\exp
\left(
-\frac{1}{4}\fint dxdy
f_{\bold{a},\vec{\bold{k}}}''(x|\epsilon_{1},\epsilon_{2})
f_{\bold{a},\vec{\bold{k}}}''(y|\epsilon_{1},\epsilon_{2})
\gamma_{\epsilon_{1},\epsilon_{2}}(x-y,\Lambda)
\right)
$$
where we have changed notation to associate a partition with its profile.

The partition function can now be reformulated as a sum over $\Gamma_{\bold{a}}^{discrete}$, the set of discrete paths of the form $f=f_{\bold{a},\vec{\bold{k}}}$:
\begin{equation}
Z(\bold{a};\epsilon_{1},\epsilon_{2},\Lambda)
=
\sum_{\Gamma_{\bold{a}}^{discrete}}
Z_{f}(\epsilon_{1},\epsilon_{2},\Lambda)
\end{equation}

Heuristically, taking $\epsilon_{1,2}\rightarrow 0$ turns this sum into an integral over paths of the form
$
f(x)
=
\sum_{l=1}^{N}
f_{l}(x-a_{l})
$
, where $f_{l}$ is a continuous Young Diagram. We call the set of such paths $\Gamma_{\bold{a}}$.

It is then clear that $Z_{f}\sim e^{\frac{1}{\epsilon_{1}\epsilon_{2}}H[f]}$, and:
$$
Z\sim\int_{\Gamma_{\bold{a}}}\D fe^{\frac{1}{\epsilon_{1}\epsilon_{2}}H[f]}
$$
where $H$ is the Hamiltonian as derived in Chapter $7$.

We have thus recovered the form of the large $k$ partition function which was assumed earlier on general grounds and also shown that $f_{*}$ corresponds to the profile of the limiting partition.
\chapter*{Conclusion and Further Directions}
\addcontentsline{toc}{chapter}{Conclusion and Further Directions}
This thesis began with a brief overview of supersymmetric quantum field theory. In particular we introduced the SUSY Poincar\'e algebra and its representations on superspace. These representations were then used to construct SUSY invariant actions including the action of $\N=2$ SYM. Effective SUSY theories were briefly discussed, in particular it was found that the low energy effective action of $\N=2$ SYM is completely determined by a holomorphic function $\F$ called the prepotential.

Seiberg-Witten theory was used to determine the low energy effective action of $\N=2$ SYM. This involved identifying and imposing various consistency conditions on $\F$. In particular the monodromies of the coordinates $a$ and $a_{D}$ about the singularities of $\M$ were determined. The monodromy about the weak coupling singularity was able to be analysed directly, while Seiberg-Witten duality allowed us to determine the monodromies about the strong coupling singularities.

The moduli space $\M$ was identified as the triply punctured complex plane endowed with a metric possessing known monodromies. From here, we were able to determine the metric and thus $\F$. This was done in two ways. Firstly by arguing that the coordinates on $\M$ were the solution to a certain ODE and secondly by identifying the metric with the modular parameter of a certain family of elliptic curves.

As a warm up for SUSY localisation we introduced the bosonic localisation of ordinary integrals with abelian symmetry. This lead to the idea of equivariant cohomology and the Atiyah-Bott-Berline-Vergne localisation formula for discrete fixed points. The localisation arguments for ordinary integrals were then generalised to SUSY QFTs and in particular to $\N=2$ SYM (via a topological twist). This reduced the partition function path integral of $\N=2$ SYM to an integral over the instanton moduli space. A model for the instanton moduli space was provided by the ADHM construction.

With a further modification of $\N=2$ SYM to the $\Omega$-background, we managed to reduce the partition function to a sum of contour integrals indexed by the instanton number, $k$. A second localisation argument then allowed us to determine the prepotential by taking the undeformed limit.

The many instanton limit was used to recover the Seiberg-Witten geometry from the localisation approach. This limit reduced the $\N=2$ SYM partition function to a quantum mechanical path integral. Solving the corresponding equations of motion recovered the Seiberg-Witten geometry.

We briefly explored the connection between $\N=2$ SYM and integer partitions. In particular the correspondence between coloured partitions and the poles of the $\N=2$ SYM partition function contour integrals 
was discussed. This correspondence allowed the partition function to be re-cast as a sum indexed by coloured integer partitions. Taking the many instanton limit then recovered the quantum mechanical problem discussed earlier, and showed that the profile function corresponded to the profile of the profile of the limiting partition.

A particular direction for further study would be to extend the localisation techniques utilised throughout this thesis to gauge groups other than $SU(N)$ and to theories including matter multiplets as in \cite{Shadchin}.

A less obvious direction is that of the AGT correspondence, a correspondence between certain four dimensional $\N=2$ $SU(2)$ SYM theories and Liouville theory on certain punctured Riemann surfaces \cite{AGT, Tachikawa:1}. 

The methods of topological field theory which we briefly encountered in Chapter $3$ have pure mathematical applications. In particular, topological QFTs can be used to calculate certain topological invariants and so are relevant to the study of $3$ and $4$-manifolds \cite{Atiyah:1, Witten:2, Witten:3}. Topologcal QFTs also find use in the field of knot theory \cite{Witten:4}.

Many of the arguments presented throughout this thesis have analogues in string theory and M-theory \cite{SW:1,Shadchin,Tachikawa:1}.
\begin{appendices}
\chapter{Definitions and Conventions}
In this appendix we provide definitions of some of the objects encountered throughout this thesis, as well as an overview of the various conventions used.
\section{Indices}
The conventions for various indices are as follows:
\begin{align*}
&
\bullet\text{Greek 3+1 dimensional spacetime indices }\mu,\nu,\rho\in\{0,1,2,3\}
\\
&
\bullet\text{Lower case latin indices }i,j,k\in\{1,2,3,4\}
\\
&
\bullet\text{Lower case latin gauge group Lie algebra indices }a,b,c\in\{1,2,\ldots,\text{rank}(G)\}
\\
&
\bullet\text{Capital latin supercharge indices }A,B,C\in\{1,2\}
\\
&
\bullet\text{Capital latin 6+1 dimensional spacetime indices }I,J,K\in\{0,1,\ldots,6\}
\\
&
\bullet\text{Greek undotted spinor indices }\alpha,\beta\in\{1,2\}
\\
&
\bullet\text{Greek dotted spinor indices }\dot{\alpha},\dot{\beta}\in\{\dot{1},\dot{2}\}
\end{align*}
\section{Spinors}
A Weyl spinor is is an element of a two dimensional irreducible representation of $SL(2;\mathbb{C})$. Throughout this work the word spinor will refer to a Weyl spinor.

So-called "undotted" spinors $\psi_{\alpha}$ and $\psi^{\alpha}$ belong to the fundamental and dual representations of $SL(2;\mathbb{C})$ respectively. On the other hand, "dotted" spinors $\psi_{\dot{\alpha}}$ and $\psi^{\dot{\alpha}}$ belong to the conjugate of the fundamental and dual representations respectively.

An undotted spinor is also called a left-handed spinor. A dotted spinor is also called a right-handed spinor. Objects with more than one dotted or undotted spinor index belong to tensor products of spinor representations.

Roughly speaking, the fundamental and dual representations of $SL(2;\mathbb{C})$ are related via the operations of raising andlowering indices. More precisely we have \cite{Bilal:2}:
\begin{align*}
&
\psi^{\alpha}=\epsilon^{\alpha\beta}\psi_{\beta}
\text{ , }\quad
\psi_{\alpha}=\epsilon_{\alpha\beta}\psi^{\beta}
\\
&
\psi^{\dot{\alpha}}=\epsilon^{\dot{\alpha}\dot{\beta}}\psi_{\dot{\beta}}
\text{ , }\quad
\psi_{\dot{\alpha}}=\epsilon_{\dot{\alpha}\dot{\beta}}\psi^{\dot{\beta}}
\end{align*}
where the matrices
$$
\epsilon^{\alpha\beta}
=
\epsilon^{\dot{\alpha}\dot{\beta}}
=\begin{pmatrix}
0& 1\\
-1& 0
\end{pmatrix}
\text{ , and }\quad
\epsilon_{\alpha\beta}
=
\epsilon_{\dot{\alpha}\dot{\beta}}
=\begin{pmatrix}
0& -1\\
1& 0
\end{pmatrix}
$$
act as a metric on spinor space.
\section{$\sigma$ and $\gamma$ matrices}
The Pauli matrices are:
$$
\tau_{1}
=
\begin{pmatrix}
0&1\\
1&0\\
\end{pmatrix}
\text{ , }\quad
\tau_{2}
=
\begin{pmatrix}
0&-i\\
i&0\\
\end{pmatrix}
\text{ , }\quad
\tau_{3}
=
\begin{pmatrix}
1&0\\
0&-1\\
\end{pmatrix}
$$
The matrices $i\tau_{i}$ form a basis for the Lie algebra $\mathfrak{sl}(2)$.

Also of interest are the Minkowski $\sigma$-matrices, defined in terms of the Pauli matrices as follows:
$$
(\sigma^{\mu})_{\alpha\dot{\alpha}}=(\mathbbm{1_{2}},-\tau_{1},-\tau_{2},-\tau_{2})_{\alpha\dot{\alpha}}
\text{ , }\quad
(\bar{\sigma}^{\mu})^{\dot{\alpha}\alpha}=(\mathbbm{1_{2}},\tau_{1},\tau_{2},\tau_{2})^{\dot{\alpha}\alpha}
$$
These matrices naturally have one dotted and one undotted spinor index as well as a single spacetime index \cite{Bilal:2}. The spinor indices will sometimes be suppressed.

We also define the generalised $\sigma$-matrices:
$$
(\sigma^{\mu\nu})^{\alpha}_{\beta}
=
\frac{1}{4}
\left(
\sigma^{\mu}\bar{\sigma}^{\nu}-\sigma^{\nu}\bar{\sigma}^{\mu}
\right)^{\alpha}_{\beta}
\text{ , }\quad
(\bar{\sigma}^{\mu\nu})^{\dot{\alpha}}_{\dot{\beta}}
=
\frac{1}{4}
\left(
\bar{\sigma}^{\mu}\sigma^{\nu}-\bar{\sigma}^{\nu}\sigma^{\mu}
\right)^{\dot{\alpha}}_{\dot{\beta}}
$$

The $\gamma$-matrices of Section $1.6$ are defined as follows:
$$
\gamma_{6}^{\mu}
=
\begin{pmatrix}
0&\gamma_{4}^{\mu}\\
\gamma_{4}^{\mu}&0
\end{pmatrix}
\text{ , }\qquad
\gamma_{6}^{4}
=
\begin{pmatrix}
0&\Gamma_{4}\\
\Gamma_{4}&0
\end{pmatrix}
\text{ , }\qquad
\gamma_{6}^{5}
=
\begin{pmatrix}
0&\mathbbm{1}_{4}\\
-\mathbbm{1}_{4}&0
\end{pmatrix}
$$
The above matrices are in turn defined by:
$$
\Gamma_{4}
=
\begin{pmatrix}
-i\mathbbm{1}_{2}&0\\
0&i\mathbbm{1}_{2}
\end{pmatrix}
\text{ , }\qquad
\gamma_{4}^{\mu}
=
\begin{pmatrix}
0&\sigma^{\mu}\\
\bar{\sigma}^{\mu}&0
\end{pmatrix}
$$
\chapter{Transformation Properties}
Throughout this thesis the transformation properties of fields under various operators have been utilised. In this appendix we show briefly the general method by which to obtain said transformations and list the results for easy reference.

Firstly note from the discussion of supersymmetry in Chapter $1$ that the SUSY Poincar\'e algebra has generators; $P_{\mu},J_{\mu\nu},Q^{A}_{\alpha},\bar{Q}^{\dot{\alpha}}_{A},\mathcal{Z}$, so a general element of this algebra can be written as follows:
\begin{equation}
-ia^{\mu}P_{\mu}-\frac{i}{2}\omega^{\mu\nu}J_{\mu\nu}+\zeta_{A}^{\alpha}Q^{A}_{\alpha}+\bar{\zeta}^{B,\dot{\beta}}\bar{Q}_{B,\dot{\beta}}-it\mathcal{Z}\label{genele}
\end{equation}
where $a^{\mu},\omega^{\mu\nu},\zeta^{\alpha}_{A},\bar{\zeta}^{B,\dot{\beta}}$, and $t$ are constant parameters. 

Restricting to the SUSY part of the algebra, a general SUSY transformation of a ($z$-independent), superfield is given by:
$$
\delta_{\zeta,\bar{\zeta}}F(x,\theta,\bar{\theta})
=
(\zeta_{A}Q^{A}+\bar{\zeta}^{A}\bar{Q}_{A})F(x,\theta,\bar{\theta})
$$
A representation of the SUSY Poincar\'e algebra on superspace is then given by identifying the generators with the following differential operators:
\begin{align*}
&
P_{\mu}=i\del_{\mu}
\text{, }
\quad
J_{\mu\nu}=ix_{\mu}\del_{\nu}-ix_{\nu}\del_{\mu}+S_{\mu\nu}
\text{, }
\quad
\mathcal{Z}
=
i\frac{\del}{\del z}
\\
&
Q_{\alpha}^{A}
=
\frac{\del}{\del \theta_{A}^{\alpha}}
+
i\sigma_{\alpha\dot{\beta}}^{\mu}\bar{\theta}^{A,\dot{\beta}}\del_{\mu}
+
\frac{i}{2}\epsilon_{\alpha\beta}Z^{AB}\theta^{\beta}_{B}\frac{\del}{\del z}
\text{, }
\quad
\bar{Q}_{A,\dot{\alpha}}
=
\frac{\del}{\del \bar{\theta}^{A,\dot{\alpha}}}
+
i\theta_{A}^{\beta}\sigma_{\beta\dot{\alpha}}^{\mu}\del_{\mu}
+
\frac{i}{2}\epsilon_{\dot{\alpha}\dot{\beta}}Z^{*}_{AB}\bar{\theta}^{B,\dot{\beta}}\frac{\del}{\del z}
\end{align*}
where $S_{\mu\nu}$ is the spin operator. From here it is easy to derive the coordinate transformations by acting with the general SUSY Poincar\'e algebra element \eqref{genele}:
\begin{align*}
&
\delta x^{\mu}
=
a^{\mu}
+
\omega^{\mu\nu}x_{\nu}
+
i\zeta^{\alpha}_{A}\sigma^{\mu}_{\alpha\dot{\beta}}\bar{\theta}^{A,\dot{\beta}}
-
i\theta^{\alpha}_{B}\sigma^{\mu}_{\alpha\dot{\beta}}\bar{\zeta}^{B,\dot{\beta}}
\\
&
\delta\theta^{\alpha}_{A}
=
\zeta^{\alpha}_{A}
+
\frac{1}{2}\omega^{\mu\nu}(\sigma_{\mu\nu})^{\alpha}_{\beta}\theta^{\beta}_{A}
\\
&
\delta\bar{\theta}^{A,\dot{\alpha}}
=
\bar{\zeta}^{A,\dot{\alpha}}
+
\frac{1}{2}\omega^{\mu\nu}(\bar{\sigma}_{\mu\nu})^{\dot{\alpha}}_{\dot{\beta}}\bar{\theta}^{A}_{\dot{\beta}}
\\
&
\delta z
=
t
+
\frac{i}{2}\zeta^{\alpha}_{A}\epsilon_{\alpha\beta}Z^{AB}\theta_{B}^{\beta}
+
\frac{i}{2}\bar{\zeta}^{A,\dot{\alpha}}\epsilon_{\dot{\alpha}\dot{\beta}}Z^{*}_{AB}\bar{\theta}^{B,\dot{\beta}}
\end{align*}
Proceeding similarly one can then deduce the SUSY transformations of the component fields of any superfield. We tabulate the result of this procedure for the $\N=2$ chiral multiplet:
\begin{align*}
&
\delta_{\zeta,\bar{\zeta}}H
=
\sqrt{2}\zeta_{A}\psi^{A}
\\
&
\delta_{\zeta,\bar{\zeta}}H^{\dagger}
=
\sqrt{2}\bar{\zeta}^{A}\bar{\psi}_{A}
\\
&
\delta_{\zeta,\bar{\zeta}}\psi^{A}_{\alpha}
=
(\sigma^{\mu\nu})_{\alpha}^{\beta}\zeta^{A}_{\beta}F_{\mu\nu}
+
i\zeta^{A}_{\alpha}[H,H^{\dagger}]
-
i\sqrt{2}\sigma^{\mu}_{\alpha,\dot{\beta}}\bar{\zeta}^{A,\dot{\beta}}\nabla_{\mu}H
\\
&
\delta_{\zeta,\bar{\zeta}}\bar{\psi}^{\dot{\alpha}}_{A}
=
(\bar{\sigma}^{\mu\nu,\dot{\alpha}})_{\dot{\beta}}\bar{\zeta}^{\dot{\beta}}_{A}F_{\mu\nu}
-
i\bar{\zeta}^{\dot{\alpha}}_{A}[H,H^{\dagger}]
-
i\sqrt{2}\bar{\sigma}^{\mu,\dot{\alpha}\beta}\zeta_{A,\beta}\nabla_{\mu}H^{\dagger}
\\
&
\delta_{\zeta,\bar{\zeta}}A_{\mu}
=
i\zeta^{A}\sigma_{\mu}\bar{\psi}_{A}-i\psi^{A}\sigma_{\mu}\bar{\zeta}_{A}
\end{align*}
These transformations can be used to derive the action of the twisted supersymmetry generators. For example, taking $\zeta=0$ in $\delta_{\zeta,\bar{\zeta}}H^{\dagger}$, we obtain $\bar{Q}_{A,\dot{\alpha}}=\sqrt{2}\bar{\psi}_{A,\dot{\alpha}}$, so that $\bar{Q}H^{\dagger}=\sqrt{2}\epsilon^{A,\dot{\alpha}}\bar{\psi}_{A,\dot{\alpha}}=\sqrt{2}\bar{\psi}$. The result of this procedure is tabulated below:
\begin{align*}
&
\bar{Q}H=0
\text{, }\quad
Q_{\mu}H=\sqrt{2}\psi_{\mu}
\text{, }\quad
\bar{Q}_{\mu\nu}H=0
\\
&
\bar{Q}H^{\dagger}=\sqrt{2}\bar{\psi}
\text{, }\quad
Q_{\mu}H^{\dagger}=0
\text{, }\quad
\bar{Q}_{\mu\nu}H^{\dagger}=\sqrt{2}\bar{\psi}_{\mu\nu}
\\
&
\bar{Q}\bar{\psi}=2i[H,H^{\dagger}]
\text{, }\quad
Q_{\mu}\bar{\psi}=2i\sqrt{2}\nabla_{\mu}H^{\dagger}
\text{, }\quad
\bar{Q}_{\mu\nu}\bar{\psi}=2(F_{\mu\nu})^{-}
\\
&
\bar{Q}\psi_{\rho}=2i\sqrt{2}\nabla_{\rho}H
\text{, }\quad
Q_{\mu}\psi_{\rho}=-4(F_{\mu\nu})^{+}+2ig_{\mu\rho}[H,H^{\dagger}]
\text{, }\quad
\bar{Q}_{\mu\nu}\psi_{\rho}=-2i\sqrt{2}(g_{\mu\rho}\nabla_{\nu}H-g_{\nu\rho}\nabla_{\mu}H)^{-}
\\
&
\bar{Q}\bar{\psi}_{\rho\tau}=-2(F_{\rho\tau})^{-}
\text{, }\quad
Q_{\mu}\bar{\psi}_{\rho\tau}=-2i\sqrt{2}(g_{\mu\rho}\nabla_{\tau}H^{\dagger}-g_{\mu\tau}\nabla_{\rho}H^{\dagger})^{-}
\text{, }\quad
\\
&
\bar{Q}_{\mu\nu}\bar{\psi}_{\rho\tau}=-(g_{\rho\mu}(F_{\tau\nu})^{-}-g_{\tau\mu}(F_{\rho\nu})^{-}+g_{\tau\nu}(F_{\rho\mu})^{-}-g_{\rho\nu}(F_{\tau\mu})^{-})^{-}+i(g_{\mu\rho}g_{\nu\tau}-g_{\mu\tau}g_{\nu\rho})^{-}[H,H^{\dagger}]
\\
&
\bar{Q}A_{\rho}=-i\psi_{\rho}
\text{, }\quad
Q_{\mu}A_{\rho}=-ig_{\mu\rho}\bar{\psi}-2i\bar{\psi}_{\mu\rho}
\text{, }\quad
\bar{Q}_{\mu\nu}A_{\rho}=-i(g_{\mu\rho}\psi_{\nu}-g_{\nu\rho}\psi_{\mu})^{-}
\end{align*}
As mentioned in Section $3.5$, the process of BV quantisation involved adding certain ghost fields to the theory. The action of $\bar{Q}$ on these fields is as follows \cite{Shadchin}:
\begin{align*}
&
\bar{Q}b=0
\text{, }\quad
\bar{Q}c=-\frac{i}{2}\{c,c\}-\phi
\text{, }\quad
\bar{Q}\bar{c}=b
\text{, }\quad
\bar{Q}\phi=-i[c,\phi]
\text{, }\quad
\bar{Q}\eta=-i[\phi,\lambda]-i\{c,\eta\}
\\
&
\bar{Q}\lambda=\eta-i[c,\lambda]
\text{, }\quad
\bar{Q}A_{\mu}=-\nabla_{\mu}c-i\psi_{\mu}
\text{, }\quad
\bar{Q}\psi_{\mu}=-i\nabla_{\mu}\phi-i\{c,\psi_{\mu}\}
\\
&
\bar{Q}H_{\mu\nu}=-i[\phi,\chi_{\mu\nu}]-i[c,H_{\mu\nu}]
\text{, }\quad
\bar{Q}\chi_{\mu\nu}=H_{\mu\nu}-i\{c,\chi_{\mu\nu}\}
\\
&
\end{align*}
\end{appendices}
\bibliographystyle{habbrv}
\bibliography{bib}
\end{document}